\documentclass[dvipsnames,table,longauth]{aa}
\usepackage{natbib,twoopt}
\usepackage[dvipsnames]{xcolor}
\usepackage{wasysym}
\usepackage[breaklinks=true]{hyperref} 
\usepackage{graphicx}
\usepackage{txfonts}
\usepackage{textcomp}
\usepackage{gensymb}
\usepackage{listings}
\usepackage{ulem}

\bibpunct{(}{)}{;}{a}{}{,}             
\makeatletter
\newcommandtwoopt{\citeads}[3][][]{\href{http://adsabs.harvard.edu/abs/#3}%
  {\def\hyper@linkstart##1##2{}%
    \let\hyper@linkend\@empty\citealp[#1][#2]{#3}}}
\newcommandtwoopt{\citepads}[3][][]{\href{http://adsabs.harvard.edu/abs/#3}%
  {\def\hyper@linkstart##1##2{}%
    \let\hyper@linkend\@empty\citep[#1][#2]{#3}}}
\newcommandtwoopt{\citetads}[3][][]{\href{http://adsabs.harvard.edu/abs/#3}%
  {\def\hyper@linkstart##1##2{}%
    \let\hyper@linkend\@empty\citet[#1][#2]{#3}}}
\newcommandtwoopt{\citeyearads}[3][][]%
{\href{http://adsabs.harvard.edu/abs/#3}
  {\def\hyper@linkstart##1##2{}%
    \let\hyper@linkend\@empty\citeyear[#1][#2]{#3}}}
\renewcommand*\aa@pageof{, page \thepage{} of \pageref*{LastPage}}
\makeatother
\hypersetup{colorlinks=true,linkcolor=blue,citecolor=blue,urlcolor=blue}

\providecommand{\orcit}[1]{\protect\href{https://orcid.org/#1}{\protect\includegraphics[width=8pt]{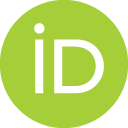}}}

\providecommand{\afe}{\ensuremath{[\alpha/\mathrm{Fe}]}\xspace}
\providecommand{\teff}{\ensuremath{T{\mathrm{_{eff}}}}\xspace} 

\providecommand{\logg}{\ensuremath{\log\,g}\xspace}
\providecommand{\lum}{\ensuremath{{L}}\xspace}

\providecommand{\loggrav}{\ensuremath{\log\,g}\xspace}
\providecommand{\mass}{\ensuremath{{M}}\xspace}
\providecommand{\radius}{\ensuremath{{R}}\xspace}

\providecommand{\mh}{\ensuremath{{[\text{M}/\text{H}]}}\xspace}

\providecommand{\azero}{\ensuremath{A_0}\xspace}

\providecommand{\ag}{\ensuremath{A_G}\xspace}
\providecommand{\bpminrp}{\ensuremath{G_\mathrm{BP}-G_\mathrm{RP}}\xspace}
\providecommand{\ebpminrp}{\ensuremath{E(G_{\rm BP} - G_{\rm RP})}\xspace}
\providecommand{\gmag}{\ensuremath{G}}
\providecommand{\bpmag}{\ensuremath{G_\mathrm{BP}}}
\providecommand{\rpmag}{\ensuremath{G_\mathrm{RP}}}
\providecommand{\mg}{M$_\gmag$}
\providecommand\gmag{\ensuremath{G}}
\providecommand\gbp{\ensuremath{G_\mathrm{BP}}}
\providecommand\grp{\ensuremath{G_\mathrm{RP}}}

\providecommand{\ra}{\ensuremath{\alpha}}
\providecommand{\dec}{\ensuremath{\delta}}

\providecommand{\plxsnr}{\ensuremath{\varpi/\sigma_\varpi}\xspace}

\providecommand{\pmra}{\ensuremath{\mu_{\ra\ast}}}
\providecommand{\pmdec}{\ensuremath{\mu_\dec}}
\providecommand{\pml}{\ensuremath{\mu_{\ell\ast}}}

\providecommand{\vtan}{\ensuremath{v_\mathrm{tan}}\xspace}
\providecommand{\vtot}{\ensuremath{v_\mathrm{tot}}\xspace}

\providecommand{\USun}{\ensuremath{U_\odot}}
\providecommand{\VSun}{\ensuremath{V_\odot}}
\providecommand{\WSun}{\ensuremath{W_\odot}}

\providecommand{\vcirc}{\ensuremath{V_\mathrm{circ}}}
\providecommand{\vcircsun}{\ensuremath{V_\mathrm{circ,\odot}}}

\providecommand{\pc}{\ensuremath{\,\rm pc}\xspace}

\providecommand{\Msun}{\ensuremath{\,{\mass}_{\odot}}\xspace}

\providecommand{\kms}{\ensuremath{\textrm{km\,s}^{-1}}\xspace}
\providecommand{\kmskpc}{\ensuremath{\textrm{km\,s}^{-1}\,\textrm{kpc}^{-1}}\xspace}
\providecommand{\maspyr}{\ensuremath{\textrm{mas\,yr}^{-1}}\xspace}
\providecommand{\msun}{\mass_\odot}

\providecommand\kms{\ensuremath{\text{\,km\,s}^{-1}}}

\providecommand\gbprpzero{\ensuremath{(G_{\rm BP} - G_{\rm RP})_0}\xspace}

\providecommand{\modulename}[1]{#1\xspace}
\providecommand{\apsis}{\modulename{Apsis}}

\providecommand{\dsc}{\modulename{DSC}}
\providecommand{\gspphot}{\modulename{GSP-Phot}}

\providecommand{\gspspec}{\modulename{GSP-Spec}}

\providecommand{\flame}{\modulename{FLAME}}
\providecommand{\espels}{\modulename{ESP-ELS}}
\providecommand{\esphs}{\modulename{ESP-HS}}

\providecommand{\espucd}{\modulename{ESP-UCD}}

\providecommand\gaia{\textit{Gaia}\xspace}

\providecommand\gdrthree{\textit{Gaia}~DR3\xspace}
\providecommand\gdr[1]{\textit{Gaia}~DR#1\xspace}
\providecommand\gedr[1]{\textit{Gaia}~EDR#1\xspace}

\definecolor{dkgreen}{rgb}{0.,0.5,0.}
\definecolor{gray}{rgb}{0.5,0.5,0.5}
\definecolor{mauve}{rgb}{0.58,0,0.82}
\definecolor{golden}{rgb}{0.86,0.65,0.01}

\providecommand{\basti}{BaSTI}

\providecommand\secref[1]{Sect.~\ref{#1}}

\providecommand\figref[1]{Fig.~\ref{#1}}

\providecommand\figrefalt[1]{Figure~\ref{#1}}

\providecommand{\secref}[1]{Sect.~\ref{#1}}

\providecommand{\linktodoc}{https://gea.esac.esa.int/archive/documentation/GDR3}
\providecommand{\linktoapparam}[2]{\href{\linktodoc/Gaia_archive/chap_datamodel/sec_dm_astrophysical_parameter_tables/ssec_dm_#1.html\##1-#2}{\fieldName{#2}\xspace}}

\providecommand{\linktogsparam}[2]{\href{\linktodoc/Gaia_archive/chap_datamodel/sec_dm_main_source_catalogue/ssec_dm_#1.html\##1-#2}{\fieldName{#2}\xspace}}

\providecommand{\linktotable}[1]{\href{\linktodoc/Gaia_archive/chap_datamodel/sec_dm_astrophysical_parameter_tables/ssec_dm_#1.html}{\fieldName{#1}\xspace}}
\providecommand{\aptable}{\linktotable{astrophysical_parameters}}

\providecommand{\apsupptable}{\linktotable{astrophysical_parameters_supp}}
\providecommand{\gstable}[1]{\href{\linktodoc/Gaia_archive/chap_datamodel/sec_dm_main_source_catalogue/ssec_dm_#1.html}{\fieldName{#1}\xspace}}
\providecommand{\varitable}[1]{\href{\linktodoc/Gaia_archive/chap_datamodel/sec_dm_variability_tables/ssec_dm_#1.html}{\fieldName{#1}\xspace}}
\providecommand{\nsstable}[1]{\href{\linktodoc/Gaia_archive/chap_datamodel/sec_dm_non--single_stars_tables/ssec_dm_#1.html}{\fieldName{#1}\xspace}}
\providecommand{\pvptable}[1]{\href{\linktodoc/Gaia_archive/chap_datamodel/sec_dm_performance_verification/ssec_dm_#1.html}{\fieldName{#1}\xspace}}
\providecommand{\ssotable}[1]{\href{\linktodoc/Gaia_archive/chap_datamodel/sec_dm_solar_system_object_tables/ssec_dm_#1.html}{\fieldName{#1}\xspace}}

\providecommand{\linksec}[2]{\href{\linktodoc/Data_analysis/chap_cu8par/#1}{#2\xspace}}
\providecommand{\linkfig}[1]{\href{\linktodoc/Data_analysis/chap_cu8par/#1}{see table\xspace}}

\makeatletter
\DeclareRobustCommand*{\fieldName}[1]{%
  \begingroup\@fieldName\scantokens{\texttt{\small {#1}}\noexpand}\endgroup}
\begingroup\lccode`\~=`\_\relax
   \lowercase{\endgroup\def\@fieldName{\catcode`\_=\active \let~\_}}
\makeatother

\makeatletter
\renewcommand*\maketitle{%
  \thispagestyle{firstpage}
\begingroup
    \if@wideboxfn
    \setlength\bibindent{1.4\parindent}
    \else
    \setlength\bibindent{\parindent}
    \fi
    \renewcommand*\thefootnote{\@fnsymbol\c@footnote}%
    \renewcommand\@makefntext[1]{%
    \ifaa@longfn\hsize\textwidth\fi
    \noindent
    \hb@xt@\bibindent{\hss\@makefnmark\enspace}##1}
  \ifaa@twocolumn
  \begingroup
    \begin{aa@strip}
          \aa@maketitle
    \end{aa@strip}
    \@thanks
  \endgroup
  \else
    \begingroup
      \let\thanks\footnote
      \aa@maketitle
    \endgroup
  \fi
\endgroup
  \setcounter{footnote}{0}%
}
\makeatother
\begin{document}

\title{\gaia\ Data Release 3: A Golden Sample of Astrophysical Parameters\thanks{This paper contains tables that are made available only through the \gaia\ archive under {\it Performance Verification}. Table 8 is only available in electronic form
at the CDS via anonymous ftp to cdsarc.u-strasbg.fr (130.79.128.5)
or via http://cdsweb.u-strasbg.fr/cgi-bin/qcat?J/A+A/}}
\titlerunning{A Golden Sample of Astrophysical Parameters}
\authorrunning{Gaia Collaboration}

\author{
{\it Gaia} Collaboration
\and       O.L.~Creevey\orcit{0000-0003-1853-6631}\inst{\ref{inst:0001}}
\and       L.M.~Sarro\orcit{0000-0002-5622-5191}\inst{\ref{inst:0002}}
\and         A.~Lobel\orcit{0000-0001-5030-019X}\inst{\ref{inst:0003}}
\and         E.~Pancino\orcit{0000-0003-0788-5879}\inst{\ref{inst:0004},\ref{inst:0005}}
\and         R.~Andrae\orcit{0000-0001-8006-6365}\inst{\ref{inst:0006}}
\and       R.L.~Smart\orcit{0000-0002-4424-4766}\inst{\ref{inst:0007}}
\and         G.~Clementini\orcit{0000-0001-9206-9723}\inst{\ref{inst:0008}}
\and         U.~Heiter\orcit{0000-0001-6825-1066}\inst{\ref{inst:0009}}
\and       A.J.~Korn\orcit{0000-0002-3881-6756}\inst{\ref{inst:0009}}
\and         M.~Fouesneau\orcit{0000-0001-9256-5516}\inst{\ref{inst:0006}}
\and         Y.~Fr\'{e}mat\orcit{0000-0002-4645-6017}\inst{\ref{inst:0003}}
\and         F.~De Angeli\orcit{0000-0003-1879-0488}\inst{\ref{inst:0013}}
\and         A.~Vallenari\orcit{0000-0003-0014-519X}\inst{\ref{inst:0014}}
\and       D.L.~Harrison\orcit{0000-0001-8687-6588}\inst{\ref{inst:0013},\ref{inst:0016}}
\and         F.~Th\'{e}venin\inst{\ref{inst:0001}}
\and         C.~Reyl\'{e}\orcit{0000-0003-2258-2403}\inst{\ref{inst:0018}}
\and         R.~Sordo\orcit{0000-0003-4979-0659}\inst{\ref{inst:0014}}
\and         A.~Garofalo\orcit{0000-0002-5907-0375}\inst{\ref{inst:0008}}
\and     A.G.A.~Brown\orcit{0000-0002-7419-9679}\inst{\ref{inst:0021}}
\and         L.~Eyer\orcit{0000-0002-0182-8040}\inst{\ref{inst:0022}}
\and         T.~Prusti\orcit{0000-0003-3120-7867}\inst{\ref{inst:0023}}
\and     J.H.J.~de Bruijne\orcit{0000-0001-6459-8599}\inst{\ref{inst:0023}}
\and         F.~Arenou\orcit{0000-0003-2837-3899}\inst{\ref{inst:0025}}
\and         C.~Babusiaux\orcit{0000-0002-7631-348X}\inst{\ref{inst:0026},\ref{inst:0025}}
\and         M.~Biermann\inst{\ref{inst:0028}}
\and         C.~Ducourant\orcit{0000-0003-4843-8979}\inst{\ref{inst:0029}}
\and       D.W.~Evans\orcit{0000-0002-6685-5998}\inst{\ref{inst:0013}}
\and         R.~Guerra\orcit{0000-0002-9850-8982}\inst{\ref{inst:0031}}
\and         A.~Hutton\inst{\ref{inst:0032}}
\and         C.~Jordi\orcit{0000-0001-5495-9602}\inst{\ref{inst:0033}}
\and       S.A.~Klioner\orcit{0000-0003-4682-7831}\inst{\ref{inst:0034}}
\and       U.L.~Lammers\orcit{0000-0001-8309-3801}\inst{\ref{inst:0031}}
\and         L.~Lindegren\orcit{0000-0002-5443-3026}\inst{\ref{inst:0036}}
\and         X.~Luri\orcit{0000-0001-5428-9397}\inst{\ref{inst:0033}}
\and         F.~Mignard\inst{\ref{inst:0001}}
\and         C.~Panem\inst{\ref{inst:0039}}
\and         D.~Pourbaix$^\dagger$\orcit{0000-0002-3020-1837}\inst{\ref{inst:0040},\ref{inst:0041}}
\and         S.~Randich\orcit{0000-0003-2438-0899}\inst{\ref{inst:0004}}
\and         P.~Sartoretti\inst{\ref{inst:0025}}
\and         C.~Soubiran\orcit{0000-0003-3304-8134}\inst{\ref{inst:0029}}
\and         P.~Tanga\orcit{0000-0002-2718-997X}\inst{\ref{inst:0001}}
\and       N.A.~Walton\orcit{0000-0003-3983-8778}\inst{\ref{inst:0013}}
\and     C.A.L.~Bailer-Jones\inst{\ref{inst:0006}}
\and         U.~Bastian\orcit{0000-0002-8667-1715}\inst{\ref{inst:0028}}
\and         R.~Drimmel\orcit{0000-0002-1777-5502}\inst{\ref{inst:0007}}
\and         F.~Jansen\inst{\ref{inst:0050}}
\and         D.~Katz\orcit{0000-0001-7986-3164}\inst{\ref{inst:0025}}
\and       M.G.~Lattanzi\orcit{0000-0003-0429-7748}\inst{\ref{inst:0007},\ref{inst:0053}}
\and         F.~van Leeuwen\inst{\ref{inst:0013}}
\and         J.~Bakker\inst{\ref{inst:0031}}
\and         C.~Cacciari\orcit{0000-0001-5174-3179}\inst{\ref{inst:0008}}
\and         J.~Casta\~{n}eda\orcit{0000-0001-7820-946X}\inst{\ref{inst:0057}}
\and         C.~Fabricius\orcit{0000-0003-2639-1372}\inst{\ref{inst:0033}}
\and         L.~Galluccio\orcit{0000-0002-8541-0476}\inst{\ref{inst:0001}}
\and         A.~Guerrier\inst{\ref{inst:0039}}
\and         E.~Masana\orcit{0000-0002-4819-329X}\inst{\ref{inst:0033}}
\and         R.~Messineo\inst{\ref{inst:0062}}
\and         N.~Mowlavi\orcit{0000-0003-1578-6993}\inst{\ref{inst:0022}}
\and         C.~Nicolas\inst{\ref{inst:0039}}
\and         K.~Nienartowicz\orcit{0000-0001-5415-0547}\inst{\ref{inst:0065},\ref{inst:0066}}
\and         F.~Pailler\orcit{0000-0002-4834-481X}\inst{\ref{inst:0039}}
\and         P.~Panuzzo\orcit{0000-0002-0016-8271}\inst{\ref{inst:0025}}
\and         F.~Riclet\inst{\ref{inst:0039}}
\and         W.~Roux\orcit{0000-0002-7816-1950}\inst{\ref{inst:0039}}
\and       G.M.~Seabroke\orcit{0000-0003-4072-9536}\inst{\ref{inst:0071}}
\and         G.~Gracia-Abril\inst{\ref{inst:0072},\ref{inst:0028}}
\and         J.~Portell\orcit{0000-0002-8886-8925}\inst{\ref{inst:0033}}
\and         D.~Teyssier\orcit{0000-0002-6261-5292}\inst{\ref{inst:0075}}
\and         M.~Altmann\orcit{0000-0002-0530-0913}\inst{\ref{inst:0028},\ref{inst:0077}}
\and         M.~Audard\orcit{0000-0003-4721-034X}\inst{\ref{inst:0022},\ref{inst:0066}}
\and         I.~Bellas-Velidis\inst{\ref{inst:0080}}
\and         K.~Benson\inst{\ref{inst:0071}}
\and         J.~Berthier\orcit{0000-0003-1846-6485}\inst{\ref{inst:0082}}
\and         R.~Blomme\orcit{0000-0002-2526-346X}\inst{\ref{inst:0003}}
\and       P.W.~Burgess\inst{\ref{inst:0013}}
\and         D.~Busonero\orcit{0000-0002-3903-7076}\inst{\ref{inst:0007}}
\and         G.~Busso\orcit{0000-0003-0937-9849}\inst{\ref{inst:0013}}
\and         H.~C\'{a}novas\orcit{0000-0001-7668-8022}\inst{\ref{inst:0075}}
\and         B.~Carry\orcit{0000-0001-5242-3089}\inst{\ref{inst:0001}}
\and         A.~Cellino\orcit{0000-0002-6645-334X}\inst{\ref{inst:0007}}
\and         N.~Cheek\inst{\ref{inst:0090}}
\and         Y.~Damerdji\orcit{0000-0002-3107-4024}\inst{\ref{inst:0091},\ref{inst:0092}}
\and         M.~Davidson\inst{\ref{inst:0093}}
\and         P.~de Teodoro\inst{\ref{inst:0031}}
\and         M.~Nu\~{n}ez Campos\inst{\ref{inst:0032}}
\and         L.~Delchambre\orcit{0000-0003-2559-408X}\inst{\ref{inst:0091}}
\and         A.~Dell'Oro\orcit{0000-0003-1561-9685}\inst{\ref{inst:0004}}
\and         P.~Esquej\orcit{0000-0001-8195-628X}\inst{\ref{inst:0098}}
\and         J.~Fern\'{a}ndez-Hern\'{a}ndez\inst{\ref{inst:0099}}
\and         E.~Fraile\inst{\ref{inst:0098}}
\and         D.~Garabato\orcit{0000-0002-7133-6623}\inst{\ref{inst:0101}}
\and         P.~Garc\'{i}a-Lario\orcit{0000-0003-4039-8212}\inst{\ref{inst:0031}}
\and         E.~Gosset\inst{\ref{inst:0091},\ref{inst:0041}}
\and         R.~Haigron\inst{\ref{inst:0025}}
\and      J.-L.~Halbwachs\orcit{0000-0003-2968-6395}\inst{\ref{inst:0106}}
\and       N.C.~Hambly\orcit{0000-0002-9901-9064}\inst{\ref{inst:0093}}
\and         J.~Hern\'{a}ndez\orcit{0000-0002-0361-4994}\inst{\ref{inst:0031}}
\and         D.~Hestroffer\orcit{0000-0003-0472-9459}\inst{\ref{inst:0082}}
\and       S.T.~Hodgkin\orcit{0000-0002-5470-3962}\inst{\ref{inst:0013}}
\and         B.~Holl\orcit{0000-0001-6220-3266}\inst{\ref{inst:0022},\ref{inst:0066}}
\and         K.~Jan{\ss}en\orcit{0000-0002-8163-2493}\inst{\ref{inst:0113}}
\and         G.~Jevardat de Fombelle\inst{\ref{inst:0022}}
\and         S.~Jordan\orcit{0000-0001-6316-6831}\inst{\ref{inst:0028}}
\and         A.~Krone-Martins\orcit{0000-0002-2308-6623}\inst{\ref{inst:0116},\ref{inst:0117}}
\and       A.C.~Lanzafame\orcit{0000-0002-2697-3607}\inst{\ref{inst:0118},\ref{inst:0119}}
\and         W.~L\"{ o}ffler\inst{\ref{inst:0028}}
\and         O.~Marchal\orcit{ 0000-0001-7461-892}\inst{\ref{inst:0106}}
\and       P.M.~Marrese\orcit{0000-0002-8162-3810}\inst{\ref{inst:0122},\ref{inst:0005}}
\and         A.~Moitinho\orcit{0000-0003-0822-5995}\inst{\ref{inst:0116}}
\and         K.~Muinonen\orcit{0000-0001-8058-2642}\inst{\ref{inst:0125},\ref{inst:0126}}
\and         P.~Osborne\inst{\ref{inst:0013}}
\and         T.~Pauwels\inst{\ref{inst:0003}}
\and         A.~Recio-Blanco\orcit{0000-0002-6550-7377}\inst{\ref{inst:0001}}
\and         M.~Riello\orcit{0000-0002-3134-0935}\inst{\ref{inst:0013}}
\and         L.~Rimoldini\orcit{0000-0002-0306-585X}\inst{\ref{inst:0066}}
\and         T.~Roegiers\orcit{0000-0002-1231-4440}\inst{\ref{inst:0132}}
\and         J.~Rybizki\orcit{0000-0002-0993-6089}\inst{\ref{inst:0006}}
\and         C.~Siopis\orcit{0000-0002-6267-2924}\inst{\ref{inst:0040}}
\and         M.~Smith\inst{\ref{inst:0071}}
\and         A.~Sozzetti\orcit{0000-0002-7504-365X}\inst{\ref{inst:0007}}
\and         E.~Utrilla\inst{\ref{inst:0032}}
\and         M.~van Leeuwen\orcit{0000-0001-9698-2392}\inst{\ref{inst:0013}}
\and         U.~Abbas\orcit{0000-0002-5076-766X}\inst{\ref{inst:0007}}
\and         P.~\'{A}brah\'{a}m\orcit{0000-0001-6015-646X}\inst{\ref{inst:0140},\ref{inst:0141}}
\and         A.~Abreu Aramburu\inst{\ref{inst:0099}}
\and         C.~Aerts\orcit{0000-0003-1822-7126}\inst{\ref{inst:0143},\ref{inst:0144},\ref{inst:0006}}
\and       J.J.~Aguado\inst{\ref{inst:0002}}
\and         M.~Ajaj\inst{\ref{inst:0025}}
\and         F.~Aldea-Montero\inst{\ref{inst:0031}}
\and         G.~Altavilla\orcit{0000-0002-9934-1352}\inst{\ref{inst:0122},\ref{inst:0005}}
\and       M.A.~\'{A}lvarez\orcit{0000-0002-6786-2620}\inst{\ref{inst:0101}}
\and         J.~Alves\orcit{0000-0002-4355-0921}\inst{\ref{inst:0152}}
\and         F.~Anders\inst{\ref{inst:0033}}
\and       R.I.~Anderson\orcit{0000-0001-8089-4419}\inst{\ref{inst:0154}}
\and         E.~Anglada Varela\orcit{0000-0001-7563-0689}\inst{\ref{inst:0099}}
\and         T.~Antoja\orcit{0000-0003-2595-5148}\inst{\ref{inst:0033}}
\and         D.~Baines\orcit{0000-0002-6923-3756}\inst{\ref{inst:0075}}
\and       S.G.~Baker\orcit{0000-0002-6436-1257}\inst{\ref{inst:0071}}
\and         L.~Balaguer-N\'{u}\~{n}ez\orcit{0000-0001-9789-7069}\inst{\ref{inst:0033}}
\and         E.~Balbinot\orcit{0000-0002-1322-3153}\inst{\ref{inst:0160}}
\and         Z.~Balog\orcit{0000-0003-1748-2926}\inst{\ref{inst:0028},\ref{inst:0006}}
\and         C.~Barache\inst{\ref{inst:0077}}
\and         D.~Barbato\inst{\ref{inst:0022},\ref{inst:0007}}
\and         M.~Barros\orcit{0000-0002-9728-9618}\inst{\ref{inst:0116}}
\and       M.A.~Barstow\orcit{0000-0002-7116-3259}\inst{\ref{inst:0167}}
\and         S.~Bartolom\'{e}\orcit{0000-0002-6290-6030}\inst{\ref{inst:0033}}
\and      J.-L.~Bassilana\inst{\ref{inst:0169}}
\and         N.~Bauchet\inst{\ref{inst:0025}}
\and         U.~Becciani\orcit{0000-0002-4389-8688}\inst{\ref{inst:0118}}
\and         M.~Bellazzini\orcit{0000-0001-8200-810X}\inst{\ref{inst:0008}}
\and         A.~Berihuete\orcit{0000-0002-8589-4423}\inst{\ref{inst:0173}}
\and         M.~Bernet\orcit{0000-0001-7503-1010}\inst{\ref{inst:0033}}
\and         S.~Bertone\orcit{0000-0001-9885-8440}\inst{\ref{inst:0175},\ref{inst:0176},\ref{inst:0007}}
\and         L.~Bianchi\orcit{0000-0002-7999-4372}\inst{\ref{inst:0178}}
\and         A.~Binnenfeld\orcit{0000-0002-9319-3838}\inst{\ref{inst:0179}}
\and         S.~Blanco-Cuaresma\orcit{0000-0002-1584-0171}\inst{\ref{inst:0180}}
\and         T.~Boch\orcit{0000-0001-5818-2781}\inst{\ref{inst:0106}}
\and         A.~Bombrun\inst{\ref{inst:0182}}
\and         D.~Bossini\orcit{0000-0002-9480-8400}\inst{\ref{inst:0183}}
\and         S.~Bouquillon\inst{\ref{inst:0077},\ref{inst:0185}}
\and         A.~Bragaglia\orcit{0000-0002-0338-7883}\inst{\ref{inst:0008}}
\and         L.~Bramante\inst{\ref{inst:0062}}
\and         E.~Breedt\orcit{0000-0001-6180-3438}\inst{\ref{inst:0013}}
\and         A.~Bressan\orcit{0000-0002-7922-8440}\inst{\ref{inst:0189}}
\and         N.~Brouillet\orcit{0000-0002-3274-7024}\inst{\ref{inst:0029}}
\and         E.~Brugaletta\orcit{0000-0003-2598-6737}\inst{\ref{inst:0118}}
\and         B.~Bucciarelli\orcit{0000-0002-5303-0268}\inst{\ref{inst:0007},\ref{inst:0053}}
\and         A.~Burlacu\inst{\ref{inst:0194}}
\and       A.G.~Butkevich\orcit{0000-0002-4098-3588}\inst{\ref{inst:0007}}
\and         R.~Buzzi\orcit{0000-0001-9389-5701}\inst{\ref{inst:0007}}
\and         E.~Caffau\orcit{0000-0001-6011-6134}\inst{\ref{inst:0025}}
\and         R.~Cancelliere\orcit{0000-0002-9120-3799}\inst{\ref{inst:0198}}
\and         T.~Cantat-Gaudin\orcit{0000-0001-8726-2588}\inst{\ref{inst:0033},\ref{inst:0006}}
\and         R.~Carballo\orcit{0000-0001-7412-2498}\inst{\ref{inst:0201}}
\and         T.~Carlucci\inst{\ref{inst:0077}}
\and       M.I.~Carnerero\orcit{0000-0001-5843-5515}\inst{\ref{inst:0007}}
\and       J.M.~Carrasco\orcit{0000-0002-3029-5853}\inst{\ref{inst:0033}}
\and         L.~Casamiquela\orcit{0000-0001-5238-8674}\inst{\ref{inst:0029},\ref{inst:0025}}
\and         M.~Castellani\orcit{0000-0002-7650-7428}\inst{\ref{inst:0122}}
\and         A.~Castro-Ginard\orcit{0000-0002-9419-3725}\inst{\ref{inst:0021}}
\and         L.~Chaoul\inst{\ref{inst:0039}}
\and         P.~Charlot\orcit{0000-0002-9142-716X}\inst{\ref{inst:0029}}
\and         L.~Chemin\orcit{0000-0002-3834-7937}\inst{\ref{inst:0211}}
\and         V.~Chiaramida\inst{\ref{inst:0062}}
\and         A.~Chiavassa\orcit{0000-0003-3891-7554}\inst{\ref{inst:0001}}
\and         N.~Chornay\orcit{0000-0002-8767-3907}\inst{\ref{inst:0013}}
\and         G.~Comoretto\inst{\ref{inst:0075},\ref{inst:0216}}
\and         G.~Contursi\orcit{0000-0001-5370-1511}\inst{\ref{inst:0001}}
\and       W.J.~Cooper\orcit{0000-0003-3501-8967}\inst{\ref{inst:0218},\ref{inst:0007}}
\and         T.~Cornez\inst{\ref{inst:0169}}
\and         S.~Cowell\inst{\ref{inst:0013}}
\and         F.~Crifo\inst{\ref{inst:0025}}
\and         M.~Cropper\orcit{0000-0003-4571-9468}\inst{\ref{inst:0071}}
\and         M.~Crosta\orcit{0000-0003-4369-3786}\inst{\ref{inst:0007},\ref{inst:0225}}
\and         C.~Crowley\inst{\ref{inst:0182}}
\and         C.~Dafonte\orcit{0000-0003-4693-7555}\inst{\ref{inst:0101}}
\and         A.~Dapergolas\inst{\ref{inst:0080}}
\and         P.~David\inst{\ref{inst:0082}}
\and         P.~de Laverny\orcit{0000-0002-2817-4104}\inst{\ref{inst:0001}}
\and         F.~De Luise\orcit{0000-0002-6570-8208}\inst{\ref{inst:0231}}
\and         R.~De March\orcit{0000-0003-0567-842X}\inst{\ref{inst:0062}}
\and         J.~De Ridder\orcit{0000-0001-6726-2863}\inst{\ref{inst:0143}}
\and         R.~de Souza\inst{\ref{inst:0234}}
\and         A.~de Torres\inst{\ref{inst:0182}}
\and       E.F.~del Peloso\inst{\ref{inst:0028}}
\and         E.~del Pozo\inst{\ref{inst:0032}}
\and         M.~Delbo\orcit{0000-0002-8963-2404}\inst{\ref{inst:0001}}
\and         A.~Delgado\inst{\ref{inst:0098}}
\and      J.-B.~Delisle\orcit{0000-0001-5844-9888}\inst{\ref{inst:0022}}
\and         C.~Demouchy\inst{\ref{inst:0241}}
\and       T.E.~Dharmawardena\orcit{0000-0002-9583-5216}\inst{\ref{inst:0006}}
\and         P.~Di Matteo\inst{\ref{inst:0025}}
\and         S.~Diakite\inst{\ref{inst:0244}}
\and         C.~Diener\inst{\ref{inst:0013}}
\and         E.~Distefano\orcit{0000-0002-2448-2513}\inst{\ref{inst:0118}}
\and         C.~Dolding\inst{\ref{inst:0071}}
\and         H.~Enke\orcit{0000-0002-2366-8316}\inst{\ref{inst:0113}}
\and         C.~Fabre\inst{\ref{inst:0249}}
\and         M.~Fabrizio\orcit{0000-0001-5829-111X}\inst{\ref{inst:0122},\ref{inst:0005}}
\and         S.~Faigler\orcit{0000-0002-8368-5724}\inst{\ref{inst:0252}}
\and         G.~Fedorets\orcit{0000-0002-8418-4809}\inst{\ref{inst:0125},\ref{inst:0254}}
\and         P.~Fernique\orcit{0000-0002-3304-2923}\inst{\ref{inst:0106},\ref{inst:0256}}
\and         F.~Figueras\orcit{0000-0002-3393-0007}\inst{\ref{inst:0033}}
\and         Y.~Fournier\orcit{0000-0002-6633-9088}\inst{\ref{inst:0113}}
\and         C.~Fouron\inst{\ref{inst:0194}}
\and         F.~Fragkoudi\orcit{0000-0002-0897-3013}\inst{\ref{inst:0260},\ref{inst:0261},\ref{inst:0262}}
\and         M.~Gai\orcit{0000-0001-9008-134X}\inst{\ref{inst:0007}}
\and         A.~Garcia-Gutierrez\inst{\ref{inst:0033}}
\and         M.~Garcia-Reinaldos\inst{\ref{inst:0031}}
\and         M.~Garc\'{i}a-Torres\orcit{0000-0002-6867-7080}\inst{\ref{inst:0266}}
\and         A.~Gavel\orcit{0000-0002-2963-722X}\inst{\ref{inst:0009}}
\and         P.~Gavras\orcit{0000-0002-4383-4836}\inst{\ref{inst:0098}}
\and         E.~Gerlach\orcit{0000-0002-9533-2168}\inst{\ref{inst:0034}}
\and         R.~Geyer\orcit{0000-0001-6967-8707}\inst{\ref{inst:0034}}
\and         P.~Giacobbe\orcit{0000-0001-7034-7024}\inst{\ref{inst:0007}}
\and         G.~Gilmore\orcit{0000-0003-4632-0213}\inst{\ref{inst:0013}}
\and         S.~Girona\orcit{0000-0002-1975-1918}\inst{\ref{inst:0273}}
\and         G.~Giuffrida\inst{\ref{inst:0122}}
\and         R.~Gomel\inst{\ref{inst:0252}}
\and         A.~Gomez\orcit{0000-0002-3796-3690}\inst{\ref{inst:0101}}
\and         J.~Gonz\'{a}lez-N\'{u}\~{n}ez\orcit{0000-0001-5311-5555}\inst{\ref{inst:0090},\ref{inst:0278}}
\and         I.~Gonz\'{a}lez-Santamar\'{i}a\orcit{0000-0002-8537-9384}\inst{\ref{inst:0101}}
\and       J.J.~Gonz\'{a}lez-Vidal\inst{\ref{inst:0033}}
\and         M.~Granvik\orcit{0000-0002-5624-1888}\inst{\ref{inst:0125},\ref{inst:0282}}
\and         P.~Guillout\inst{\ref{inst:0106}}
\and         J.~Guiraud\inst{\ref{inst:0039}}
\and         R.~Guti\'{e}rrez-S\'{a}nchez\inst{\ref{inst:0075}}
\and       L.P.~Guy\orcit{0000-0003-0800-8755}\inst{\ref{inst:0066},\ref{inst:0287}}
\and         D.~Hatzidimitriou\orcit{0000-0002-5415-0464}\inst{\ref{inst:0288},\ref{inst:0080}}
\and         M.~Hauser\inst{\ref{inst:0006},\ref{inst:0291}}
\and         M.~Haywood\orcit{0000-0003-0434-0400}\inst{\ref{inst:0025}}
\and         A.~Helmer\inst{\ref{inst:0169}}
\and         A.~Helmi\orcit{0000-0003-3937-7641}\inst{\ref{inst:0160}}
\and         T.~Hilger\orcit{0000-0003-1646-0063}\inst{\ref{inst:0034}}
\and       M.H.~Sarmiento\orcit{0000-0003-4252-5115}\inst{\ref{inst:0032}}
\and       S.L.~Hidalgo\orcit{0000-0002-0002-9298}\inst{\ref{inst:0296},\ref{inst:0297}}
\and         N.~H\l{}adczuk\orcit{0000-0001-9163-4209}\inst{\ref{inst:0031},\ref{inst:0299}}
\and         D.~Hobbs\orcit{0000-0002-2696-1366}\inst{\ref{inst:0036}}
\and         G.~Holland\inst{\ref{inst:0013}}
\and       H.E.~Huckle\inst{\ref{inst:0071}}
\and         K.~Jardine\inst{\ref{inst:0303}}
\and         G.~Jasniewicz\inst{\ref{inst:0304}}
\and         A.~Jean-Antoine Piccolo\orcit{0000-0001-8622-212X}\inst{\ref{inst:0039}}
\and     \'{O}.~Jim\'{e}nez-Arranz\orcit{0000-0001-7434-5165}\inst{\ref{inst:0033}}
\and         J.~Juaristi Campillo\inst{\ref{inst:0028}}
\and         F.~Julbe\inst{\ref{inst:0033}}
\and         L.~Karbevska\inst{\ref{inst:0066},\ref{inst:0310}}
\and         P.~Kervella\orcit{0000-0003-0626-1749}\inst{\ref{inst:0311}}
\and         S.~Khanna\orcit{0000-0002-2604-4277}\inst{\ref{inst:0160},\ref{inst:0007}}
\and         G.~Kordopatis\orcit{0000-0002-9035-3920}\inst{\ref{inst:0001}}
\and      \'{A}~K\'{o}sp\'{a}l\orcit{ikl\'{o}s \'{u}t 15}\inst{\ref{inst:0140},\ref{inst:0006},\ref{inst:0141}}
\and         Z.~Kostrzewa-Rutkowska\inst{\ref{inst:0021},\ref{inst:0319}}
\and         K.~Kruszy\'{n}ska\orcit{0000-0002-2729-5369}\inst{\ref{inst:0320}}
\and         M.~Kun\orcit{0000-0002-7538-5166}\inst{\ref{inst:0140}}
\and         P.~Laizeau\inst{\ref{inst:0322}}
\and         S.~Lambert\orcit{0000-0001-6759-5502}\inst{\ref{inst:0077}}
\and       A.F.~Lanza\orcit{0000-0001-5928-7251}\inst{\ref{inst:0118}}
\and         Y.~Lasne\inst{\ref{inst:0169}}
\and      J.-F.~Le Campion\inst{\ref{inst:0029}}
\and         Y.~Lebreton\orcit{0000-0002-4834-2144}\inst{\ref{inst:0311},\ref{inst:0328}}
\and         T.~Lebzelter\orcit{0000-0002-0702-7551}\inst{\ref{inst:0152}}
\and         S.~Leccia\orcit{0000-0001-5685-6930}\inst{\ref{inst:0330}}
\and         N.~Leclerc\inst{\ref{inst:0025}}
\and         I.~Lecoeur-Taibi\orcit{0000-0003-0029-8575}\inst{\ref{inst:0066}}
\and         S.~Liao\orcit{0000-0002-9346-0211}\inst{\ref{inst:0333},\ref{inst:0007},\ref{inst:0335}}
\and       E.L.~Licata\orcit{0000-0002-5203-0135}\inst{\ref{inst:0007}}
\and     H.E.P.~Lindstr{\o}m\inst{\ref{inst:0007},\ref{inst:0338},\ref{inst:0339}}
\and       T.A.~Lister\orcit{0000-0002-3818-7769}\inst{\ref{inst:0340}}
\and         E.~Livanou\orcit{0000-0003-0628-2347}\inst{\ref{inst:0288}}
\and         A.~Lorca\inst{\ref{inst:0032}}
\and         C.~Loup\inst{\ref{inst:0106}}
\and         P.~Madrero Pardo\inst{\ref{inst:0033}}
\and         A.~Magdaleno Romeo\inst{\ref{inst:0194}}
\and         S.~Managau\inst{\ref{inst:0169}}
\and       R.G.~Mann\orcit{0000-0002-0194-325X}\inst{\ref{inst:0093}}
\and         M.~Manteiga\orcit{0000-0002-7711-5581}\inst{\ref{inst:0348}}
\and       J.M.~Marchant\orcit{0000-0002-3678-3145}\inst{\ref{inst:0349}}
\and         M.~Marconi\orcit{0000-0002-1330-2927}\inst{\ref{inst:0330}}
\and         J.~Marcos\inst{\ref{inst:0075}}
\and     M.M.S.~Marcos Santos\inst{\ref{inst:0090}}
\and         D.~Mar\'{i}n Pina\orcit{0000-0001-6482-1842}\inst{\ref{inst:0033}}
\and         S.~Marinoni\orcit{0000-0001-7990-6849}\inst{\ref{inst:0122},\ref{inst:0005}}
\and         F.~Marocco\orcit{0000-0001-7519-1700}\inst{\ref{inst:0356}}
\and       D.J.~Marshall\orcit{0000-0003-3956-3524}\inst{\ref{inst:0357}}
\and         L.~Martin Polo\inst{\ref{inst:0090}}
\and       J.M.~Mart\'{i}n-Fleitas\orcit{0000-0002-8594-569X}\inst{\ref{inst:0032}}
\and         G.~Marton\orcit{0000-0002-1326-1686}\inst{\ref{inst:0140}}
\and         N.~Mary\inst{\ref{inst:0169}}
\and         A.~Masip\orcit{0000-0003-1419-0020}\inst{\ref{inst:0033}}
\and         D.~Massari\orcit{0000-0001-8892-4301}\inst{\ref{inst:0008}}
\and         A.~Mastrobuono-Battisti\orcit{0000-0002-2386-9142}\inst{\ref{inst:0025}}
\and         T.~Mazeh\orcit{0000-0002-3569-3391}\inst{\ref{inst:0252}}
\and       P.J.~McMillan\orcit{0000-0002-8861-2620}\inst{\ref{inst:0036}}
\and         S.~Messina\orcit{0000-0002-2851-2468}\inst{\ref{inst:0118}}
\and         D.~Michalik\orcit{0000-0002-7618-6556}\inst{\ref{inst:0023}}
\and       N.R.~Millar\inst{\ref{inst:0013}}
\and         A.~Mints\orcit{0000-0002-8440-1455}\inst{\ref{inst:0113}}
\and         D.~Molina\orcit{0000-0003-4814-0275}\inst{\ref{inst:0033}}
\and         R.~Molinaro\orcit{0000-0003-3055-6002}\inst{\ref{inst:0330}}
\and         L.~Moln\'{a}r\orcit{0000-0002-8159-1599}\inst{\ref{inst:0140},\ref{inst:0374},\ref{inst:0141}}
\and         G.~Monari\orcit{0000-0002-6863-0661}\inst{\ref{inst:0106}}
\and         M.~Mongui\'{o}\orcit{0000-0002-4519-6700}\inst{\ref{inst:0033}}
\and         P.~Montegriffo\orcit{0000-0001-5013-5948}\inst{\ref{inst:0008}}
\and         A.~Montero\inst{\ref{inst:0032}}
\and         R.~Mor\orcit{0000-0002-8179-6527}\inst{\ref{inst:0033}}
\and         A.~Mora\inst{\ref{inst:0032}}
\and         R.~Morbidelli\orcit{0000-0001-7627-4946}\inst{\ref{inst:0007}}
\and         T.~Morel\orcit{0000-0002-8176-4816}\inst{\ref{inst:0091}}
\and         D.~Morris\inst{\ref{inst:0093}}
\and         T.~Muraveva\orcit{0000-0002-0969-1915}\inst{\ref{inst:0008}}
\and       C.P.~Murphy\inst{\ref{inst:0031}}
\and         I.~Musella\orcit{0000-0001-5909-6615}\inst{\ref{inst:0330}}
\and         Z.~Nagy\orcit{0000-0002-3632-1194}\inst{\ref{inst:0140}}
\and         L.~Noval\inst{\ref{inst:0169}}
\and         F.~Oca\~{n}a\inst{\ref{inst:0075},\ref{inst:0390}}
\and         A.~Ogden\inst{\ref{inst:0013}}
\and         C.~Ordenovic\inst{\ref{inst:0001}}
\and       J.O.~Osinde\inst{\ref{inst:0098}}
\and         C.~Pagani\orcit{0000-0001-5477-4720}\inst{\ref{inst:0167}}
\and         I.~Pagano\orcit{0000-0001-9573-4928}\inst{\ref{inst:0118}}
\and         L.~Palaversa\orcit{0000-0003-3710-0331}\inst{\ref{inst:0396},\ref{inst:0013}}
\and       P.A.~Palicio\orcit{0000-0002-7432-8709}\inst{\ref{inst:0001}}
\and         L.~Pallas-Quintela\orcit{0000-0001-9296-3100}\inst{\ref{inst:0101}}
\and         A.~Panahi\orcit{0000-0001-5850-4373}\inst{\ref{inst:0252}}
\and         S.~Payne-Wardenaar\inst{\ref{inst:0028}}
\and         X.~Pe\~{n}alosa Esteller\inst{\ref{inst:0033}}
\and         A.~Penttil\"{ a}\orcit{0000-0001-7403-1721}\inst{\ref{inst:0125}}
\and         B.~Pichon\orcit{0000 0000 0062 1449}\inst{\ref{inst:0001}}
\and       A.M.~Piersimoni\orcit{0000-0002-8019-3708}\inst{\ref{inst:0231}}
\and      F.-X.~Pineau\orcit{0000-0002-2335-4499}\inst{\ref{inst:0106}}
\and         E.~Plachy\orcit{0000-0002-5481-3352}\inst{\ref{inst:0140},\ref{inst:0374},\ref{inst:0141}}
\and         G.~Plum\inst{\ref{inst:0025}}
\and         E.~Poggio\orcit{0000-0003-3793-8505}\inst{\ref{inst:0001},\ref{inst:0007}}
\and         A.~Pr\v{s}a\orcit{0000-0002-1913-0281}\inst{\ref{inst:0412}}
\and         L.~Pulone\orcit{0000-0002-5285-998X}\inst{\ref{inst:0122}}
\and         E.~Racero\orcit{0000-0002-6101-9050}\inst{\ref{inst:0090},\ref{inst:0390}}
\and         S.~Ragaini\inst{\ref{inst:0008}}
\and         M.~Rainer\orcit{0000-0002-8786-2572}\inst{\ref{inst:0004},\ref{inst:0418}}
\and       C.M.~Raiteri\orcit{0000-0003-1784-2784}\inst{\ref{inst:0007}}
\and         P.~Ramos\orcit{0000-0002-5080-7027}\inst{\ref{inst:0033},\ref{inst:0106}}
\and         M.~Ramos-Lerate\inst{\ref{inst:0075}}
\and         P.~Re Fiorentin\orcit{0000-0002-4995-0475}\inst{\ref{inst:0007}}
\and         S.~Regibo\inst{\ref{inst:0143}}
\and       P.J.~Richards\inst{\ref{inst:0425}}
\and         C.~Rios Diaz\inst{\ref{inst:0098}}
\and         V.~Ripepi\orcit{0000-0003-1801-426X}\inst{\ref{inst:0330}}
\and         A.~Riva\orcit{0000-0002-6928-8589}\inst{\ref{inst:0007}}
\and      H.-W.~Rix\orcit{0000-0003-4996-9069}\inst{\ref{inst:0006}}
\and         G.~Rixon\orcit{0000-0003-4399-6568}\inst{\ref{inst:0013}}
\and         N.~Robichon\orcit{0000-0003-4545-7517}\inst{\ref{inst:0025}}
\and       A.C.~Robin\orcit{0000-0001-8654-9499}\inst{\ref{inst:0018}}
\and         C.~Robin\inst{\ref{inst:0169}}
\and         M.~Roelens\orcit{0000-0003-0876-4673}\inst{\ref{inst:0022}}
\and     H.R.O.~Rogues\inst{\ref{inst:0241}}
\and         L.~Rohrbasser\inst{\ref{inst:0066}}
\and         M.~Romero-G\'{o}mez\orcit{0000-0003-3936-1025}\inst{\ref{inst:0033}}
\and         N.~Rowell\orcit{0000-0003-3809-1895}\inst{\ref{inst:0093}}
\and         F.~Royer\orcit{0000-0002-9374-8645}\inst{\ref{inst:0025}}
\and         D.~Ruz Mieres\orcit{0000-0002-9455-157X}\inst{\ref{inst:0013}}
\and       K.A.~Rybicki\orcit{0000-0002-9326-9329}\inst{\ref{inst:0320}}
\and         G.~Sadowski\orcit{0000-0002-3411-1003}\inst{\ref{inst:0040}}
\and         A.~S\'{a}ez N\'{u}\~{n}ez\inst{\ref{inst:0033}}
\and         A.~Sagrist\`{a} Sell\'{e}s\orcit{0000-0001-6191-2028}\inst{\ref{inst:0028}}
\and         J.~Sahlmann\orcit{0000-0001-9525-3673}\inst{\ref{inst:0098}}
\and         E.~Salguero\inst{\ref{inst:0099}}
\and         N.~Samaras\orcit{0000-0001-8375-6652}\inst{\ref{inst:0003},\ref{inst:0448}}
\and         V.~Sanchez Gimenez\orcit{0000-0003-1797-3557}\inst{\ref{inst:0033}}
\and         N.~Sanna\orcit{0000-0001-9275-9492}\inst{\ref{inst:0004}}
\and         R.~Santove\~{n}a\orcit{0000-0002-9257-2131}\inst{\ref{inst:0101}}
\and         M.~Sarasso\orcit{0000-0001-5121-0727}\inst{\ref{inst:0007}}
\and       M.~Schultheis\orcit{0000-0002-6590-1657}\inst{\ref{inst:0001}}
\and         E.~Sciacca\orcit{0000-0002-5574-2787}\inst{\ref{inst:0118}}
\and         M.~Segol\inst{\ref{inst:0241}}
\and       J.C.~Segovia\inst{\ref{inst:0090}}
\and         D.~S\'{e}gransan\orcit{0000-0003-2355-8034}\inst{\ref{inst:0022}}
\and         D.~Semeux\inst{\ref{inst:0249}}
\and         S.~Shahaf\orcit{0000-0001-9298-8068}\inst{\ref{inst:0459}}
\and       H.I.~Siddiqui\orcit{0000-0003-1853-6033}\inst{\ref{inst:0460}}
\and         A.~Siebert\orcit{0000-0001-8059-2840}\inst{\ref{inst:0106},\ref{inst:0256}}
\and         L.~Siltala\orcit{0000-0002-6938-794X}\inst{\ref{inst:0125}}
\and         A.~Silvelo\orcit{0000-0002-5126-6365}\inst{\ref{inst:0101}}
\and         E.~Slezak\inst{\ref{inst:0001}}
\and         I.~Slezak\inst{\ref{inst:0001}}
\and       O.N.~Snaith\inst{\ref{inst:0025}}
\and         E.~Solano\inst{\ref{inst:0468}}
\and         F.~Solitro\inst{\ref{inst:0062}}
\and         D.~Souami\orcit{0000-0003-4058-0815}\inst{\ref{inst:0311},\ref{inst:0471}}
\and         J.~Souchay\inst{\ref{inst:0077}}
\and         A.~Spagna\orcit{0000-0003-1732-2412}\inst{\ref{inst:0007}}
\and         L.~Spina\orcit{0000-0002-9760-6249}\inst{\ref{inst:0014}}
\and         F.~Spoto\orcit{0000-0001-7319-5847}\inst{\ref{inst:0180}}
\and       I.A.~Steele\orcit{0000-0001-8397-5759}\inst{\ref{inst:0349}}
\and         H.~Steidelm\"{ u}ller\inst{\ref{inst:0034}}
\and       C.A.~Stephenson\inst{\ref{inst:0075},\ref{inst:0479}}
\and         M.~S\"{ u}veges\orcit{0000-0003-3017-5322}\inst{\ref{inst:0480}}
\and         J.~Surdej\orcit{0000-0002-7005-1976}\inst{\ref{inst:0091},\ref{inst:0482}}
\and         L.~Szabados\orcit{0000-0002-2046-4131}\inst{\ref{inst:0140}}
\and         E.~Szegedi-Elek\orcit{0000-0001-7807-6644}\inst{\ref{inst:0140}}
\and         F.~Taris\inst{\ref{inst:0077}}
\and       M.B.~Taylor\orcit{0000-0002-4209-1479}\inst{\ref{inst:0486}}
\and         R.~Teixeira\orcit{0000-0002-6806-6626}\inst{\ref{inst:0234}}
\and         L.~Tolomei\orcit{0000-0002-3541-3230}\inst{\ref{inst:0062}}
\and         N.~Tonello\orcit{0000-0003-0550-1667}\inst{\ref{inst:0273}}
\and         F.~Torra\orcit{0000-0002-8429-299X}\inst{\ref{inst:0057}}
\and         J.~Torra$^\dagger$\inst{\ref{inst:0033}}
\and         G.~Torralba Elipe\orcit{0000-0001-8738-194X}\inst{\ref{inst:0101}}
\and         M.~Trabucchi\orcit{0000-0002-1429-2388}\inst{\ref{inst:0493},\ref{inst:0022}}
\and       A.T.~Tsounis\inst{\ref{inst:0495}}
\and         C.~Turon\orcit{0000-0003-1236-5157}\inst{\ref{inst:0025}}
\and         A.~Ulla\orcit{0000-0001-6424-5005}\inst{\ref{inst:0497}}
\and         N.~Unger\orcit{0000-0003-3993-7127}\inst{\ref{inst:0022}}
\and       M.V.~Vaillant\inst{\ref{inst:0169}}
\and         E.~van Dillen\inst{\ref{inst:0241}}
\and         W.~van Reeven\inst{\ref{inst:0501}}
\and         O.~Vanel\orcit{0000-0002-7898-0454}\inst{\ref{inst:0025}}
\and         A.~Vecchiato\orcit{0000-0003-1399-5556}\inst{\ref{inst:0007}}
\and         Y.~Viala\inst{\ref{inst:0025}}
\and         D.~Vicente\orcit{0000-0002-1584-1182}\inst{\ref{inst:0273}}
\and         S.~Voutsinas\inst{\ref{inst:0093}}
\and         M.~Weiler\inst{\ref{inst:0033}}
\and         T.~Wevers\orcit{0000-0002-4043-9400}\inst{\ref{inst:0013},\ref{inst:0509}}
\and      \L{}.~Wyrzykowski\orcit{0000-0002-9658-6151}\inst{\ref{inst:0320}}
\and         A.~Yoldas\inst{\ref{inst:0013}}
\and         P.~Yvard\inst{\ref{inst:0241}}
\and         H.~Zhao\orcit{0000-0003-2645-6869}\inst{\ref{inst:0001}}
\and         J.~Zorec\inst{\ref{inst:0514}}
\and         S.~Zucker\orcit{0000-0003-3173-3138}\inst{\ref{inst:0179}}
\and         T.~Zwitter\orcit{0000-0002-2325-8763}\inst{\ref{inst:0516}}
}
\institute{
     Universit\'{e} C\^{o}te d'Azur, Observatoire de la C\^{o}te d'Azur, CNRS, Laboratoire Lagrange, Bd de l'Observatoire, CS 34229, 06304 Nice Cedex 4, France\relax                                                                                                                                                                                              \label{inst:0001}
\and Dpto. de Inteligencia Artificial, UNED, c/ Juan del Rosal 16, 28040 Madrid, Spain\relax                                                                                                                                                                                                                                                                       \label{inst:0002}\vfill
\and Royal Observatory of Belgium, Ringlaan 3, 1180 Brussels, Belgium\relax                                                                                                                                                                                                                                                                                        \label{inst:0003}\vfill
\and INAF - Osservatorio Astrofisico di Arcetri, Largo Enrico Fermi 5, 50125 Firenze, Italy\relax                                                                                                                                                                                                                                                                  \label{inst:0004}\vfill
\and Space Science Data Center - ASI, Via del Politecnico SNC, 00133 Roma, Italy\relax                                                                                                                                                                                                                                                                             \label{inst:0005}\vfill
\and Max Planck Institute for Astronomy, K\"{ o}nigstuhl 17, 69117 Heidelberg, Germany\relax                                                                                                                                                                                                                                                                       \label{inst:0006}\vfill
\and INAF - Osservatorio Astrofisico di Torino, via Osservatorio 20, 10025 Pino Torinese (TO), Italy\relax                                                                                                                                                                                                                                                         \label{inst:0007}\vfill
\and INAF - Osservatorio di Astrofisica e Scienza dello Spazio di Bologna, via Piero Gobetti 93/3, 40129 Bologna, Italy\relax                                                                                                                                                                                                                                      \label{inst:0008}\vfill
\and Observational Astrophysics, Division of Astronomy and Space Physics, Department of Physics and Astronomy, Uppsala University, Box 516, 751 20 Uppsala, Sweden\relax                                                                                                                                                                                           \label{inst:0009}\vfill
\and Institute of Astronomy, University of Cambridge, Madingley Road, Cambridge CB3 0HA, United Kingdom\relax                                                                                                                                                                                                                                                      \label{inst:0013}\vfill
\and INAF - Osservatorio astronomico di Padova, Vicolo Osservatorio 5, 35122 Padova, Italy\relax                                                                                                                                                                                                                                                                   \label{inst:0014}\vfill
\and Kavli Institute for Cosmology Cambridge, Institute of Astronomy, Madingley Road, Cambridge, CB3 0HA\relax                                                                                                                                                                                                                                                     \label{inst:0016}\vfill
\and Institut UTINAM CNRS UMR6213, Universit\'{e} Bourgogne Franche-Comt\'{e}, OSU THETA Franche-Comt\'{e} Bourgogne, Observatoire de Besan\c{c}on, BP1615, 25010 Besan\c{c}on Cedex, France\relax                                                                                                                                                                 \label{inst:0018}\vfill
\and Leiden Observatory, Leiden University, Niels Bohrweg 2, 2333 CA Leiden, The Netherlands\relax                                                                                                                                                                                                                                                                 \label{inst:0021}\vfill
\and Department of Astronomy, University of Geneva, Chemin Pegasi 51, 1290 Versoix, Switzerland\relax                                                                                                                                                                                                                                                              \label{inst:0022}\vfill
\and European Space Agency (ESA), European Space Research and Technology Centre (ESTEC), Keplerlaan 1, 2201AZ, Noordwijk, The Netherlands\relax                                                                                                                                                                                                                    \label{inst:0023}\vfill
\and GEPI, Observatoire de Paris, Universit\'{e} PSL, CNRS, 5 Place Jules Janssen, 92190 Meudon, France\relax                                                                                                                                                                                                                                                      \label{inst:0025}\vfill
\and Univ. Grenoble Alpes, CNRS, IPAG, 38000 Grenoble, France\relax                                                                                                                                                                                                                                                                                                \label{inst:0026}\vfill
\and Astronomisches Rechen-Institut, Zentrum f\"{ u}r Astronomie der Universit\"{ a}t Heidelberg, M\"{ o}nchhofstr. 12-14, 69120 Heidelberg, Germany\relax                                                                                                                                                                                                         \label{inst:0028}\vfill
\and Laboratoire d'astrophysique de Bordeaux, Univ. Bordeaux, CNRS, B18N, all{\'e}e Geoffroy Saint-Hilaire, 33615 Pessac, France\relax                                                                                                                                                                                                                             \label{inst:0029}\vfill
\and European Space Agency (ESA), European Space Astronomy Centre (ESAC), Camino bajo del Castillo, s/n, Urbanizacion Villafranca del Castillo, Villanueva de la Ca\~{n}ada, 28692 Madrid, Spain\relax                                                                                                                                                             \label{inst:0031}\vfill
\and Aurora Technology for European Space Agency (ESA), Camino bajo del Castillo, s/n, Urbanizacion Villafranca del Castillo, Villanueva de la Ca\~{n}ada, 28692 Madrid, Spain\relax                                                                                                                                                                               \label{inst:0032}\vfill
\and Institut de Ci\`{e}ncies del Cosmos (ICCUB), Universitat  de  Barcelona  (IEEC-UB), Mart\'{i} i  Franqu\`{e}s  1, 08028 Barcelona, Spain\relax                                                                                                                                                                                                                \label{inst:0033}\vfill
\and Lohrmann Observatory, Technische Universit\"{ a}t Dresden, Mommsenstra{\ss}e 13, 01062 Dresden, Germany\relax                                                                                                                                                                                                                                                 \label{inst:0034}\vfill
\and Lund Observatory, Department of Astronomy and Theoretical Physics, Lund University, Box 43, 22100 Lund, Sweden\relax                                                                                                                                                                                                                                          \label{inst:0036}\vfill
\and CNES Centre Spatial de Toulouse, 18 avenue Edouard Belin, 31401 Toulouse Cedex 9, France\relax                                                                                                                                                                                                                                                                \label{inst:0039}\vfill
\and Institut d'Astronomie et d'Astrophysique, Universit\'{e} Libre de Bruxelles CP 226, Boulevard du Triomphe, 1050 Brussels, Belgium\relax                                                                                                                                                                                                                       \label{inst:0040}\vfill
\and F.R.S.-FNRS, Rue d'Egmont 5, 1000 Brussels, Belgium\relax                                                                                                                                                                                                                                                                                                     \label{inst:0041}\vfill
\and European Space Agency (ESA, retired)\relax                                                                                                                                                                                                                                                                                                                    \label{inst:0050}\vfill
\and University of Turin, Department of Physics, Via Pietro Giuria 1, 10125 Torino, Italy\relax                                                                                                                                                                                                                                                                    \label{inst:0053}\vfill
\and DAPCOM for Institut de Ci\`{e}ncies del Cosmos (ICCUB), Universitat  de  Barcelona  (IEEC-UB), Mart\'{i} i  Franqu\`{e}s  1, 08028 Barcelona, Spain\relax                                                                                                                                                                                                     \label{inst:0057}\vfill
\and ALTEC S.p.a, Corso Marche, 79,10146 Torino, Italy\relax                                                                                                                                                                                                                                                                                                       \label{inst:0062}\vfill
\and S\`{a}rl, Geneva, Switzerland\relax                                                                                                                                                                                                                                                                                                                           \label{inst:0065}\vfill
\and Department of Astronomy, University of Geneva, Chemin d'Ecogia 16, 1290 Versoix, Switzerland\relax                                                                                                                                                                                                                                                            \label{inst:0066}\vfill
\and Mullard Space Science Laboratory, University College London, Holmbury St Mary, Dorking, Surrey RH5 6NT, United Kingdom\relax                                                                                                                                                                                                                                  \label{inst:0071}\vfill
\and Gaia DPAC Project Office, ESAC, Camino bajo del Castillo, s/n, Urbanizacion Villafranca del Castillo, Villanueva de la Ca\~{n}ada, 28692 Madrid, Spain\relax                                                                                                                                                                                                  \label{inst:0072}\vfill
\and Telespazio UK S.L. for European Space Agency (ESA), Camino bajo del Castillo, s/n, Urbanizacion Villafranca del Castillo, Villanueva de la Ca\~{n}ada, 28692 Madrid, Spain\relax                                                                                                                                                                              \label{inst:0075}\vfill
\and SYRTE, Observatoire de Paris, Universit\'{e} PSL, CNRS,  Sorbonne Universit\'{e}, LNE, 61 avenue de l'Observatoire 75014 Paris, France\relax                                                                                                                                                                                                                  \label{inst:0077}\vfill
\and National Observatory of Athens, I. Metaxa and Vas. Pavlou, Palaia Penteli, 15236 Athens, Greece\relax                                                                                                                                                                                                                                                         \label{inst:0080}\vfill
\and IMCCE, Observatoire de Paris, Universit\'{e} PSL, CNRS, Sorbonne Universit{\'e}, Univ. Lille, 77 av. Denfert-Rochereau, 75014 Paris, France\relax                                                                                                                                                                                                             \label{inst:0082}\vfill
\and Serco Gesti\'{o}n de Negocios for European Space Agency (ESA), Camino bajo del Castillo, s/n, Urbanizacion Villafranca del Castillo, Villanueva de la Ca\~{n}ada, 28692 Madrid, Spain\relax                                                                                                                                                                   \label{inst:0090}\vfill
\and Institut d'Astrophysique et de G\'{e}ophysique, Universit\'{e} de Li\`{e}ge, 19c, All\'{e}e du 6 Ao\^{u}t, B-4000 Li\`{e}ge, Belgium\relax                                                                                                                                                                                                                    \label{inst:0091}\vfill
\and CRAAG - Centre de Recherche en Astronomie, Astrophysique et G\'{e}ophysique, Route de l'Observatoire Bp 63 Bouzareah 16340 Algiers, Algeria\relax                                                                                                                                                                                                             \label{inst:0092}\vfill
\and Institute for Astronomy, University of Edinburgh, Royal Observatory, Blackford Hill, Edinburgh EH9 3HJ, United Kingdom\relax                                                                                                                                                                                                                                  \label{inst:0093}\vfill
\and RHEA for European Space Agency (ESA), Camino bajo del Castillo, s/n, Urbanizacion Villafranca del Castillo, Villanueva de la Ca\~{n}ada, 28692 Madrid, Spain\relax                                                                                                                                                                                            \label{inst:0098}\vfill
\and ATG Europe for European Space Agency (ESA), Camino bajo del Castillo, s/n, Urbanizacion Villafranca del Castillo, Villanueva de la Ca\~{n}ada, 28692 Madrid, Spain\relax                                                                                                                                                                                      \label{inst:0099}\vfill
\and CIGUS CITIC - Department of Computer Science and Information Technologies, University of A Coru\~{n}a, Campus de Elvi\~{n}a s/n, A Coru\~{n}a, 15071, Spain\relax                                                                                                                                                                                             \label{inst:0101}\vfill
\and Universit\'{e} de Strasbourg, CNRS, Observatoire astronomique de Strasbourg, UMR 7550,  11 rue de l'Universit\'{e}, 67000 Strasbourg, France\relax                                                                                                                                                                                                            \label{inst:0106}\vfill
\and Leibniz Institute for Astrophysics Potsdam (AIP), An der Sternwarte 16, 14482 Potsdam, Germany\relax                                                                                                                                                                                                                                                          \label{inst:0113}\vfill
\and CENTRA, Faculdade de Ci\^{e}ncias, Universidade de Lisboa, Edif. C8, Campo Grande, 1749-016 Lisboa, Portugal\relax                                                                                                                                                                                                                                            \label{inst:0116}\vfill
\and Department of Informatics, Donald Bren School of Information and Computer Sciences, University of California, Irvine, 5226 Donald Bren Hall, 92697-3440 CA Irvine, United States\relax                                                                                                                                                                        \label{inst:0117}\vfill
\and INAF - Osservatorio Astrofisico di Catania, via S. Sofia 78, 95123 Catania, Italy\relax                                                                                                                                                                                                                                                                       \label{inst:0118}\vfill
\and Dipartimento di Fisica e Astronomia ""Ettore Majorana"", Universit\`{a} di Catania, Via S. Sofia 64, 95123 Catania, Italy\relax                                                                                                                                                                                                                               \label{inst:0119}\vfill
\and INAF - Osservatorio Astronomico di Roma, Via Frascati 33, 00078 Monte Porzio Catone (Roma), Italy\relax                                                                                                                                                                                                                                                       \label{inst:0122}\vfill
\and Department of Physics, University of Helsinki, P.O. Box 64, 00014 Helsinki, Finland\relax                                                                                                                                                                                                                                                                     \label{inst:0125}\vfill
\and Finnish Geospatial Research Institute FGI, Geodeetinrinne 2, 02430 Masala, Finland\relax                                                                                                                                                                                                                                                                      \label{inst:0126}\vfill
\and HE Space Operations BV for European Space Agency (ESA), Keplerlaan 1, 2201AZ, Noordwijk, The Netherlands\relax                                                                                                                                                                                                                                                \label{inst:0132}\vfill
\and Konkoly Observatory, Research Centre for Astronomy and Earth Sciences, E\"{ o}tv\"{ o}s Lor{\'a}nd Research Network (ELKH), MTA Centre of Excellence, Konkoly Thege Mikl\'{o}s \'{u}t 15-17, 1121 Budapest, Hungary\relax                                                                                                                                     \label{inst:0140}\vfill
\and ELTE E\"{ o}tv\"{ o}s Lor\'{a}nd University, Institute of Physics, 1117, P\'{a}zm\'{a}ny P\'{e}ter s\'{e}t\'{a}ny 1A, Budapest, Hungary\relax                                                                                                                                                                                                                 \label{inst:0141}\vfill
\and Instituut voor Sterrenkunde, KU Leuven, Celestijnenlaan 200D, 3001 Leuven, Belgium\relax                                                                                                                                                                                                                                                                      \label{inst:0143}\vfill
\and Department of Astrophysics/IMAPP, Radboud University, P.O.Box 9010, 6500 GL Nijmegen, The Netherlands\relax                                                                                                                                                                                                                                                   \label{inst:0144}\vfill
\and University of Vienna, Department of Astrophysics, T\"{ u}rkenschanzstra{\ss}e 17, A1180 Vienna, Austria\relax                                                                                                                                                                                                                                                 \label{inst:0152}\vfill
\and Institute of Physics, Laboratory of Astrophysics, Ecole Polytechnique F\'ed\'erale de Lausanne (EPFL), Observatoire de Sauverny, 1290 Versoix, Switzerland\relax                                                                                                                                                                                              \label{inst:0154}\vfill
\and Kapteyn Astronomical Institute, University of Groningen, Landleven 12, 9747 AD Groningen, The Netherlands\relax                                                                                                                                                                                                                                               \label{inst:0160}\vfill
\and School of Physics and Astronomy / Space Park Leicester, University of Leicester, University Road, Leicester LE1 7RH, United Kingdom\relax                                                                                                                                                                                                                     \label{inst:0167}\vfill
\and Thales Services for CNES Centre Spatial de Toulouse, 18 avenue Edouard Belin, 31401 Toulouse Cedex 9, France\relax                                                                                                                                                                                                                                            \label{inst:0169}\vfill
\and Depto. Estad\'istica e Investigaci\'on Operativa. Universidad de C\'adiz, Avda. Rep\'ublica Saharaui s/n, 11510 Puerto Real, C\'adiz, Spain\relax                                                                                                                                                                                                             \label{inst:0173}\vfill
\and Center for Research and Exploration in Space Science and Technology, University of Maryland Baltimore County, 1000 Hilltop Circle, Baltimore MD, USA\relax                                                                                                                                                                                                    \label{inst:0175}\vfill
\and GSFC - Goddard Space Flight Center, Code 698, 8800 Greenbelt Rd, 20771 MD Greenbelt, United States\relax                                                                                                                                                                                                                                                      \label{inst:0176}\vfill
\and EURIX S.r.l., Corso Vittorio Emanuele II 61, 10128, Torino, Italy\relax                                                                                                                                                                                                                                                                                       \label{inst:0178}\vfill
\and Porter School of the Environment and Earth Sciences, Tel Aviv University, Tel Aviv 6997801, Israel\relax                                                                                                                                                                                                                                                      \label{inst:0179}\vfill
\and Harvard-Smithsonian Center for Astrophysics, 60 Garden St., MS 15, Cambridge, MA 02138, USA\relax                                                                                                                                                                                                                                                             \label{inst:0180}\vfill
\and HE Space Operations BV for European Space Agency (ESA), Camino bajo del Castillo, s/n, Urbanizacion Villafranca del Castillo, Villanueva de la Ca\~{n}ada, 28692 Madrid, Spain\relax                                                                                                                                                                          \label{inst:0182}\vfill
\and Instituto de Astrof\'{i}sica e Ci\^{e}ncias do Espa\c{c}o, Universidade do Porto, CAUP, Rua das Estrelas, PT4150-762 Porto, Portugal\relax                                                                                                                                                                                                                    \label{inst:0183}\vfill
\and LFCA/DAS,Universidad de Chile,CNRS,Casilla 36-D, Santiago, Chile\relax                                                                                                                                                                                                                                                                                        \label{inst:0185}\vfill
\and SISSA - Scuola Internazionale Superiore di Studi Avanzati, via Bonomea 265, 34136 Trieste, Italy\relax                                                                                                                                                                                                                                                        \label{inst:0189}\vfill
\and Telespazio for CNES Centre Spatial de Toulouse, 18 avenue Edouard Belin, 31401 Toulouse Cedex 9, France\relax                                                                                                                                                                                                                                                 \label{inst:0194}\vfill
\and University of Turin, Department of Computer Sciences, Corso Svizzera 185, 10149 Torino, Italy\relax                                                                                                                                                                                                                                                           \label{inst:0198}\vfill
\and Dpto. de Matem\'{a}tica Aplicada y Ciencias de la Computaci\'{o}n, Univ. de Cantabria, ETS Ingenieros de Caminos, Canales y Puertos, Avda. de los Castros s/n, 39005 Santander, Spain\relax                                                                                                                                                                   \label{inst:0201}\vfill
\and Centro de Astronom\'{i}a - CITEVA, Universidad de Antofagasta, Avenida Angamos 601, Antofagasta 1270300, Chile\relax                                                                                                                                                                                                                                          \label{inst:0211}\vfill
\and DLR Gesellschaft f\"{ u}r Raumfahrtanwendungen (GfR) mbH M\"{ u}nchener Stra{\ss}e 20 , 82234 We{\ss}ling\relax                                                                                                                                                                                                                                               \label{inst:0216}\vfill
\and Centre for Astrophysics Research, University of Hertfordshire, College Lane, AL10 9AB, Hatfield, United Kingdom\relax                                                                                                                                                                                                                                         \label{inst:0218}\vfill
\and University of Turin, Mathematical Department ""G.Peano"", Via Carlo Alberto 10, 10123 Torino, Italy\relax                                                                                                                                                                                                                                                     \label{inst:0225}\vfill
\and INAF - Osservatorio Astronomico d'Abruzzo, Via Mentore Maggini, 64100 Teramo, Italy\relax                                                                                                                                                                                                                                                                     \label{inst:0231}\vfill
\and Instituto de Astronomia, Geof\`{i}sica e Ci\^{e}ncias Atmosf\'{e}ricas, Universidade de S\~{a}o Paulo, Rua do Mat\~{a}o, 1226, Cidade Universitaria, 05508-900 S\~{a}o Paulo, SP, Brazil\relax                                                                                                                                                                \label{inst:0234}\vfill
\and APAVE SUDEUROPE SAS for CNES Centre Spatial de Toulouse, 18 avenue Edouard Belin, 31401 Toulouse Cedex 9, France\relax                                                                                                                                                                                                                                        \label{inst:0241}\vfill
\and M\'{e}socentre de calcul de Franche-Comt\'{e}, Universit\'{e} de Franche-Comt\'{e}, 16 route de Gray, 25030 Besan\c{c}on Cedex, France\relax                                                                                                                                                                                                                  \label{inst:0244}\vfill
\and ATOS for CNES Centre Spatial de Toulouse, 18 avenue Edouard Belin, 31401 Toulouse Cedex 9, France\relax                                                                                                                                                                                                                                                       \label{inst:0249}\vfill
\and School of Physics and Astronomy, Tel Aviv University, Tel Aviv 6997801, Israel\relax                                                                                                                                                                                                                                                                          \label{inst:0252}\vfill
\and Astrophysics Research Centre, School of Mathematics and Physics, Queen's University Belfast, Belfast BT7 1NN, UK\relax                                                                                                                                                                                                                                        \label{inst:0254}\vfill
\and Centre de Donn\'{e}es Astronomique de Strasbourg, Strasbourg, France\relax                                                                                                                                                                                                                                                                                    \label{inst:0256}\vfill
\and Institute for Computational Cosmology, Department of Physics, Durham University, Durham DH1 3LE, UK\relax                                                                                                                                                                                                                                                     \label{inst:0260}\vfill
\and European Southern Observatory, Karl-Schwarzschild-Str. 2, 85748 Garching, Germany\relax                                                                                                                                                                                                                                                                       \label{inst:0261}\vfill
\and Max-Planck-Institut f\"{ u}r Astrophysik, Karl-Schwarzschild-Stra{\ss}e 1, 85748 Garching, Germany\relax                                                                                                                                                                                                                                                      \label{inst:0262}\vfill
\and Data Science and Big Data Lab, Pablo de Olavide University, 41013, Seville, Spain\relax                                                                                                                                                                                                                                                                       \label{inst:0266}\vfill
\and Barcelona Supercomputing Center (BSC), Pla\c{c}a Eusebi G\"{ u}ell 1-3, 08034-Barcelona, Spain\relax                                                                                                                                                                                                                                                          \label{inst:0273}\vfill
\and ETSE Telecomunicaci\'{o}n, Universidade de Vigo, Campus Lagoas-Marcosende, 36310 Vigo, Galicia, Spain\relax                                                                                                                                                                                                                                                   \label{inst:0278}\vfill
\and Asteroid Engineering Laboratory, Space Systems, Lule\aa{} University of Technology, Box 848, S-981 28 Kiruna, Sweden\relax                                                                                                                                                                                                                                    \label{inst:0282}\vfill
\and Vera C Rubin Observatory,  950 N. Cherry Avenue, Tucson, AZ 85719, USA\relax                                                                                                                                                                                                                                                                                  \label{inst:0287}\vfill
\and Department of Astrophysics, Astronomy and Mechanics, National and Kapodistrian University of Athens, Panepistimiopolis, Zografos, 15783 Athens, Greece\relax                                                                                                                                                                                                  \label{inst:0288}\vfill
\and TRUMPF Photonic Components GmbH, Lise-Meitner-Stra{\ss}e 13,  89081 Ulm, Germany\relax                                                                                                                                                                                                                                                                        \label{inst:0291}\vfill
\and IAC - Instituto de Astrofisica de Canarias, Via L\'{a}ctea s/n, 38200 La Laguna S.C., Tenerife, Spain\relax                                                                                                                                                                                                                                                   \label{inst:0296}\vfill
\and Department of Astrophysics, University of La Laguna, Via L\'{a}ctea s/n, 38200 La Laguna S.C., Tenerife, Spain\relax                                                                                                                                                                                                                                          \label{inst:0297}\vfill
\and Faculty of Aerospace Engineering, Delft University of Technology, Kluyverweg 1, 2629 HS Delft, The Netherlands\relax                                                                                                                                                                                                                                          \label{inst:0299}\vfill
\and Radagast Solutions\relax                                                                                                                                                                                                                                                                                                                                      \label{inst:0303}\vfill
\and Laboratoire Univers et Particules de Montpellier, CNRS Universit\'{e} Montpellier, Place Eug\`{e}ne Bataillon, CC72, 34095 Montpellier Cedex 05, France\relax                                                                                                                                                                                                 \label{inst:0304}\vfill
\and Universit\'{e} de Caen Normandie, C\^{o}te de Nacre Boulevard Mar\'{e}chal Juin, 14032 Caen, France\relax                                                                                                                                                                                                                                                     \label{inst:0310}\vfill
\and LESIA, Observatoire de Paris, Universit\'{e} PSL, CNRS, Sorbonne Universit\'{e}, Universit\'{e} de Paris, 5 Place Jules Janssen, 92190 Meudon, France\relax                                                                                                                                                                                                   \label{inst:0311}\vfill
\and SRON Netherlands Institute for Space Research, Niels Bohrweg 4, 2333 CA Leiden, The Netherlands\relax                                                                                                                                                                                                                                                         \label{inst:0319}\vfill
\and Astronomical Observatory, University of Warsaw,  Al. Ujazdowskie 4, 00-478 Warszawa, Poland\relax                                                                                                                                                                                                                                                             \label{inst:0320}\vfill
\and Scalian for CNES Centre Spatial de Toulouse, 18 avenue Edouard Belin, 31401 Toulouse Cedex 9, France\relax                                                                                                                                                                                                                                                    \label{inst:0322}\vfill
\and Universit\'{e} Rennes, CNRS, IPR (Institut de Physique de Rennes) - UMR 6251, 35000 Rennes, France\relax                                                                                                                                                                                                                                                      \label{inst:0328}\vfill
\and INAF - Osservatorio Astronomico di Capodimonte, Via Moiariello 16, 80131, Napoli, Italy\relax                                                                                                                                                                                                                                                                 \label{inst:0330}\vfill
\and Shanghai Astronomical Observatory, Chinese Academy of Sciences, 80 Nandan Road, Shanghai 200030, People's Republic of China\relax                                                                                                                                                                                                                             \label{inst:0333}\vfill
\and University of Chinese Academy of Sciences, No.19(A) Yuquan Road, Shijingshan District, Beijing 100049, People's Republic of China\relax                                                                                                                                                                                                                       \label{inst:0335}\vfill
\and Niels Bohr Institute, University of Copenhagen, Juliane Maries Vej 30, 2100 Copenhagen {\O}, Denmark\relax                                                                                                                                                                                                                                                    \label{inst:0338}\vfill
\and DXC Technology, Retortvej 8, 2500 Valby, Denmark\relax                                                                                                                                                                                                                                                                                                        \label{inst:0339}\vfill
\and Las Cumbres Observatory, 6740 Cortona Drive Suite 102, Goleta, CA 93117, USA\relax                                                                                                                                                                                                                                                                            \label{inst:0340}\vfill
\and CIGUS CITIC, Department of Nautical Sciences and Marine Engineering, University of A Coru\~{n}a, Paseo de Ronda 51, 15071, A Coru\~{n}a, Spain\relax                                                                                                                                                                                                          \label{inst:0348}\vfill
\and Astrophysics Research Institute, Liverpool John Moores University, 146 Brownlow Hill, Liverpool L3 5RF, United Kingdom\relax                                                                                                                                                                                                                                  \label{inst:0349}\vfill
\and IPAC, Mail Code 100-22, California Institute of Technology, 1200 E. California Blvd., Pasadena, CA 91125, USA\relax                                                                                                                                                                                                                                           \label{inst:0356}\vfill
\and IRAP, Universit\'{e} de Toulouse, CNRS, UPS, CNES, 9 Av. colonel Roche, BP 44346, 31028 Toulouse Cedex 4, France\relax                                                                                                                                                                                                                                        \label{inst:0357}\vfill
\and MTA CSFK Lend\"{ u}let Near-Field Cosmology Research Group, Konkoly Observatory, MTA Research Centre for Astronomy and Earth Sciences, Konkoly Thege Mikl\'{o}s \'{u}t 15-17, 1121 Budapest, Hungary\relax                                                                                                                                                    \label{inst:0374}\vfill
\and Departmento de F\'{i}sica de la Tierra y Astrof\'{i}sica, Universidad Complutense de Madrid, 28040 Madrid, Spain\relax                                                                                                                                                                                                                                        \label{inst:0390}\vfill
\and Ru{\dj}er Bo\v{s}kovi\'{c} Institute, Bijeni\v{c}ka cesta 54, 10000 Zagreb, Croatia\relax                                                                                                                                                                                                                                                                     \label{inst:0396}\vfill
\and Villanova University, Department of Astrophysics and Planetary Science, 800 E Lancaster Avenue, Villanova PA 19085, USA\relax                                                                                                                                                                                                                                 \label{inst:0412}\vfill
\and INAF - Osservatorio Astronomico di Brera, via E. Bianchi, 46, 23807 Merate (LC), Italy\relax                                                                                                                                                                                                                                                                  \label{inst:0418}\vfill
\and STFC, Rutherford Appleton Laboratory, Harwell, Didcot, OX11 0QX, United Kingdom\relax                                                                                                                                                                                                                                                                         \label{inst:0425}\vfill
\and Charles University, Faculty of Mathematics and Physics, Astronomical Institute of Charles University, V Holesovickach 2, 18000 Prague, Czech Republic\relax                                                                                                                                                                                                   \label{inst:0448}\vfill
\and Department of Particle Physics and Astrophysics, Weizmann Institute of Science, Rehovot 7610001, Israel\relax                                                                                                                                                                                                                                                 \label{inst:0459}\vfill
\and Department of Astrophysical Sciences, 4 Ivy Lane, Princeton University, Princeton NJ 08544, USA\relax                                                                                                                                                                                                                                                         \label{inst:0460}\vfill
\and Departamento de Astrof\'{i}sica, Centro de Astrobiolog\'{i}a (CSIC-INTA), ESA-ESAC. Camino Bajo del Castillo s/n. 28692 Villanueva de la Ca\~{n}ada, Madrid, Spain\relax                                                                                                                                                                                      \label{inst:0468}\vfill
\and naXys, University of Namur, Rempart de la Vierge, 5000 Namur, Belgium\relax                                                                                                                                                                                                                                                                                   \label{inst:0471}\vfill
\and CGI Deutschland B.V. \& Co. KG, Mornewegstr. 30, 64293 Darmstadt, Germany\relax                                                                                                                                                                                                                                                                               \label{inst:0479}\vfill
\and Institute of Global Health, University of Geneva\relax                                                                                                                                                                                                                                                                                                        \label{inst:0480}\vfill
\and Astronomical Observatory Institute, Faculty of Physics, Adam Mickiewicz University, Pozna\'{n}, Poland\relax                                                                                                                                                                                                                                                  \label{inst:0482}\vfill
\and H H Wills Physics Laboratory, University of Bristol, Tyndall Avenue, Bristol BS8 1TL, United Kingdom\relax                                                                                                                                                                                                                                                    \label{inst:0486}\vfill
\and Department of Physics and Astronomy G. Galilei, University of Padova, Vicolo dell'Osservatorio 3, 35122, Padova, Italy\relax                                                                                                                                                                                                                                  \label{inst:0493}\vfill
\and CERN, Geneva, Switzerland\relax                                                                                                                                                                                                                                                                                                                               \label{inst:0495}\vfill
\and Applied Physics Department, Universidade de Vigo, 36310 Vigo, Spain\relax                                                                                                                                                                                                                                                                                     \label{inst:0497}\vfill
\and Association of Universities for Research in Astronomy, 1331 Pennsylvania Ave. NW, Washington, DC 20004, USA\relax                                                                                                                                                                                                                                             \label{inst:0501}\vfill
\and European Southern Observatory, Alonso de C\'ordova 3107, Casilla 19, Santiago, Chile\relax                                                                                                                                                                                                                                                                    \label{inst:0509}\vfill
\and Sorbonne Universit\'{e}, CNRS, UMR7095, Institut d'Astrophysique de Paris, 98bis bd. Arago, 75014 Paris, France\relax                                                                                                                                                                                                                                         \label{inst:0514}\vfill
\and Faculty of Mathematics and Physics, University of Ljubljana, Jadranska ulica 19, 1000 Ljubljana, Slovenia\relax                                                                                                                                                                                                                                               \label{inst:0516}\vfill
}

\date{Received ; accepted }

\abstract
{\gaia\ Data Release 3 (DR3) provides a wealth of new data products for the astronomical community to exploit, including astrophysical parameters for a half billion stars.  In this work we demonstrate
the high quality of these data products and illustrate their use in different astrophysical contexts.  }
{We produce homogeneous samples of  stars with high quality astrophysical parameters by exploiting \gdr{3} while focusing on many regimes across the Hertzsprung-Russell (HR) diagram; spectral types OBA, FGKM, and ultra-cool dwarfs (UCDs).  We also focus on specific sub-samples which are  of particular interest to the community: solar analogues, carbon stars, and the Spectro Photometric Standard Stars (SPSS).}
{We query the astrophysical parameter tables along with other tables in \gdr{3}  to derive the samples of the stars of interest.   
We validate our results by using the \gaia\ catalogue itself and by comparison with external data.
}
{We have produced six homogeneous samples of stars with high quality astrophysical parameters across the HR diagram for the community to exploit.  
We first focus on three samples that span a large parameter space: young massive disk stars (OBA, $\sim$3M), FGKM spectral type stars ($\sim$3M), and UCDs ($\sim$20~K). 
We provide these sources along with additional information (either a flag or complementary parameters) as tables that are made available in the  \gaia\ archive. 
We furthermore identify 15\,740 bone fide carbon stars, 5\,863 solar-analogues, and provide the first homogeneous set of stellar parameters of the SPSS sample.
We demonstrate some applications of these samples in different astrophysical contexts.  
 {We} use a subset of the OBA sample to illustrate its usefulness to analyse the Milky Way rotation curve.
We then use the properties of the FGKM stars to analyse known exoplanet systems.  We also analyse the ages of some unseen UCD-companions to the FGKM stars.    We additionally predict the colours of the Sun in various passbands (\gaia, 2MASS, WISE) using the solar-analogue sample.
}
{
\gdr{3} contains a wealth of new high quality astrophysical parameters for the community to exploit.}


\maketitle

\section{Introduction}
\label{sec:introduction}

The knowledge of astrophysical parameters of stars (APs; effective temperatures, radii etc., see Sect.~\ref{sec:data}) is fundamental for understanding the structure, formation, and evolution of astrophysical systems.  
For example,  exploring chemical distributions of populations of our Galaxy requires well-constrained stellar effective temperatures (\teff) and surface gravities (\loggrav) in order to derive precise and accurate abundances, see e.g. \cite{nissen2018,jofre2019} for reviews. If we want to place our solar system in the context of exoplanet system formation and evolution, we need to determine the radius, mass, and age of many exoplanets and their host stars, e.g. \cite{kaltenegger2015,plato2014,cheops2020}. 
\gdr{3} contains a wealth of new data products. In particular, it provides us with stellar parameters derived from the analysis of the \gaia\  RVS spectra
\citep{DR2-DPACP-47}, the low-resolution spectra produced by the Blue Photometer and the Red Photometer (BP and RP) \citep{2021A&A...652A..86C, EDR3-DPACP-118}, astrometry \citep{2021A&A...649A...2L}, and integrated photometry \citep{2021A&A...649A...3R} for up to 470 million stars \citep{DR3-DPACP-156,DR3-DPACP-157,DR3-DPACP-160,DR3-DPACP-175,DR3-DPACP-186}. The accuracy and precision of these parameters {vary}, as expected, with brightness, distance, stellar types and {the} number of observations. These parameters can be exploited in many ways from detailed studies of individual stars, to statistical studies of large samples of stars or populations. 
This catalogue, based uniquely on \gaia\ data, also  has a long-term legacy value as a rich database for target selection for future follow-up studies and missions. 

In this work we demonstrate the potential of the new data products in \gdr{3} by 
producing very high quality samples of astrophysical parameters of stars all across the HR diagram.  
We aim to make clean samples of stars based on severe quality cuts.  We consider these  sources to have the most accurate and precise stellar properties in this catalogue that can be used on a star-by-star basis.  
These quality cuts, however, have an important impact on the selection function and completeness. 
Our selection criteria will not be optimal for many user's specific scientific cases, and we fully encourage the exploration of the full catalogue of APs in \gdr{3}.

This paper is laid out as follows.  In Sect.~\ref{sec:data} we describe the data products that are used in this work.  Then, in the first part of this analysis, we focus on three main stellar regimes and we produce large high quality samples of stars covering the hot O-, B-, and A-type stars (OBA, Sect.~\ref{sec:oba}), the cooler F-, G-, K, and M-type stars (FGKM, Sect.~\ref{sec:fgkm}), and the sub-stellar ultra-cool dwarfs (UCDs, Sect.~\ref{sec:ucd}).   
We then focus on specific objects of interest:  carbon stars (Sect.~\ref{sec:carbon}), solar analogues (Sect.~\ref{sec:solaranalogues}), and finally the \gaia\  Spectro-Photometric Standards (SPSS, Sect.~\ref{sec:spss}, \citealt{2021MNRAS.503.3660P}).  
Sects.~\ref{sec:oba} -- \ref{sec:spss} are entirely independent sections and a reader can choose to focus only on their section of choice without missing important information for the rest of the paper.  
In Sect.~\ref{sec:summarytables} we describe the six tables from this work that are made available in \gdr{3}, and then in 
Sect.~\ref{sec:exploitation} we illustrate some applications of the various samples in different astrophysical domains.

\section{Data description}
\label{sec:data}

To define our samples of stars, we use primarily the astrophysical parameters (APs) from the \gdrthree\ catalogue.
These data provide us with a uniformly-derived, all-sky catalogue of APs: atmospheric properties (\teff, \logg, \mh, \afe, activity index, emission lines, rotation), abundance estimates for 13 chemical species, evolution characteristics (radius, age, mass, bolometric luminosity), distance, and dust extinction.
The APs are found in two tables of the archive: 
\aptable\ and \apsupptable, and a subset of these are also copied to the \gstable{gaia_source} table for convenience to the user.

These data were produced by the \gaia Data Processing and Analysis Consortium (DPAC) -- Coordination Unit 8 (CU8) using the Astrophysical Parameters Inference Software (\apsis,  \citealt{Apsis2013, DR3-DPACP-157}) and 
a series of three papers describe the methodologies and content of the APs in \gdr{3}.  
\citet{DR3-DPACP-157} presents an overview of the processing, the architecture and the modules of \apsis, along with a summary of the data products. \citet{DR3-DPACP-160} focuses on the stellar content, its description, and quality assessments, and \citet{DR3-DPACP-158} details the non-stellar content, in particular object classification, extinction and extra-galactic objects.

The DPAC data processing chain also uses these APs, for example, to identify the best template spectrum for the  extraction of the radial velocities from the RVS spectra, the identification of quasars used to fix the astrometric reference frame, and the optimization of the BP and RP calibration.

In this work we focus on the  data products produced by six modules of the \apsis\ chain; the General Stellar Parametrizer from Photometry, \gspphot, the General Stellar Parametrizer from Spectroscopy, \gspspec, Extended Stellar Parametrizer for Emission-Line Stars, \espels, 
Extended Stellar Parametrizer for Hot Stars, \esphs, Extended Stellar Parametrizer for Ultra-Cool Dwarfs, \espucd, and
the Final Luminosity Age Mass Estimator, \flame.  
These are described in detail in \cite{DR3-DPACP-157} and \linksec{}{in the online documentation}.  Further details on \gspphot and \gspspec are also found in the dedicated module papers \citep{DR3-DPACP-156,DR3-DPACP-186}.

Briefly, \gspphot\ processes all sources with mean BP and RP spectra \citep{EDR3-DPACP-118,EDR3-DPACP-120} to produce spectroscopic parameters and extinction estimates.  It also uses parallaxes and photometry\footnote{Within the \apsis\ software, the parallaxes are corrected for the known zero-point biases as a function of latitude, magnitude and colour, see \cite{EDR3-DPACP-132}.}.  It processes the sources considering four stellar libraries and the individual results {for each of these libraries are found in the \apsupptable\ table.
The results from the library responsible for the highest log posterior for that source (see   \linktoapparam{astrophysical_parameters}{libname_gspphot}) are those that appear in the main \aptable\ table}.  
\gspspec\ processes sources with mean RVS spectra \citep{DR3-DPACP-154} and produces not only atmospheric parameters but also chemical abundances and the diffuse interstellar band characterisation. These latter products are not the focus of this work, we instead refer readers to \cite{DR3-DPACP-104} and \cite{DR3-DPACP-144} respectively.
The results from \gspspec\ used in this work are found in the \aptable\ table.  
\esphs\ processes both the BP and RP and the RVS spectra when available and by default just the BP and RP spectra.  It produces stellar parameters for stars hotter than 7\,500~K along with a spectral type for all stars.  The \espels module analyses emission-line stars and provides class probabilities and labels, along with a measurement of the H-$\alpha$ equivalent width.  
\espucd\ is a module dedicated to the analysis of UCDs and it produces a \teff.  All of these results are found in the
\aptable\ table.
Finally, \flame\ processes the output spectroscopic parameters from \gspphot\ and \gspspec\ along with astrometry and photometry to derive evolutionary parameters (\radius, \lum, \mass, age).
The \flame\ results based on the \gspphot\ input are found in the \aptable\ table, while those based on the \gspspec\ input are found in the \apsupptable\ table.
These six modules (\gspphot, \gspspec, \espels, \esphs, \espucd, and \flame) produce the data that are the focus of this paper.  For further details on the methods we refer readers to the above references.

This work also exploits other data products from \gdr{3}; the astrometry (parallaxes errors and proper motions) and properties of the photometry and spectroscopy are found in the main \gstable{gaia_source} table and these were also available in \gedr{3}, see also \citealt{DR3-DPACP-161,EDR3-DPACP-128,EDR3-DPACP-132,EDR3-DPACP-117,DR3-DPACP-154}. 
We additionally exploit the variability analysis performed by the Coordination Unit 7 \citep{DR3-DPACP-162,DR3-DPACP-168,DR3-DPACP-170} and the analysis of binary and multiple systems by the Coordination Unit 4 \citep{DR3-DPACP-100,DR3-DPACP-163,DR3-DPACP-176,DR3-DPACP-179} to further define our samples.

\section{OBA stars}\label{sec:oba}

\begin{figure*}[!ht]
    \centering
    \includegraphics[width=\linewidth]{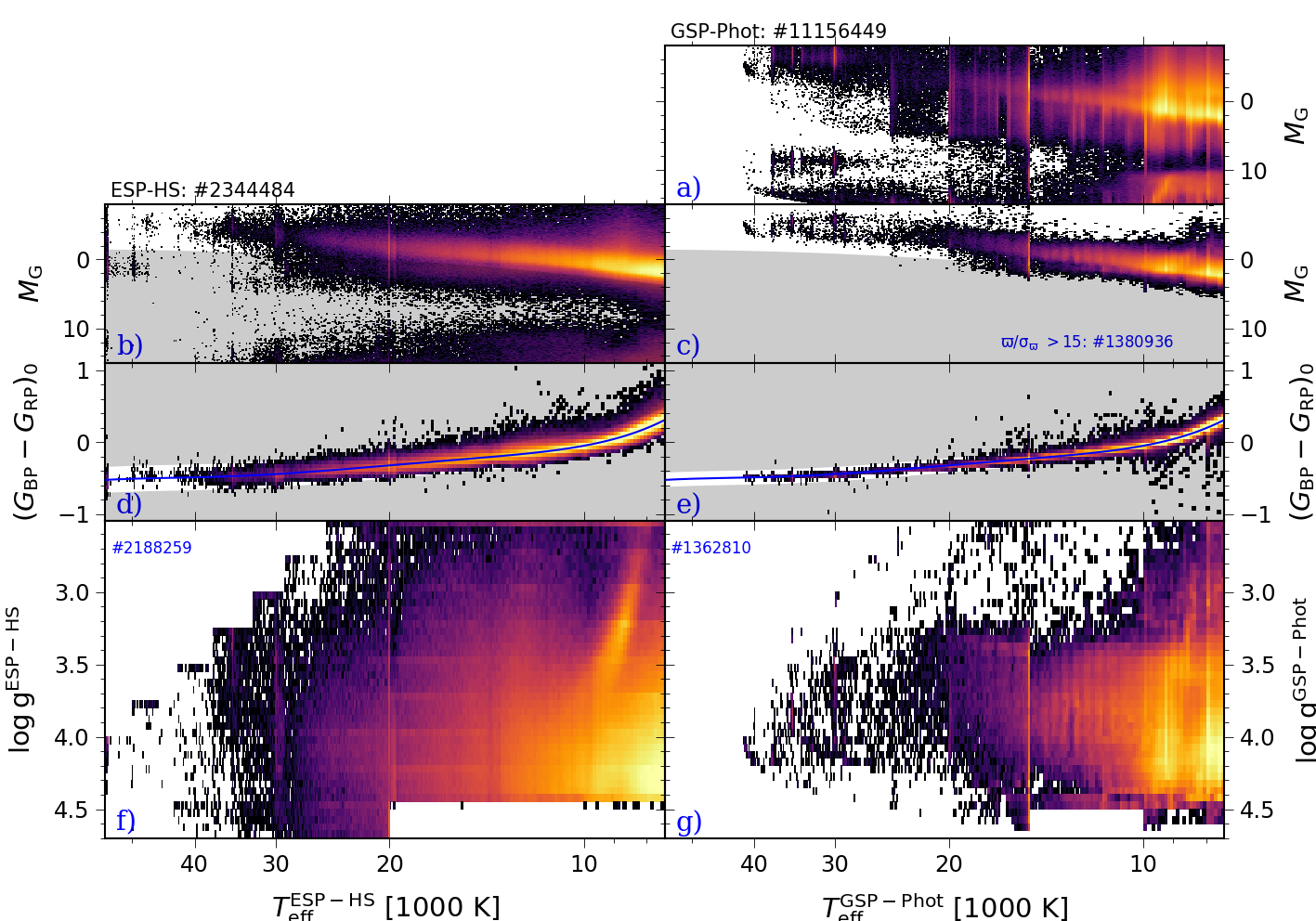}
    \caption{Selection of the OBA sample. Left and right panels show the OBA samples from \esphs and \gspphot, respectively. A first filter on the parallax SNR is applied to the \gspphot targets (from panel a to c). For both samples sub-luminous targets are removed (gray shading in panels b and c), and then the outliers at 6 standard deviations from  the expected colour vs.\ \teff relation (blue line) are filtered out (gray shading in panels d and e). The absolute magnitude $M_\mathrm{G}$ is computed using the measured parallax and the estimated interstellar extinction $A_\mathrm{G}$ provided by both modules. The de-reddened colour, \gbprpzero, is derived using the value of \ebpminrp. The resulting Kiel diagrams are shown in the bottom rows (panels f and g). The over-densities seen at \teff\ = 15\,000~K, 20\,000 K, and 30\,000 K are linked to the temperature limits of the adopted synthetic spectra libraries.
}
    \label{fig:oba.luminosity.limit}
\end{figure*}

\subsection{Scientific motivation}
\label{ssec:hotmotivate}

O- and B- and A-type (OBA) stars are intermediate to large mass stars that evolve rapidly and usually do not migrate very far away from their birth association or cluster. For this reason, they are the best targets to study the structure and dynamics of star forming regions, as well as of the Galactic spiral arms \citep[e.g.][]{DR3-DPACP-75}. Hot stars also play an important role in the evolution of the Galaxy: they are the main contributors to its enrichment in elements heavier than carbon and their strong ultra-violet radiation is the main source of ionisation of the interstellar medium. The most massive OB stars have strong stellar winds and explode as supernovae at the end of their lives, and therefore are an important contributor to the chemical composition of the interstellar medium. Here we focus on the construction of a sample of OBA-type stars which are part of the Milky Way young disk population. Hence older stellar populations covering the same effective temperature range (e.g., white dwarfs and blue horizontal branch stars) are excluded from this sample as much as possible. Because young OBA stars are significantly less numerous than cooler stars, their identification can be considered as a key issue in large surveys. Their spectral energy distribution is less sensitive to effective temperature in comparison to later type stars which hampers the accurate and non-biased determination of \teff. We fixed the lower \teff\ threshold at 7500~K. Moreover, although we have taken care to eliminate the most significant contaminants (e.g. white dwarfs, RR Lyrae, Sect.\,\ref{ssec:hotselection}), our sample still includes stars that are not young OBA stars, such as blue horizontal branch (HB) stars (Sect.\,\ref{sec:kinematics}). In the rest of this section we will continue to refer to the `OBA star sample' as a shorthand for young OBA stars in the disk of the Milky Way.

\subsection{Sample selection}
\label{ssec:hotselection}
%
\gspphot and \esphs are the two main \apsis\ modules that derive the APs of OBA stars. While \gspphot processed all targets with $G\le 19$, \esphs only processed OBA stars brighter than $G=17.65$\footnote{This limitation was imposed during operations in order to remain within the processing schedule, see \linksec{sec_cu8par_intro/ssec_cu8par_intro_softhardware.html}{Sect. 11.1.4 of the online documentation} for details}, and additionally it only processed those stars that 
received a \linktoapparam{astrophysical_parameters}{spectraltype_esphs} tag of $\in$~[`A',`B',`O']. This tag is derived from a random forest classification of the BP and RP spectra, see \linksec{sec_cu8par_apsis/ssec_cu8par_apsis_espels.html}{Sect.~11.3.7 of the online documentation} for details. In terms of effective temperature this is equivalent to selecting targets hotter than $7500$~K. The same lower \teff\ limit is applied to the \gspphot stellar sample. Because \gspphot processes targets down to $G=19$, the corresponding sample initially contains more ($11\,156\,449$ stars) candidate-OBA targets than \esphs ($2\,344\,484$). The \gspphot parametrisation partly relies on the use of parallax, and more outliers (e.g.\ misclassified cool objects, white dwarfs, \ldots) are included when the astrometry is less reliable \citep{DR3-DPACP-160}. To exclude a significant fraction of these we removed all targets based on the ratio of the parallax $\varpi$ to its uncertainty $\sigma_\varpi$ \linktogsparam{gaia_source}{parallax_over_error} $\le 15$, 
as illustrated in Fig.~\ref{fig:oba.luminosity.limit} (panels a and c). \esphs does not use information that allows, for example, to remove white dwarfs. Therefore, we applied a lower luminosity threshold to both samples and removed all sub-luminous objects. The limit was fixed by computing the dispersion around the running median of $M_\mathrm{G}$ as a function of \teff and by using the AP determinations obtained by \esphs in its BP/RP+RVS processing mode. The gray shading in Fig.~\ref{fig:oba.luminosity.limit} (panels b and c) shows the area of the HR diagram from which targets were excluded. Ideally the observed de-reddened $(\bpminrp)$ color vs.\ \teff follows the same relation as the one found from synthetic spectra (Fig.~\ref{fig:oba.luminosity.limit}, panels d and e, blue curve) used to derive the APs. All outliers at more than 6 standard deviations from the theoretical relation were discarded from the sample, as shown by the gray shading in Fig.\,\ref{fig:oba.luminosity.limit} (panels d and e). The Kiel diagram of each sample is shown in the bottom row of the same figure, with the corresponding number of remaining stars. We noticed that the modules were misclassifiying some RR-Lyrae stars as OBA stars, therefore the list was cross-matched with the RR-Lyrae table \varitable{vari_rrlyrae} in \gdr{3} \citep{DR3-DPACP-168}. After filtering, $3\,023\,388$ unique sources remained in the list of candidate-OBA stars. Among these, 1\,661\,459 and 843\,324 have \esphs or \gspphot APs, respectively, while 518\,605 have both. Among those targets with \gspphot parameters, all but 889 received a \linktoapparam{astrophysical_parameters}{spectraltype_esphs} tag.


The corresponding \pvptable{gold_sample_oba_stars} table (this will appear as {\tt gaiadr3.gold\_sample\_oba\_stars}) has two columns: one lists the \texttt{source\_id} and the other a {\tt flag} that provides information on the kinematics of the targets (Sect.~\ref{sec:kinematics}).

\begin{figure}[!ht]
    \centering
    \includegraphics[width=\linewidth]{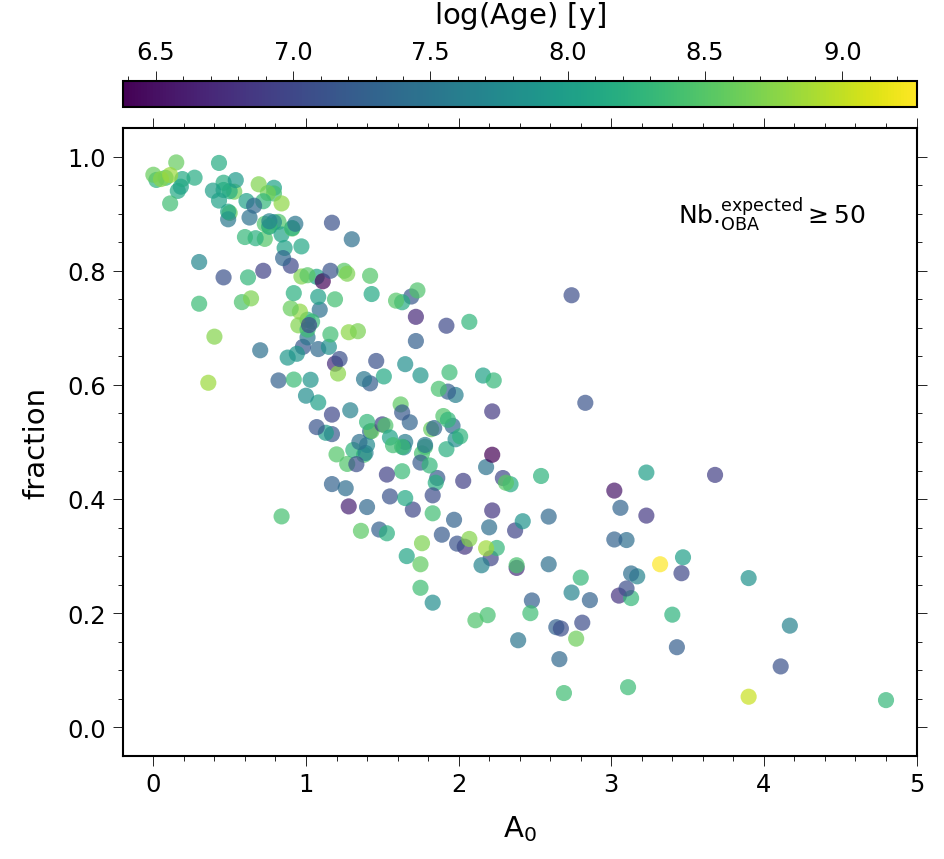}
    \caption{Completeness of the OBA list in various open clusters \citep{2020A&A...640A...1C} as a function of the interstellar extinction. The fraction corresponds to the ratio between the number of cluster members present in our list, and the number of expected OBA stars. The color code follows the cluster age provided by \citet{2020A&A...640A...1C}.
}
    \label{fig:oba.completeness.clusters}
\end{figure}

We tested the completeness of the GDR3 OBA sample by cross-matching it with the Galactic open cluster members identified by \citet{2020A&A...640A...1C}. The selection of the {\it expected} OBA stars in each cluster is based on the $(\bpminrp)_0$ color at $\teff=7500$~K,  estimated by taking into account the published cluster extinction $A_\mathrm{0}$. Their number, $N^\mathrm{expected}_\mathrm{OBA}$, was used to estimate the completeness fraction as follows

\begin{equation}
\mathrm{fraction} = \frac{N^\mathrm{GDR3}_\mathrm{OBA}}{N^\mathrm{expected}_\mathrm{OBA}}
\end{equation}

\noindent where $N^\mathrm{GDR3}_\mathrm{OBA}$ is the number of the OBA open cluster targets found in our sample. We expect the fraction to vary with magnitude and, due to the extinction/temperature degeneracy, with interstellar extinction. We show in Fig.\,\ref{fig:oba.completeness.clusters} how the completeness varies with $A_\mathrm{0}$. The fraction of targets we have in common with the LAMOST OBA \citep{2021arXiv210802878X} and GOSC \citep{2013msao.confE.198M} catalogues are 0.55 and 0.41, respectively. The \teff\ distributions provided by both modules confirm that above 10\,000 K, the \esphs\ APs should be preferred over the {\gspphot\ estimates in the \aptable\ table, whose temperature scale tends to be underestimated in this regime}. This is especially true at higher interstellar extinction.

A Simbad query of the proposed OBA sample provides 34\,055 targets with a confirmed main object type not equal to "*". Among these, 27\% have types not compatible with what would be expected for hot young stars, and of which 79\% are known HB stars. This high density of hot HB stars can be seen, for example, in the bottom panels of Fig.\,\ref{fig:oba.luminosity.limit} where their presence produces a significant overdensity of stars with \teff\ ranging from 8,000 to 10\,000 K and \logg\ lower than 3.5. As explained in the following section, a number of these lower mass evolved targets can be flagged by studying their kinematics.
Furthermore, 134\,498 targets in our list have a spectral type recorded in Simbad, which in 96\% of the cases starts with the letter "O", "B" or "A".


\subsection{Using kinematics to remove halo contaminants}
\label{sec:kinematics}

To further clean the sample of (young) OBA stars from contaminating populations we propose a simple kinematic filter which removes what are presumably blue horizontal branch stars from the halo, which occupy the same colour-brightness space in the colour-magnitude diagram as the OBA stars, as well as the same \teff-\loggrav space in the Kiel diagram. We filter on the tangential velocity $\vtan=A_v(\pmra^2+\pmdec^2)^{1/2}/\varpi$, where \pmra\ and \pmdec\ are the proper motions in the right ascension and declination, and $A_v=4.74074...$~km~yr~s$^{-1}$, using similar limits for the thin disk, thick disk, and halo as in \cite{2018A&A...616A..10G}. Thin disk stars are defined as having $\vtan<40$~\kms, thick disk stars as having $40\leq\vtan\leq180$~\kms, and halo stars have $\vtan>180$~\kms. We next illustrate the effects of this kinematic selection and thereby focus on stars for which $\varpi/\sigma_\varpi>10$. This parallax quality cut ensures a reliable calculation of the tangential velocities.

\begin{figure}[h]
    \centering
    \includegraphics[width=\linewidth]{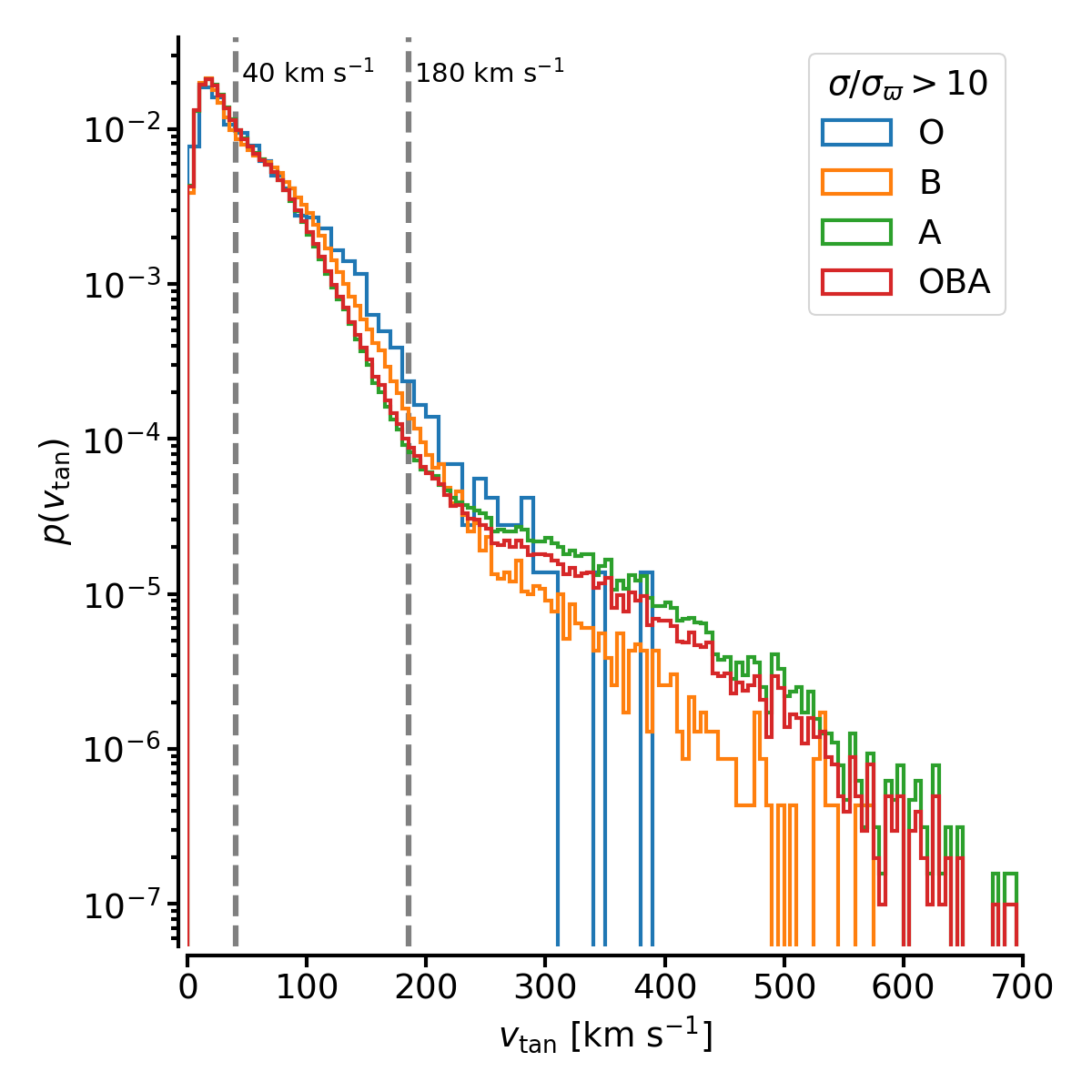}
    \caption{Histogram of tangential velocities of the stars in the OBA sample with $\varpi/\sigma_\varpi>10$. The combined OBA star sample is shown as well as the individual O, B, and A star samples (based on the classifications from the \esphs module). The limits in tangential velocity separating the thin disk, thick disk, and halo populations are shown as vertical dashed lines.}
    \label{fig:obavtan}
\end{figure}

\begin{figure}[h]
    \centering
    \includegraphics[width=\linewidth]{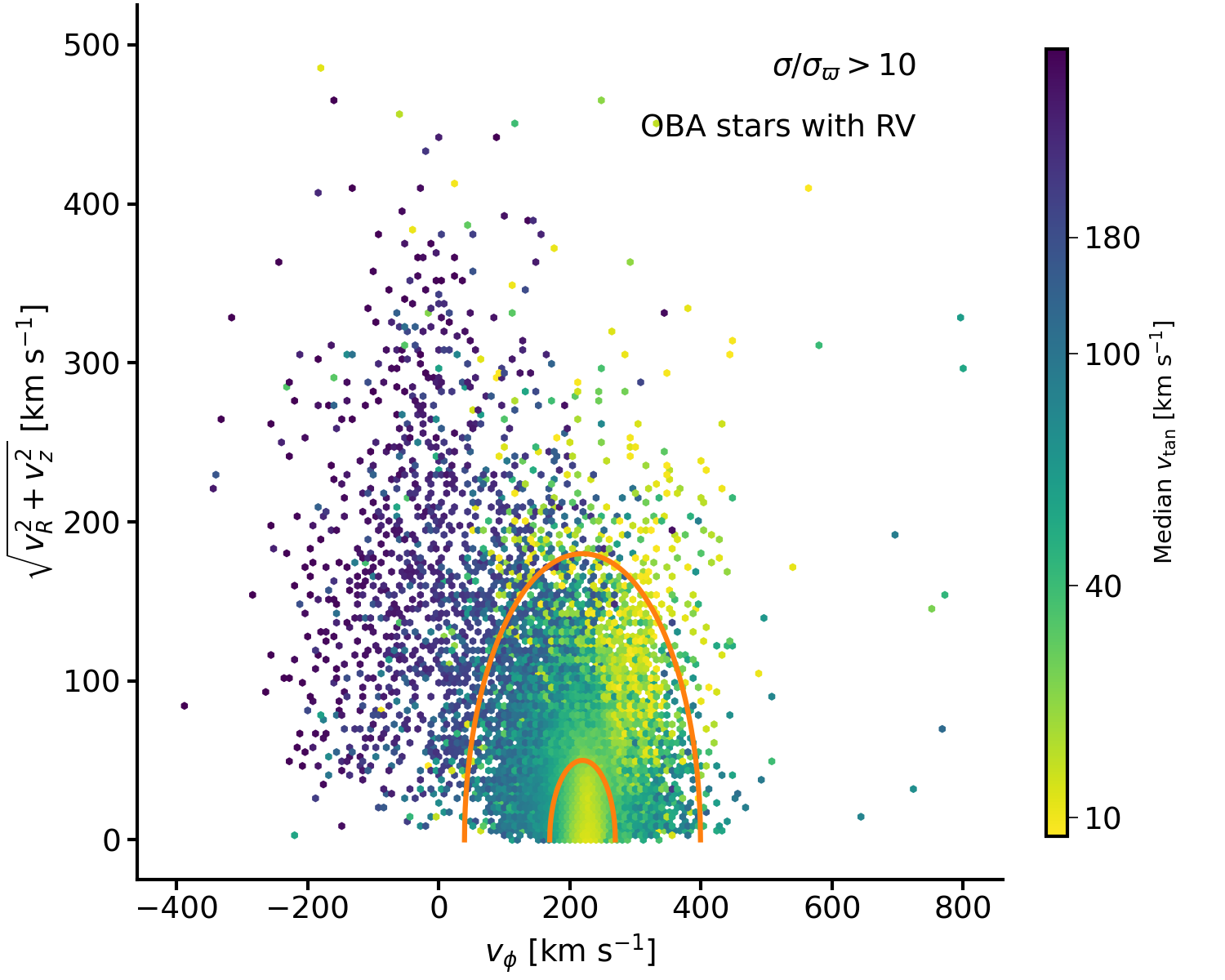}
    \caption{Toomre diagram for the OBA stars for which a radial velocity is available in \gdr{3}. See text for explanations on the diagram. The colour coding indicates the median value of \vtan at a given location on this diagram. The half-circles indicate limits on the total velocity with respect to the local circular velocity of $50$ and $180$~\kms.}
    \label{fig:obatoomre}
\end{figure}

\figrefalt{fig:obavtan} shows the distribution of tangential velocities for the OBA star sub-sample for which $\varpi/\sigma_\varpi>10$. The vertical dashed lines indicate the above limits on \vtan and these correspond well to the inflections in the histograms for the full, B, and A star samples. The O star sample contains almost no sources with $\vtan>180$~\kms. To further explore the tangential velocity selection we show in \figref{fig:obatoomre} the Toomre diagram, which shows $\sqrt{V_R^2+V_z^2}$ along the vertical axis and $V_\phi$ along the horizontal axis, where $(V_R, V_\phi, V_z)$ are the velocity components of the stars in the Galactocentric cylindrical coordinate system, with $R$ pointing from the Galactic centre to the Sun, $z$ along the axis perpendicular to the Galactic plane, and $\phi$ along the azimuthal direction in the Milky Way disk plane (where a left handed coordinate system is used such that the value of $V_\phi$ is positive for prograde stars in the disk). The values of $(V_R, V_\phi, V_z)$ are calculated assuming the local circular velocity from the \texttt{MWPotential2014} Milky Way model \citep{2015ApJS..216...29B}, which is $219$~\kms at the distance of the Sun from the Galactic center \citep[$8277$~pc,][]{2022A&A...657L..12G}. The height of the Sun above the disk plane is assumed to be $20.8$~pc \citep{2019MNRAS.482.1417B} and the peculiar motion of the Sun is assumed to be $(U,V,W)=(11.1, 12.24, 7.25)$~\kms \citep{2010MNRAS.403.1829S}. \figrefalt{fig:obatoomre} only contains stars for which the radial velocity is available in \gdr{3} and the colour coding indicates the value of $\vtan$. The two half-circles indicate the limits on total velocity $\vtot=\sqrt{V_R^2+V_\phi^2+V_z^2}$ of $50$ and $180$~\kms which separate thin disk, thick disk, and halo populations \citep{2018A&A...616A..10G}. In this figure a population of stars can be seen at total velocities of more than $180$~\kms from the local circular velocity and these are most probably halo stars, in particular the population at negative $V_\phi$ which is associated with merger debris in the halo \citep[e.g.,][]{2018Natur.563...85H}.

\begin{figure}[h]
    \centering
    \includegraphics[width=\linewidth]{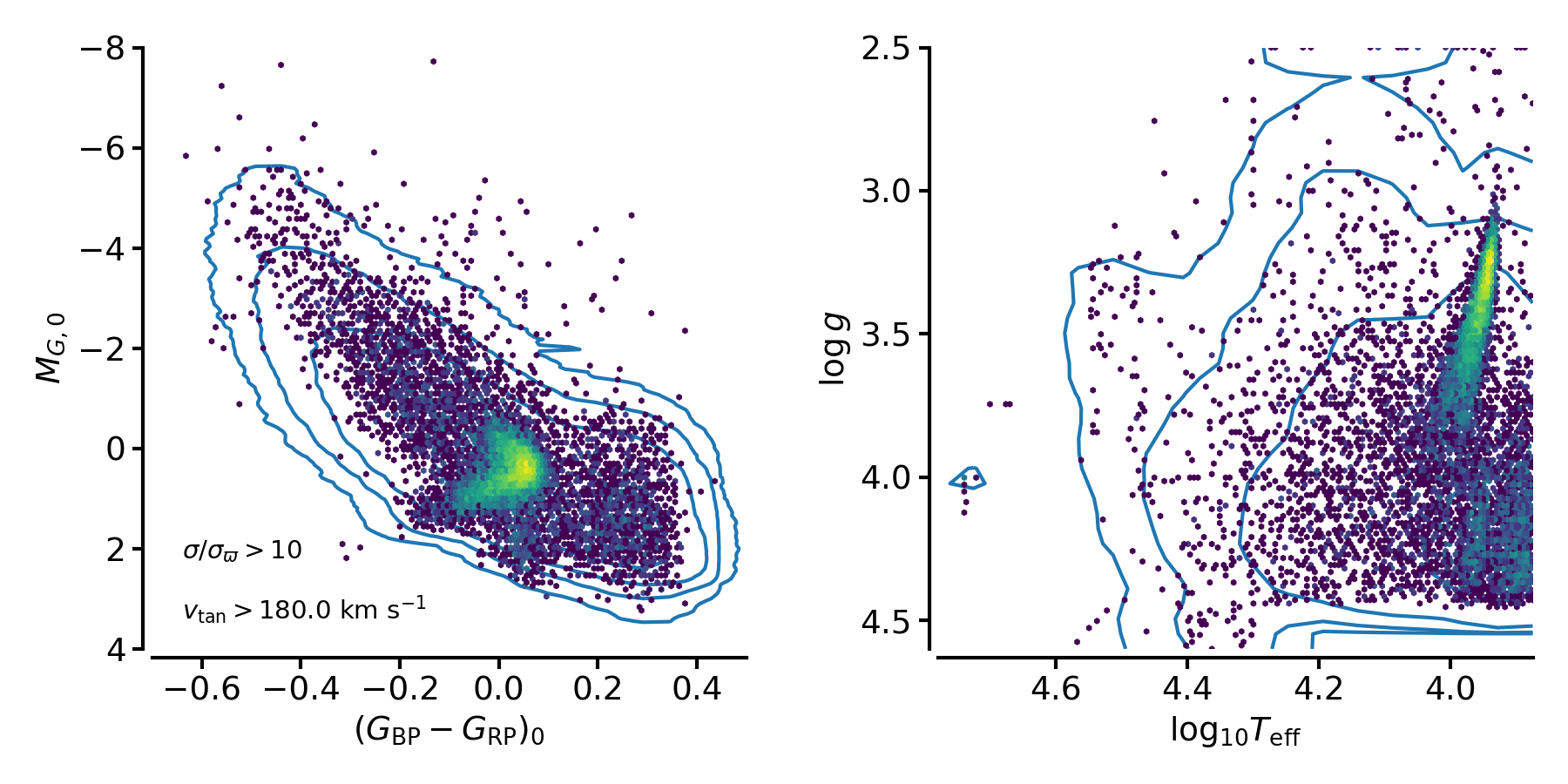}
    \caption{Left: observational Hertzsprung-Russell diagram for the stars in the OBA sample with $\plxsnr>10$. Right: Kiel diagram for the same sample of stars. The contours indicate the distribution of the full sample. The colour coded density images show the stars for which $\vtan>180$~\kms.}
    \label{fig:obacmdv180}
\end{figure}

The colour coding in \figref{fig:obatoomre} suggests that the halo contaminants in the OBA sample can be filtered out by demanding $v_\mathrm{tan}<180$~\kms, although clearly there will be stars left at low tangential velocities which have a total velocity which puts them in the halo. \figrefalt{fig:obacmdv180} shows the observational Hertzsprung-Russell and the Kiel diagrams for the sample of OBA stars with $\plxsnr>10$. Extinction corrections using $A_G$ and $E(\bpminrp)$ from the \esphs module were applied. The contours show the distribution of the full sample, while the colour coded density images show the distribution of stars selected according to $\vtan>180$~\kms. The  velocity-filtered sample mostly occupies the colour magnitude space where blue horizontal branch stars are expected, around $(\bpminrp)_0\sim0.05$ and $M_{G,0}\sim0.5$ \citep[compare to the rightmost panel of figure 21 in][]{2018A&A...616A..10G}. The Kiel diagram shows a prominent feature at $\log_{10}\teff\sim4$, from $\loggrav\sim4$ to $\loggrav\sim2$, corresponding to the known location of the horizontal branch stars in this diagram. These same stars are also primarily located at high galactic latitude as expected for a halo population. A search in SIMBAD \citep{2000A&AS..143....9W} results in 8124 matches for which there is information on the stellar type, of which 5770 are incompatible with stellar types corresponding to hot young stars, including 5499 sources classified as horizontal branch star. This further supports using $\vtan>180$~\kms as a filter to clean the OBA sample from halo star contamination.

\begin{figure*}[ht]
    \centering
    \includegraphics[width=\linewidth]{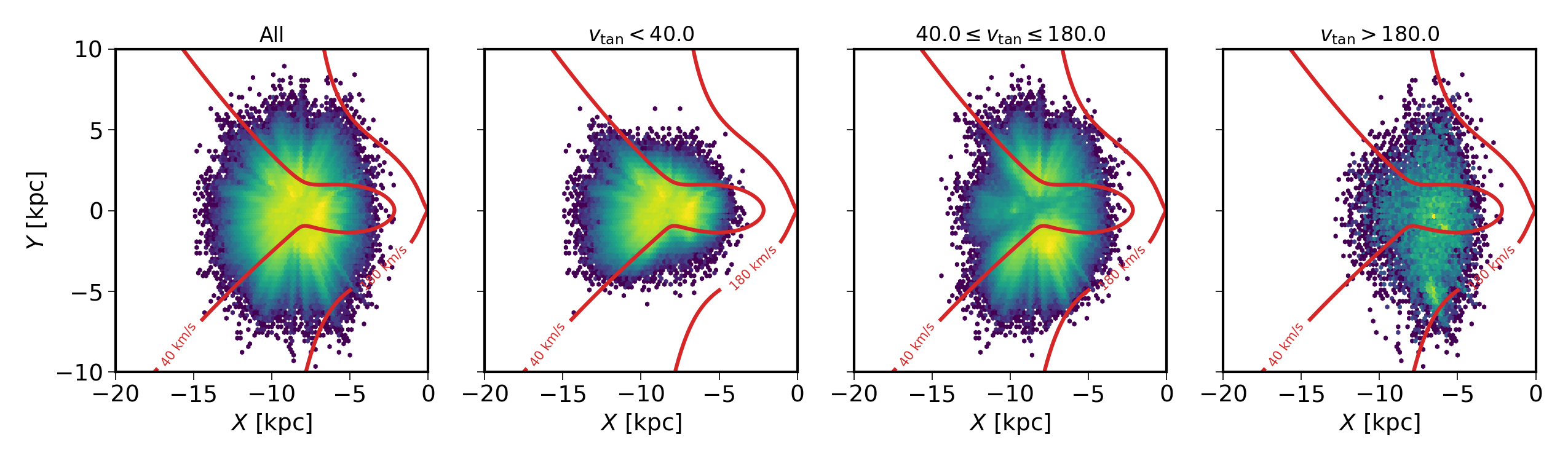}
    \caption{Distribution of stars in the OBA sample projected on the Galactic plane. The galactic centre is to the right at $(X,Y)=(0,0)$ and the Sun is at $(-8,0)$. The panels show from left to right the full sample (with $\plxsnr>10$) and the samples selected according to the \vtan ranges indicated. The red contours indicate lines of constant \vtan, calculated with the simple kinematic disk model as explained in the text.}
    \label{fig:obavtanbias}
\end{figure*}

One might consider further filtering on \vtan, however we caution that because of the large reach of the OBA sample this can lead to significant spatial selection effects. This is illustrated in \figref{fig:obavtanbias}. The figure shows the OBA stars with  $\varpi/\sigma_\varpi>10$ projected on the Galactic plane. The full sample is shown in the leftmost panel and the other panels show the effects of filtering on \vtan. The star positions in Galactocentric coordinates were calculated using the same Milky Way parameters as listed above. The red contours show the limits of $40$~\kms and $180$~\kms on the observed tangential velocities, predicted from a simplistic model of the Milky Way disk kinematics. In this model it is assumed that all stars are located in the disk and follow perfectly circular orbits according to the rotation curve from the \texttt{MWPotential2014} Milky Way model \citep{2015ApJS..216...29B}. The expected values of \vtan are then calculated over a grid of $(X,Y)$ positions, using the method outlined in \cite{2010A&A...510A..34B}. The contours indicate the boundaries between smaller \vtan values to the left and larger ones to the right. The contours show that due to the large reach of \gaia even for stars moving at zero velocity dispersion on circular orbits in the disk one can still expect to observe tangential velocities out to values normally associated with the thick disk and halo. The rightmost panel in \figref{fig:obavtanbias} again confirms that $\vtan>180$~\kms can be used to clean the OBA sample from halo stars, as the stars are all located to the left of the $180$~\kms contour, where these would be expected on the right (in the simple model used) if they were disk stars. The middle panels illustrate the spatial selection bias introduced when further restricting the tangential velocity. The second panel from the right shows that limiting the sample to $\vtan<40$~\kms leads to the exclusion of a significant fraction of young OBA stars which occupy regions of the Galactic disk where the values of \vtan are expected to be larger than $40$~\kms. In addition there is a lack of stars in the sample along the $X=0$ line which roughly follows the shape of the $40$~\kms contour. The simple disk model predicts zero stars there, thus the shape of the gap shows that the model is useful in assessing the spatial selection biases induced by the kinematic selection. In the third panel from the right the shape of the sample distribution also roughly follows the $40$~\kms contour.

In conclusion, we provide a table of  3\,023\,388 young OBA disk stars, cleaned as much as possible to remove older stellar populations, for exploitation by the community.
We recommend to further clean the OBA star sample by applying the kinematic filter $\vtan\leq180$~\kms. Sources with $\vtan>180$~\kms have the flag in the table \pvptable{gold_sample_oba_stars} set to 1, all other sources have the flag set to 0. We have only used the simple Galactic disk kinematic model to make the point that one should be careful not to introduce spatial biases when selecting on kinematics. \cite{2021A&A...650A.112Z} describe a more sophisticated way of employing a simple disk kinematic model to select a clean sample of OBA stars. By assuming the stars follow disk kinematics they can use the observed proper motions to infer distances. Stars with kinematic distances inconsistent with distances based on the parallaxes and photometric information can then be analyzed further to see if they should be removed from the OBA star sample. Further filtering can of course be done on the various data quality indicators available in \gdr{3} (see the following section for examples), and one can also use the astrometric fidelity indicator from \cite{2022MNRAS.510.2597R}.

\section{FGKM stars}\label{sec:fgkm}

\subsection{Scientific motivation}\label{ssec:fgkmmotivation}
F, G, K, and M stars form the majority of the stars of our Milky Way. These stars inform us of how our Galaxy was formed and how it has evolved and are thus the targets of many Milky Way surveys. 
These stars are also the targets of the future ESA PLATO mission \citep{plato2014} which promises to help answer our questions about the formation and evolution of our own Solar System by studying other exoplanet systems.  
In this section we focus on F, G, K and M star types (FGKM) to provide a clean sample of  stars with the following astrophysical parameters : \teff, \logg, \mh, \radius, \mass, age, evolutionary stage, and spectral type.  
Our final sample contains 3\,273\,041 stars after vigorous quality cuts based on astrometric, photometric, and astrophysical parameters, along with other \gaia-based criteria.  

Our sample selection is described in Sects.~\ref{ssec:fgkmsampleselection_phot} and \ref{ssec:fgkmsampleselection_spec} where we analyse the \gspphot-based and \gspspec-based {atmospheric} parameters individually.  
 For both samples we also report on {evolutionary} parameters from \flame and the spectral type from \esphs.   We then perform some additional filtering by removing variables and binaries.  We also further filter on individual parameters from \flame\ and \esphs\ for some sources.
 We validate the target list using open clusters and comparisons with external survey catalogues.
In Sect.~\ref{sec:exploitation} we illustrate two applications of this sample by analysing known transiting exoplanets and studying unseen UCD-companions in the \gaia\ data.

\begin{figure}
    \centering
\includegraphics[width=0.4\textwidth]{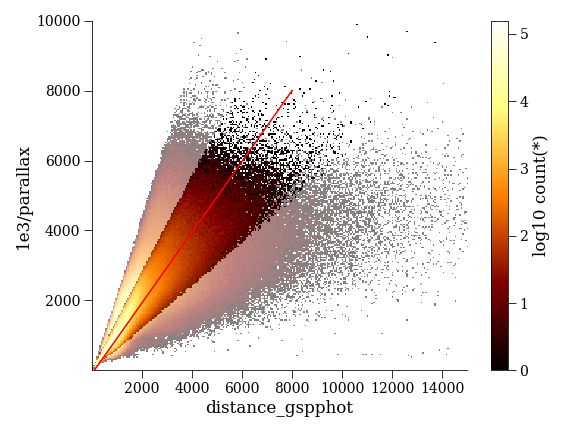}
\includegraphics[width=0.4\textwidth]{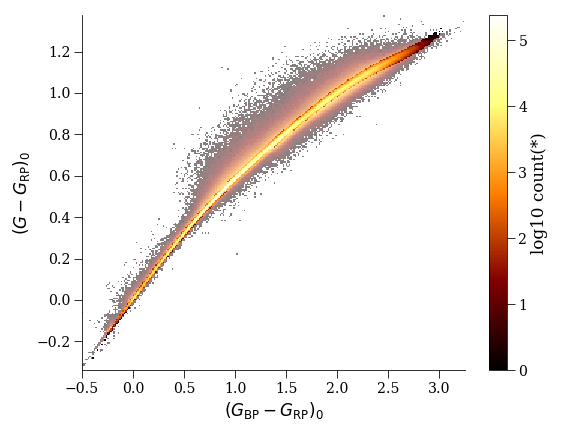}
\caption{Comparisons between sample fgkm\_1 and intermediate samples based on some of the criteria used to define sample fgkm\_2. The top panel illustrates the distance-parallax-error constraint and the lower panel shows the ($\gmag-\grp)_0-\gbprpzero$ relation after imposing the colour-\teff and colour-colour cuts described in Sect.~\ref{ssec:fgkmsampleselection_phot}.  In both panels the sources in fgkm\_1 are shown in the background, while those satisfying the criteria are illustrated in the foreground, colour-coded by logarithmic count.  }
\label{fig:fgk_hrdiag_selection_examples}
\end{figure}

\subsection{\gspphot\ sample selection}\label{ssec:fgkmsampleselection_phot}
\gspphot\ provides stellar and extinction parameters, along with distances, radii, and an absolute magnitude for 470 million stars with $G \le 19$.   
We performed our initial query on the full \gaia\  archive by selecting sources with a parallax signal-to-noise ratio (SNR) better than 10, along with a number of other initial quality cuts based on astrometric and photometric parameters.   
These criteria were based on an analysis of a random set of 2 million sources.
This resulted in a total of 70.4 million stars which we refer to as sample fgkm\_1, and which is described by the following Astronomical Data Query Language (ADQL) query:
\begin{verbatim}
parallax_over_error > 10
ipd_frac_multi_peak < 6
phot_bp_n_blended_transits < 10
teff_gspphot > 2500
\end{verbatim}
in addition to a quality cut on \verb|bp_rp_error < 0.06| ($\sigma_{({\rm BP} - {\rm RP})}$). 
This latter quantity is calculated from a standard propagation of errors using the parameters \verb|phot_bp_mean_flux_over_error| $\left (\frac{f_{\rm BP}}{\sigma{f_{\rm BP}}}\right )$ and \verb|phot_rp_mean_flux_over_error| $\left (\frac{f_{\rm RP}}{\sigma{f_{\rm RP}}}\right )$ from the archive\footnote{
$\sigma_{({\rm BP} - {\rm RP})}^2 = \sqrt{\sigma_{\rm BP}^2 + \sigma_{\rm RP}^2} $, where 
$\sigma_{\rm BP}^2 = \sqrt{\left (\frac{-2.5}{\ln(10)}\frac{ {\sigma f_{\rm BP}}}{f_{\rm BP}} \right )^2 + \sigma_{\rm BP,0}^2}$ and
$\sigma_{\rm RP}^2 = \sqrt{\left (\frac{-2.5}{\ln(10)}\frac{ \sigma{f_{\rm RP}}}{f_{\rm RP}} \right )^2 + \sigma_{\rm RP,0}^2}$, and \href{https://www.cosmos.esa.int/web/gaia/edr3-passbands}{the \gaia\  EDR3 passband zeropoint errors} are 
$\sigma_{\rm BP,0} = 0.00279$ and $\sigma_{\rm RP,0} =  0.00231$.}.  
\begin{table*}
    \centering
    \caption{The coefficients of the polynomials used to fit the \teff\ versus (\bpmag $-$ \gmag)$_0$ and (\gmag $-$ \rpmag)$_0$ versus (\bpmag $-$ \gmag)$_0$ relations in order to remove outliers from the fgkm\_1 sample.
    }
    \label{tab:colourrelations}
    \begin{tabular}{ccccccccccccc}
    \hline\hline
         $y$ & $a_0$ & $a_1$ & $a_2$ & $a_3$ & $a_4$ & $a_5$ & fitted range & $\Delta y$  \\
         \hline
         \teff\ &      9255.55 &     -17911.0 & 27241.4 &  -23103.4
 &      9659.18 &     -1480.37 & 3\,500-7\,500 & 203 \\
         $(\gmag-\rpmag)_0$ &
         0.000373747 &   2.19368 &  -2.95122 &  3.21155 & -1.87172 &     0.418337 & & 0.0267\\
         \hline\hline
    \end{tabular}
    \tablefoot{
    The independent variable $x$ is (\bpmag $-$ \gmag)$_0$, $y$ is the fitted parameter, and $a_i$ are the coefficients of the fit: $y = \sum_{i=0}^5 a_i x^i$. $\Delta y$ is the difference between the fit and the values within which we kept the source.
    }
\end{table*}

\begin{figure}
    \centering
\includegraphics[width=0.48\textwidth]{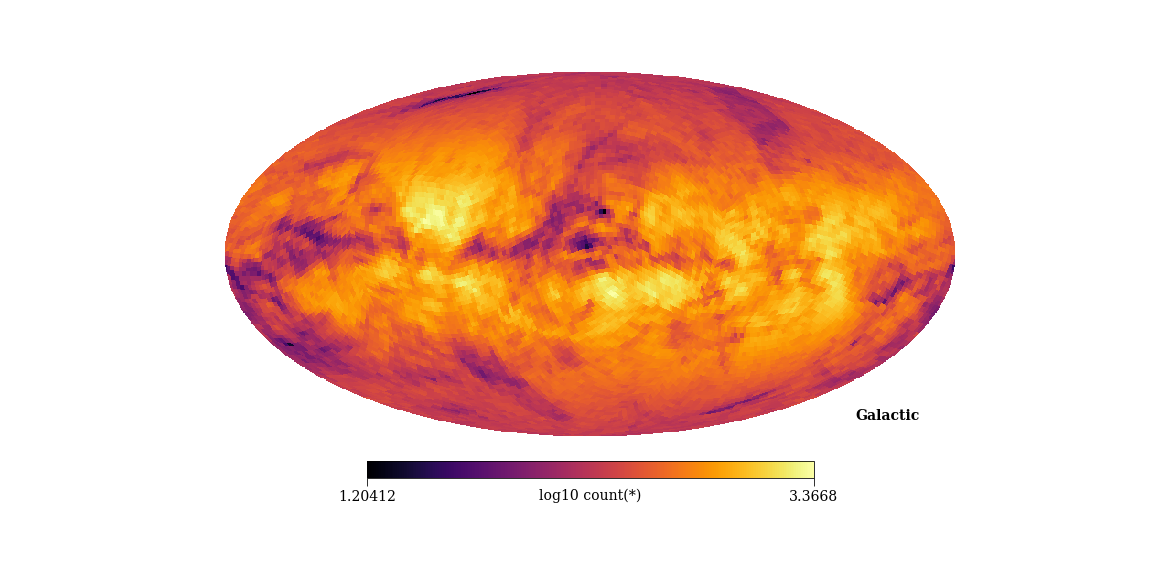}
\includegraphics[width=0.48\textwidth]{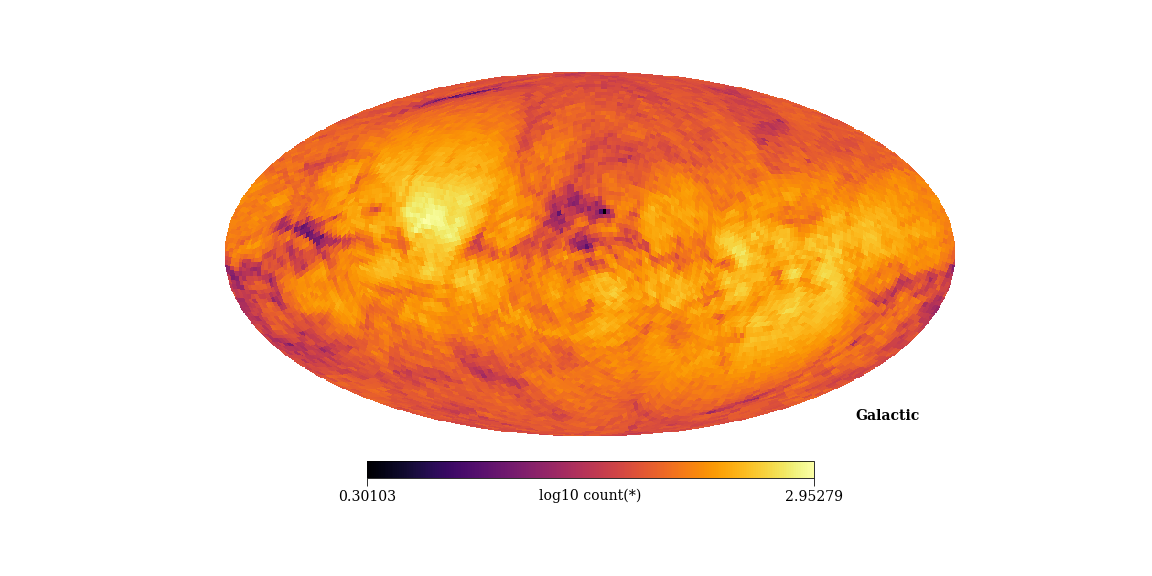}
\caption{ 
Galactic plane projections illustrating the density of sources of the samples fgkm\_2 (top) and fgkm\_3 (bottom).}
\label{fig:fgkm_galactic_nsources}
\end{figure}

We then refined this selection by considering the number of photometric transits, colour-colour  and colour-\teff correlations,  ensuring that the source is classified as a star by \dsc, along with further constraints based on the \gspphot\ parameters themselves. 
These are described in the following paragraphs.

We retained sources whose parameters are within the FGKM regime: \teff $<$ 7500~K, \mg $<$ 12, \radius $<$ 100~$R_\odot$, and had a $\log$ posterior $> -4000$ (goodness-of-fit indicator). We also retained sources with \mh\ $>$ --0.8 which excludes low-metallicity sources with unreliable metallicities \citep{DR3-DPACP-156,DR3-DPACP-157,DR3-DPACP-160}.

\gspphot\ provides four results for each source based on different stellar libraries: MARCS, PHOENIX, A and OB.  Only MARCS and PHOENIX are applicable to the stellar regime considered here.  The results for all libraries are found in the \aptable\ table, and we used the difference between \linktoapparam{astrophysical_parameters_supp}{teff_gspphot_marcs} and 
\linktoapparam{astrophysical_parameters_supp}{teff_gspphot_phoenix} (below called $\Delta \teff$ or {\tt dteff}) as a criterion to further refine the sample.  There is a small bias of up to 100~K between results from these two libraries, due in part to the different spectral energy distributions (SEDs) of the different models and this bias varies with stellar parameters.
We therefore selected those sources where the two values modulo the peak offset were in agreement, i.e. $|\Delta\teff + 65| < 150$~K.
Additionally, we only retained sources when their uncertainties (upper $-$ lower) are $<$ 150~K, and the sources for which the
``best'' model is the MARCS one (75\% of sample), i.e. 
\linktoapparam{astrophysical_parameters}{libname_gspphot}
= ``MARCS'' in the \aptable\ table.  
These strict criteria based on \teff\ removed about 70\% of the sources.
We also imposed that 
\linktoapparam{astrophysical_parameters}{distance_gspphot}
was less than the distance corresponding to the \linktogsparam{gaia_source}{parallax} decreased by 4 times 
\linktogsparam{gaia_source}{parallax_error}
(and vice versa).  
Fig.~\ref{fig:fgk_hrdiag_selection_examples} top panel illustrates the impact of the cut based on distance.  The sources in fgkm\_1 are shown in the background, and those with the distance criteria applied (40\%) are shown in the foreground.  We also show the one-to-one line to guide the eye.

We corrected the \bpmag, \rpmag, and \gmag\ observed colours for the interstellar extinction provided by \gspphot: \bpmag$_0$ = \bpmag\ $-$ A$_{\rm BP}$, \rpmag$_0$ = \rpmag\ $-$ A$_{\rm RP}$, and \gmag$_0$ = \gmag\ $-$ A$_{G}$.  Then we fitted polynomials to the \teff\ versus (\bpmag $-$ \gmag)$_0$ (difference between fit and values denoted as {\tt dtb}) and (\gmag $-$ \rpmag)$_0$ versus (\bpmag $-$ \gmag)$_0$ (difference denoted as {\tt dgb}), and used these polynomial fits to remove sources further than $3\sigma$ ($\sim$7\% of fgkm\_1).  
The coefficients of the polynomials are given in Table~\ref{tab:colourrelations}.  
The bottom panel of  Fig.~\ref{fig:fgk_hrdiag_selection_examples} illustrates the ($\gmag-\grp)_0-\gbprpzero$ relation for sample fgkm\_1 in the background and the sample with the 3$\sigma$ constraints on the colour-colour and the colour-\teff\ relations in the foreground.

All of the above criteria along with further constraints on \dsc\ class probabilities and number of transits ({\tt n\_obs} below) were used to define the sample fgkm\_2 which resulted in a total of 6.3M sources i.e. 12.5\% of the fgkm\_1 sample.  
A projection of the retained sources on the Galactic plane is shown in Fig.~\ref{fig:fgkm_galactic_nsources}. We note that the criteria on the number of transits were adjusted in order to ensure a full-sky coverage. The full list of constraints for sample fgkm\_2 is summarised as follows:  
\begin{verbatim}
|dgb| < 203,
|dtb| < 0.0267
|dteff + 65| < 150
libname_gspphot = "MARCS"
teff_gspphot_upper-teff_gspphot_lower < 150
teff_gspphot < 7500 
mh_gspphot > -0.8 
distance_gspphot < 1e3/(parallax-4*parallax_error) 
distance_gspphot > 1e3/(parallax+4*parallax_error)
radius_gspphot < 100 
mg_gspphot < 12
logposterior_gspphot > -4000
classprob_dsc_combmod_star > 0.9
phot_bp_n_obs > 19
phot_rp_n_obs > 19
phot_g_n_obs > 150
\end{verbatim}

\begin{figure*}
    \centering

\includegraphics[width=0.48\textwidth]{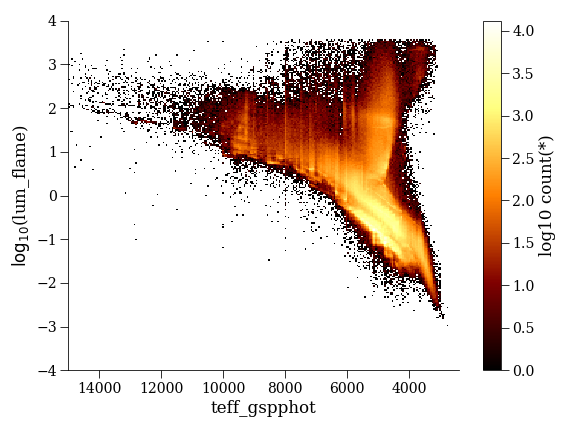}
\includegraphics[width=0.48\textwidth]{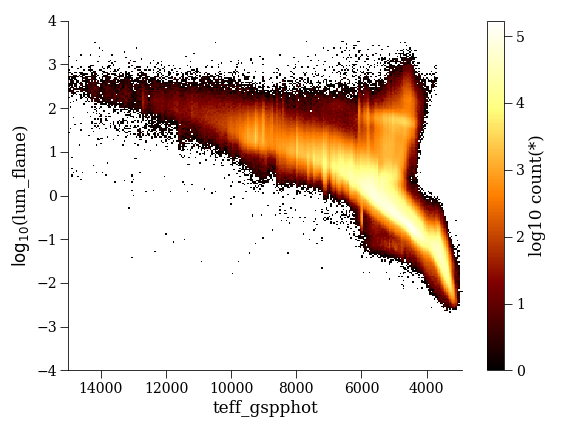}
\includegraphics[width=0.48\textwidth]{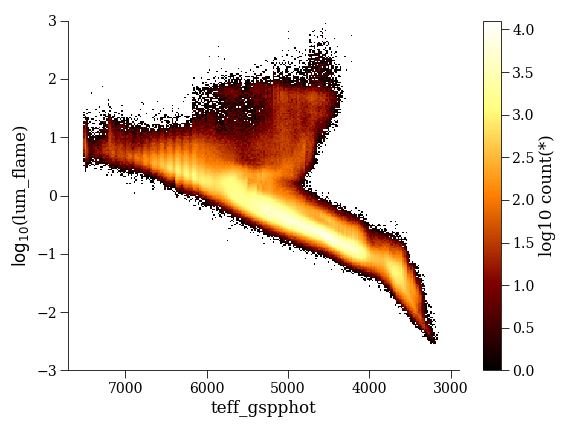}
\includegraphics[width=0.48\textwidth]{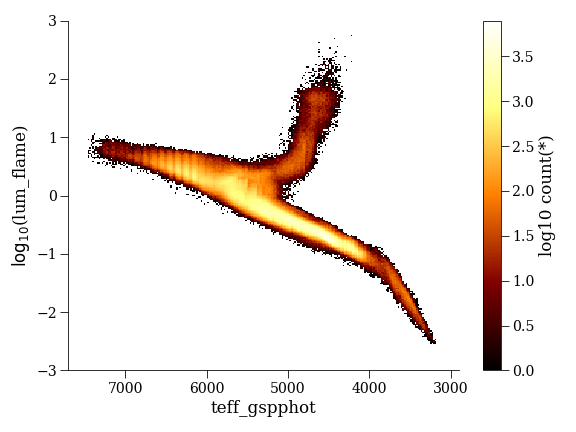}
\caption{HR diagram based on \gspphot\ and \flame\ for the definition of the FGKM sample.  The top left panel illustrates the HR diagram before any selection is made using a random sample of 2M stars. The rest of the panels show the various quality cuts.  Top right is fgkm\_1, bottom left is fgkm\_2, and bottom right is fgkm\_3 before cleaning for variables and binaries.}
\label{fig:fgk_hrdiag_selectionprocess}
\end{figure*}

\begin{figure}
    \centering

\includegraphics[width=0.48\textwidth]{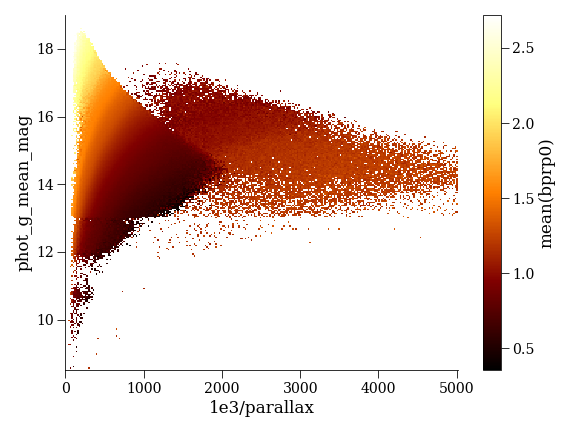}
\caption{Distribution of the final sample fgkm\_3 of the observed parameters \gmag\ and parallax, colour-coded by \gbprpzero.}
\label{fig:fgkm_plx_gmag_bprp0}
\end{figure}

The final selection, fgkm\_3, was done by applying different quality cuts based on the position of the star in the HR diagram. 
Giants were defined as $\logg < 3.6$ and $\teff < 5900$~K, and outliers were removed by retaining sources with $\logg < 0.34 M_G + 2.45$.
Subgiants were defined as  $3.6 \le \logg \le 4.0$ and $\teff < 5900$~K, and outliers were removed by retaining sources with $\logg < 0.75 M_G + 1.13$.  
Main sequence stars were defined as $\logg > 4.0$ and $\teff < 7450$~K, and we imposed a further constraint of 
\linktogsparam{gaia_source}{parallax_over_error} $> 33.34$ in order to have sources with relative errors on \radius and \lum with contributions of parallax errors at 3\% or less.
To further refine the main sequence sample we applied different criteria in three different  colour regimes.   
For $x < 0.98$ where $x = \gbprpzero$ no further selection was done.
For $0.98 \le x \le 1.8$ we removed the sequence of young pre-main sequence stars and binaries by retaining sources that satisfied $\log L < 2.32 - 3.20x + 0.78x^2 $ where $L$ is {\tt lum\_flame}.
For $x>1.8$ we retained sources that satisfied $\logg < 8.525-6.950x+3.680x^2-0.584x^3$.
This final refinement resulted in a total sample size of {3\,530\,174} sources.

We illustrate the different selection criteria in the HR diagram in  Fig.~\ref{fig:fgk_hrdiag_selectionprocess}.  The top left panel shows the HR diagram using a random sample of data from the \gaia\  archive and imposing only that \teff\ and \lum\ exist.  The top right panel shows the selection of sources after applying the ADQL search criteria (fgkm\_1) which is dominated by the criterion on parallax SNR.  One can see that already many outliers and artefacts have been removed with this cut.  
The bottom left panel shows the sample fgkm\_2 where constraints were based on the \gspphot\ parameters, along with further constraints on the photometry and \dsc\ class probabilities. The HR diagram has not changed drastically, but the quality of the data in fgkm\_2 is much higher than in fgkm\_1.
Finally, the bottom right panel illustrates sample fgkm\_3 where the sample was separated into five parts (giants, subgiants, upper, middle, and lower main sequence as described above) and different polynomial cuts were applied based on \logg, \mg, \lum, and \gbprpzero. 

The galactic projection of density of sources is illustrated in the bottom panel of
Fig.~\ref{fig:fgkm_galactic_nsources}.  We also illustrate the distribution of the observable parameters, \gmag, parallax, and \gbprpzero\ in Fig.~\ref{fig:fgkm_plx_gmag_bprp0}.
The main sequence stars occupy the dense triangular region and extend to approximately 1900 parsec for the hottest stars.

\subsection{\gspspec\ sample selection}
\label{ssec:fgkmsampleselection_spec}
The selection described in the previous section relies entirely on the BP and RP spectra and their parametrisation, apart from a few criteria on astrometric and photometric parameters. BP and RP spectra have important degeneracies between \teff and \ag, and by imposing our strict selection criteria, we not only inevitably remove sources with excellent parameters derived from the RVS spectra by \gspspec, but we can not guarantee either that they fulfill the 'gold' criteria.
We therefore made an independent selection by first querying the archive for sources with {\tt flags\_gspspec LIKE '0000000000000\%'}, i.e., sources for which the first 13 characters of the 41-character long quality flag provided by \gspspec are equal to '0', see \cite{DR3-DPACP-186}. These flag settings indicate low potential biases on \teff, \logg, \mh, and to some extent \afe due to rotational velocity, macroturbulence, uncertainties in the radial velocity shift correction and in the RVS flux, and extrapolation, absence of undefined or negative flux values or emission lines, non-zero uncertainties in the parameters, as well as high quality parameters for KM-type giants (see online documentation). The remaining flag characters are related to element abundances and CN and Diffuse Interstellar Band (DIB) features and were not taken into account for the current selection. 
This resulted in about 1.9 million sources.


For the further selection we considered the quality parameters 
\linktogsparam{gaia_source}{parallax_over_error}
and 
\linktogsparam{gaia_source}{rvs_spec_sig_to_noise}. The latter contains the signal-to-noise ratio in the mean RVS spectrum and is provided only for stars for which the mean RVS spectrum is published in \gdrthree.
We produced HR diagrams (\linktoapparam{astrophysical_parameters_supp}{lum_flame_spec}
versus \linktoapparam{astrophysical_parameters}{teff_gspspec}) 
and Kiel diagrams (\linktoapparam{astrophysical_parameters}{logg_gspspec} versus \linktoapparam{astrophysical_parameters}{teff_gspspec}) by imposing different lower limits on 
\linktogsparam{gaia_source}{rvs_spec_sig_to_noise}\footnote{0, 50, 100, 150, 200, 250, 400, 500}.
Visual inspection of the HR diagrams showed a group of sources at \teff$\sim$4000~K clustered around unrealistically high luminosities. Applying the criterion 
\linktogsparam{gaia_source}{rvs_spec_sig_to_noise} $\ge$ 150 removed 99\% of these sources. We combined this with the criterion 
\linktogsparam{gaia_source}{parallax_over_error} $>$ 33.34, similarly to what was applied to main sequence stars in the \gspphot\ based sample, resulting in 22\,143 sources ($\sim$1\% of the flag-selected sources), hereafter ``fgkm\_spec''.

\begin{figure*}
\centering
\includegraphics[width=0.48\textwidth]{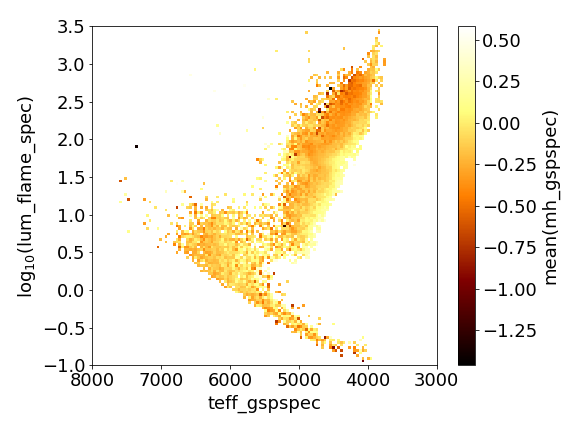}
\includegraphics[width=0.48\textwidth]{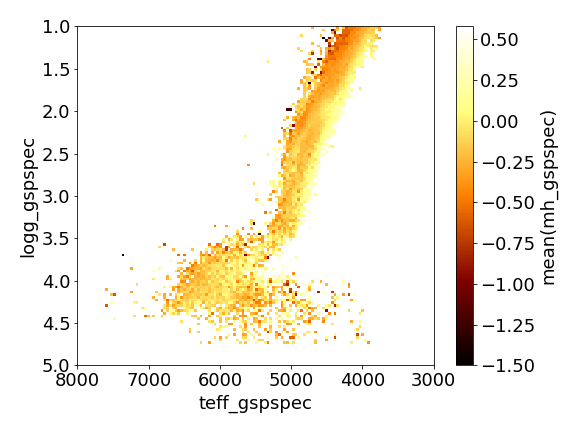}
\caption{HR and Kiel diagrams using \gspspec-based parameters for the fgkm\_spec sample described in Sect.~\ref{ssec:fgkmsampleselection_spec} colour-coded by the metallicity from \gspspec, with {\tt parallax\_over\_error} $\ge 33.34$ and
{\tt rvs\_spec\_sig\_to\_noise} $\ge$ 150.}
\label{fig:fgkm_spec_selection}
\end{figure*}

The HR and Kiel diagrams for this selection are shown in Fig.~\ref{fig:fgkm_spec_selection}.
The HR diagram displays a distinct giant branch and red clump as well as a region with turn-off stars and a clear main-sequence. 
However, as can be seen in the Kiel diagram, the \logg values for main sequence stars show a large spread. This is addressed by further filtering described in the next section.

We also compared the distributions of uncertainties in \teff, \logg, \mh, and \afe  from \gspspec for the flag-selected sample and the fgkm\_spec sample, where the uncertainty was defined as half of the difference between the upper and lower confidence levels. We found that the distributions for the latter sample have a smaller width by a factor 3 to 9 and peak at about half the uncertainty compared with the former sample. 

\subsection{Final sample and table description}
We merged the two samples described in Sects.~\ref{ssec:fgkmsampleselection_phot} and \ref{ssec:fgkmsampleselection_spec} with the objective to provide one unique FGKM gold sample.  As both sample definitions contain criteria that are not applicable to the other sample, we publish an independent table in the \gaia\  archive, \pvptable{gold_sample_fgkm_stars}, that also accounts for additional filtering on specific parameters. The description of the published table is given in Table~\ref{tab:fgkm_table}.  Further filtering is also done based on other archive products.  This is described in this section, and it results in a total of {3\,273\,041} sources. 

\begin{table*}
\begin{center}
\caption{Content of the table {\tt gold\_sample\_fgkm\_stars} in the \gaia\  archive.  The final column lists the section where the sample is defined in this work.  Further filtering on all of these samples is described in Sects.~\ref{ssec:fgkm_filter} and \ref{ssec:fgkm_varbin}.}  
\label{tab:fgkm_table}
\begin{tabular}{llllll}
\hline\hline
table & CU8 module & fields & sample definition \\
\hline
ap & \gspphot & {\tt teff, logg, mh, ag, ebpminrp, mg} &
Sect.~\ref{ssec:fgkmsampleselection_phot}\\ 
ap & \gspspec & {\tt teff, logg, mh, alphafe} & Sect.~\ref{ssec:fgkmsampleselection_spec}\\ 
ap & \flame & {\tt radius, lum, mass, age, evolstage}& Sect.~\ref{ssec:fgkmsampleselection_phot} \\
aps & \flame & {\tt mass, age, evolstage} {(\tt \_spec)} & Sect.~\ref{ssec:fgkmsampleselection_spec} \\
ap & \esphs & {\tt spectraltype, flags\_esphs} & Sects.~\ref{ssec:fgkmsampleselection_phot} -- \ref{ssec:fgkmsampleselection_spec} \\
\hline\hline
\end{tabular}
\tablefoot{
The first column indicates the archive table from which the parameters were taken, where ap = \aptable\ and aps = \apsupptable.  
The second column indicates the CU8 module responsible for producing the data. 
The third column indicates the parameter name that is copied from the ap and aps tables to the {\tt gold\_sample\_fgkm\_stars} table.  
The fourth column indicates the section where the sample definition is described.
}
\end{center}
\end{table*}

\subsubsection{Filtering of \flame, \gspspec\ and \esphs\ parameters in samples fgkm\_3 and fgkm\_spec}\label{ssec:fgkm_filter}
\begin{figure}
\includegraphics[width=0.45\textwidth]{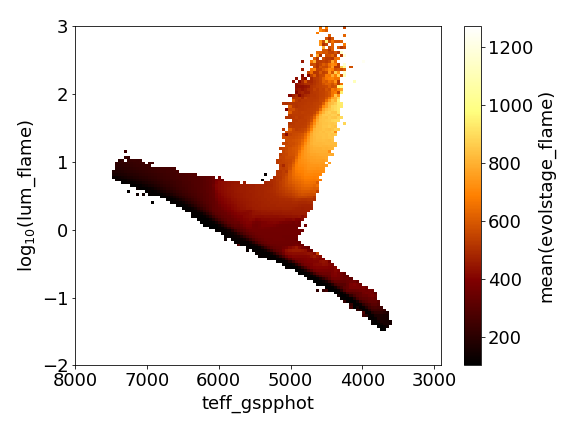}
\caption{HR diagram using sample fgkm\_3 colour-coded by {\tt evolstage\_flame}.  The low values of evolution stage on the giant branch correspond to the \flame\ parameters that were removed from the table, see Sect.~\ref{ssec:fgkm_filter} for details.}
\label{fig:fgkm_filterflame_phot}
\end{figure}

The fgkm\_3 \gspphot\ sample includes {3\,529\,613} sources with \flame\ parameters, and {3\,313\,190} 
with at least one model-dependent parameter (mass, age, evolutionary stage).  
Figure~\ref{fig:fgkm_filterflame_phot} shows an HR diagram using \teff\ and \lum, colour-coded by evolutionary stage ($\epsilon$).  
There is a region on the giant branch that has low evolutionary stages compared to the bulk of the giant branch.  These could be red clump stars that have been incorrectly assigned, because the models that were used to produce these parameters only span from the Zero-Age-Main-Sequence (ZAMS) to the tip of the giant branch.
These targets also have masses larger than 2~\Msun.  
Validation of \flame\ parameters has shown that the model values are inaccurate when \mass $>$ 2~\Msun for giants \citep{DR3-DPACP-127,DR3-DPACP-157}.  
We therefore retained 
\linktoapparam{astrophysical_parameters}{mass_flame}, 
\linktoapparam{astrophysical_parameters}{age_flame}, and
\linktoapparam{astrophysical_parameters}{evolstage_flame} for giants,  only if the following conditions were met: $\logg < 3.5$ and $\mass < 2.0$~\Msun and age $>1.0$~Gyr.   For $\logg > 3.5$ no filtering was done.
This same criteria was applied to the \flame\ parameters in the fgkm\_spec sample.

The fgkm\_spec sample shown in Fig.~\ref{fig:fgkm_spec_selection} shows some problems with \logg, below a certain threshold.  Validation of these values have  indicated a systematic offset on the order of 0.3 with respect to external catalogues for main sequence stars, see e.g. \cite{DR3-DPACP-157,DR3-DPACP-160,DR3-DPACP-186}.
We therefore removed \logg\ when $\logg > 4.0$ in order to retain a 'gold' status, and kept all of the other parameters.
As explained in the above references, a calibration of this parameter has been provided and a user can safely use the archive values with or without the calibration, depending on their use case.  

We retained the {\tt spectraltype} tag from \esphs\ in our table only if it had a quality flag of rank 1 or 2 (out of 5). This is given in the 
\linktoapparam{astrophysical_parameters}{flags_esphs} field in the 
\aptable\ table as the second character in that string field.

\subsubsection{Further filtering of the merged sample} 
\label{ssec:fgkm_varbin}
To ensure that our samples are as clean as possible, we further exploited other \gaia\ DR3 products.  
We removed all sources that were considered variable or non-single stars, by cross-matching our final source list with the
source lists given in the \varitable{vari_summary} table, 
which removed 249\,020 sources, 4\,873 of which are eclipsing binaries.
We also removed the sources appearing in any of the non-single star tables \nsstable{nss_two_body_orbit}, \nsstable{nss_acceleration_astro}, \nsstable{nss_non_linear_spectro} or \nsstable{nss_vim_fl}
which removed a further 28\,896 sources.
We then used the DPAC-Source Environment Analysis Pipeline (SEAPipe) to further check for any new binary contaminants, and this removed a further 16 sources\footnote{ 
The aim of SEAPipe is to combine the transit data for each source and to identify any additional sources in the local vicinity. 
Its first operation is image reconstruction, where a two-dimensional image is formed from the mostly one-dimensional transit data ($G>13$~mag), see  
\cite{harrison11}. 
These images are then analysed and classified, based on whether the source is extended, whether additional sources are present or whether the source is an isolated point source within the reconstructed image area (radius of $\sim 2^{\prime\prime}$).
It is this classification which is used to reject sources not found to be isolated point sources, from our sample.
The full SEAPipe analysis will be described in Harrison et. al. (in preparation).}.

\subsection{Validation of the sample}\label{ssec:validation}

\subsubsection{Validation with clusters}

We take advantage of the properties of open clusters to assess the  global quality of the FGKM sample.
From the FGKM sample, we selected those stars classified as cluster members in the \citet{2020A&A...640A...1C} catalogue as refined by
\citet{DR3-DPACP-75}.  The cross-match between those stars and our sample corresponds to $4\,132$ stars and contains only cross-matches with the \gspphot\ sample.
Using the full set of cluster members, we approximated each cluster with  an isochrone and derived reference values of \teff and \logg. Using this \teff we derived $A_G$ adopting the literature value of $A_V$ as a proxy of \azero. We made use of the PARSEC isochrone data set \citep{parsec2012}.
Differential extinction was assumed to be negligible inside the clusters for this validation work. This is justified by the fact that our sample excluded clusters younger than  $100$\,Myr.

\begin{figure*}
    \begin{center}
        \includegraphics[width=0.49\textwidth]{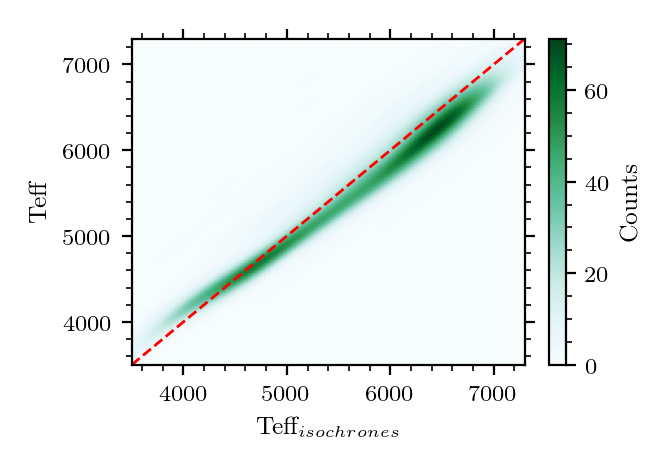}
        \includegraphics[width=0.49\textwidth]{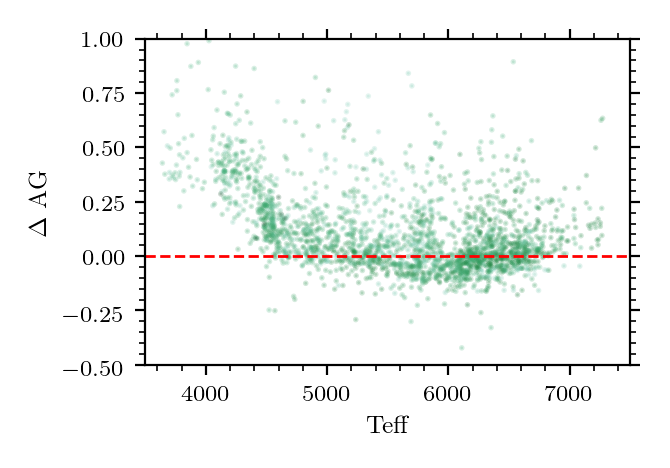}
        \caption{Comparison of \teff\ and \ag\ from \gspphot\  compared to the reference values from isochrones for stars of the FGKM sample in clusters. {\sl Left:} Comparison of \teff, with colour indicating the density of sources.
        The red line indicates the one-to-one values.
         {\sl Right:} $\Delta A_G =A_{G,{\rm \gspphot}} - A_{G,{\rm isochrones}}$  versus $\teff_{\rm \gspphot}$.  }
        \label{fig:Clu}
    \end{center}
\end{figure*}

\begin{figure}
    \begin{center}
        \includegraphics[width=0.49\textwidth]{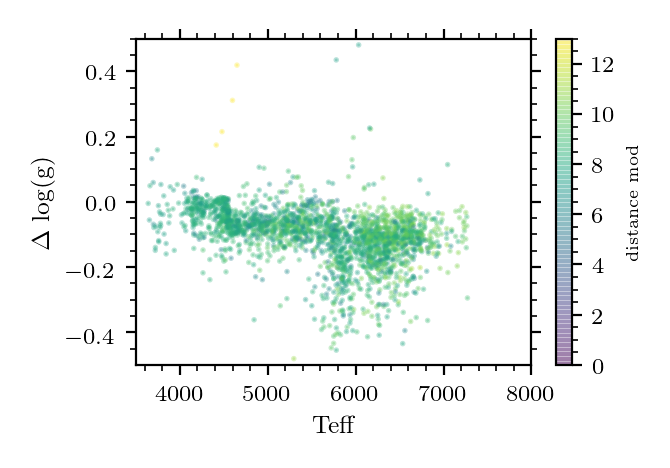}
     \caption{$\Delta \logg = \logg_{\rm \gspphot} - \logg_{\rm isochrones}$
      versus $\teff_{\rm \gspphot}$ for stars of the FGKM sample in clusters. 
      The color indicates the distance modulus ($m-M$) as derived from the \gspphot distance. }
        \label{fig:Clu2}
    \end{center}
\end{figure}

We compared the \teff, \logg, \mass, $A_G$, and [M/H] reference values with those from \gspphot\ and \flame\ in our sample. 
We adopted the average values of the member's \mh\ as the cluster \mh.
Table~\ref{tab:cluValid}  and Figs.~\ref{fig:Clu} and \ref{fig:Clu2} present the results, which show 
good agreement with the reference values.
Stars cooler than \teff $\sim$ 4500~K have \gspphot parameters that show the largest differences with reference values.  This overestimation of \teff\ at low temperatures often has higher increased extinction in this regime.

\begin{table}
    \caption{Differences in \gspphot\ and \flame\ parameters from isochrone-fitted values for stars of the FGKM sample in clusters.  $\Delta P$ is given in the sense of \gdr{3} value minus the cluster value. MD and MAD indicate the median and median absolute deviation of the differences, respectively. \label{tab:cluValid}}
    \begin{center}
        \begin{tabular}{cccc}
            \hline
            $P$   & $\Delta P$.MD$^{(1)}$     & $\Delta P$.MAD$^{(2)}$  & Units \\
            \hline\hline
            \teff       & -94    & 136  & K      \\
            \logg       & -0.09  & 0.04 & dex   \\
         \mh      &-0.20&0.07&dex\\
            $A_{G}$       & 0.05  & 0.09 & mag   \\
            \mass      &-0.04&0.05&\Msun\\
            \hline
        \end{tabular}
    \end{center}
\end{table}

\subsubsection{Validation with other galactic surveys}

\begin{figure*}
    \centering
    \includegraphics[width=0.32\textwidth]{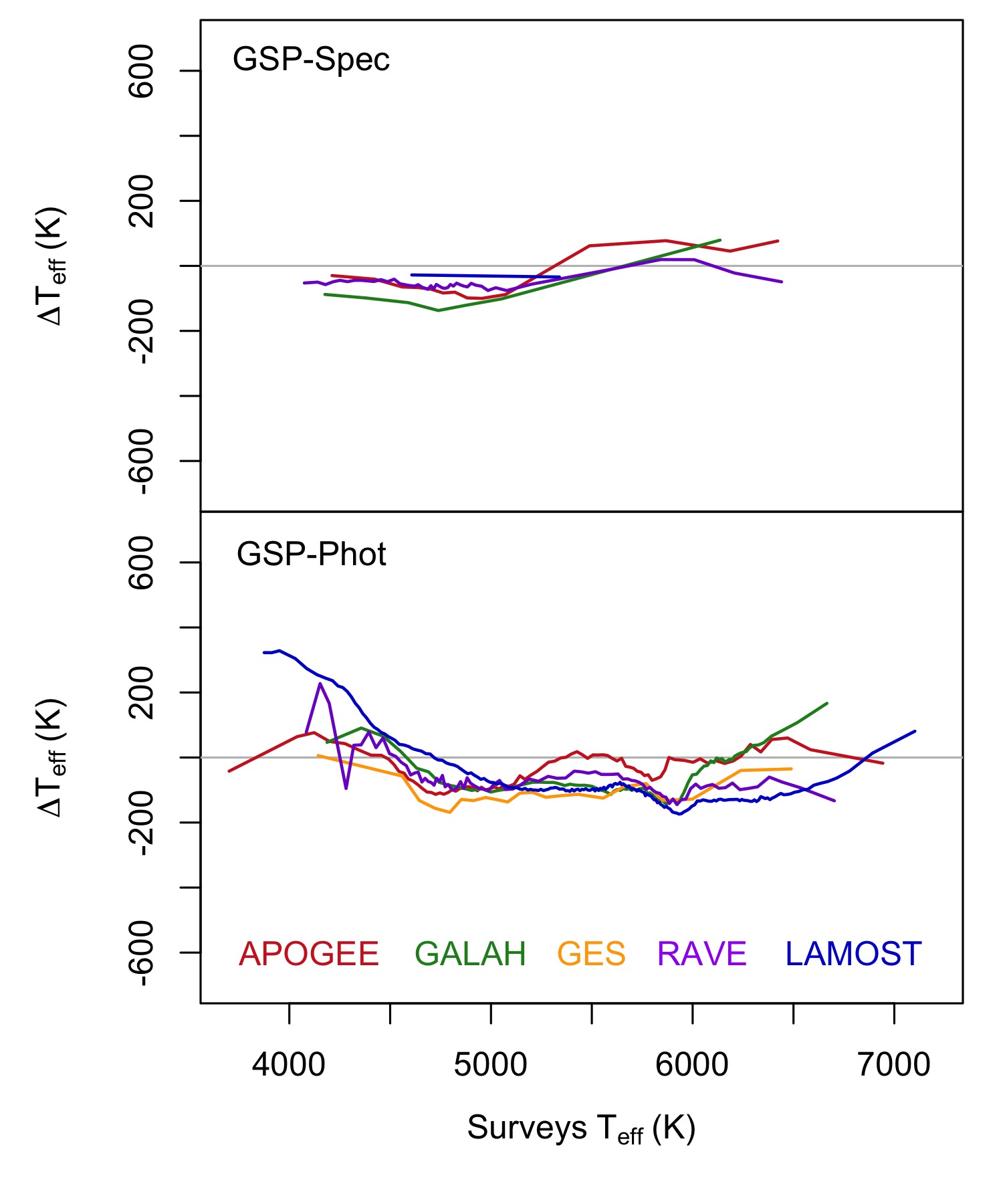}
    \includegraphics[width=0.32\textwidth]{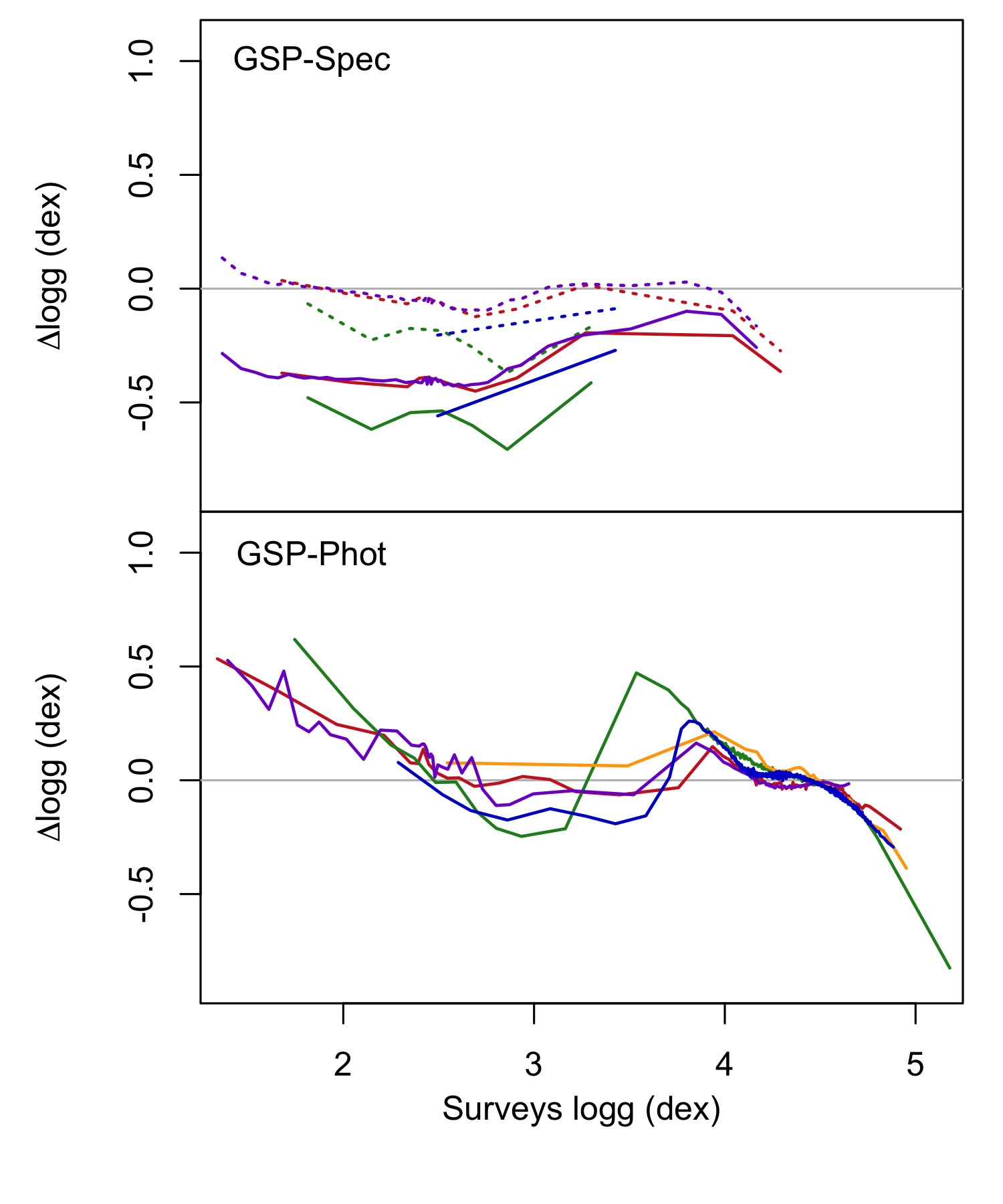}
    \includegraphics[width=0.32\textwidth]{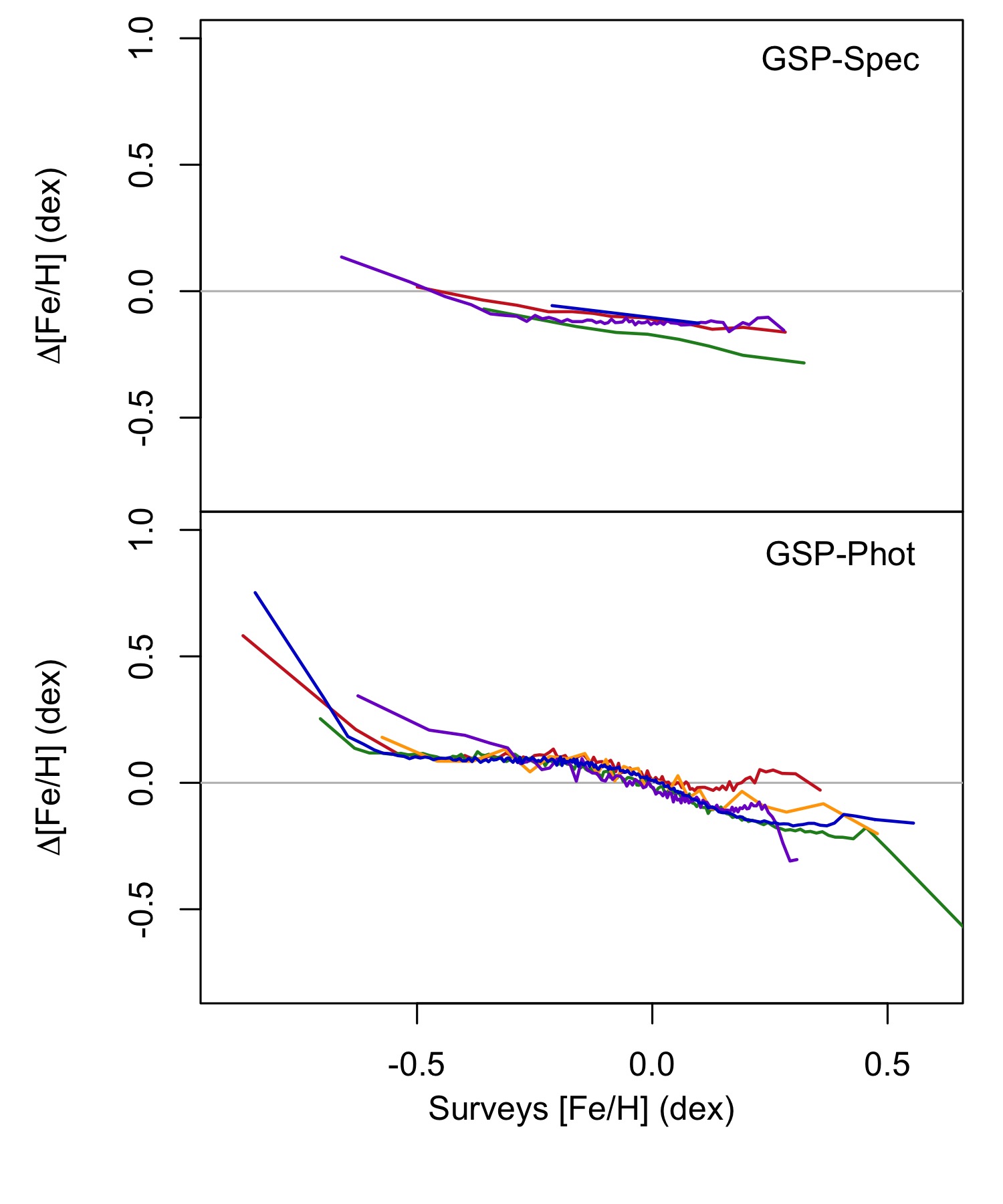}
    \caption{Comparison of atmospheric parameters with the spectroscopic surveys for the FGKM sample. The top panels show the comparison of \gspspec\ parameters and the bottom panels \gspphot. The left panels show the case of \teff, the middle ones of \logg\ and the right ones of [Fe/H]. The differences on the y-axes are in the sense of \gaia\ minus the surveys. Each survey is represented as the medians of equally populated bins (solid lines, colored according to the legend in the bottom-left panel). The dotted lines for the \gspspec \logg\ are obtained after the corrections recommended by \citet{DR3-DPACP-186}.}
    \label{fig:sos}
\end{figure*}

\begin{table*}
    \centering
    \caption{Comparison of the \gspphot and \gspspec parameters with the ones from the five main spectroscopic surveys, for the FGKM sample.}
    \label{tab:sos}
    \begin{tabular}{llrrrrrr}
    \hline\hline
    Survey & Set & $\Delta$\teff & \# \teff & $\Delta$\logg & \# \logg & $\Delta$[Fe/H] & \# [Fe/H]\\
    & & (K) & & (dex) & &  (dex) & \\
         \hline
APOGEE   & \gspspec &  --45 $\pm$  79 &   2942 & --0.38 $\pm$ 0.13 &   2484 & --0.08 $\pm$ 0.10 &   2942 \\
         & \gspphot &  --33 $\pm$ 100 &  22160 & --0.03 $\pm$ 0.09 &  22155 &   0.04 $\pm$ 0.14 &  22149 \\
GALAH    & \gspspec & --105 $\pm$  73 &   1572 & --0.57 $\pm$ 0.31 &   1498 & --0.18 $\pm$ 0.10 &   1571 \\
         & \gspphot &  --64 $\pm$ 108 &  28545 &   0.05 $\pm$ 0.13 &  28545 &   0.00 $\pm$ 0.16 &  28545 \\
Gaia-ESO & \gspspec &    55 $\pm$   0 &      1 & --0.03 $\pm$ 0.00 &      1 &   0.00 $\pm$ 0.00 &      1 \\
         & \gspphot & --115 $\pm$  92 &    745 & --0.01 $\pm$ 0.12 &    650 &   0.01 $\pm$ 0.15 &    680 \\
RAVE     & \gspspec &  --55 $\pm$  77 &  15108 & --0.38 $\pm$ 0.13 &  13955 & --0.12 $\pm$ 0.09 &  15108 \\
         & \gspphot &  --81 $\pm$ 117 &  21374 & --0.01 $\pm$ 0.04 &  21374 & --0.07 $\pm$ 0.14 &  21374 \\
LAMOST   & \gspspec &  --29 $\pm$  71 &   1260 & --0.50 $\pm$ 0.25 &   1072 & --0.09 $\pm$ 0.10 &   1260 \\
         & \gspphot &  --83 $\pm$ 105 & 299148 & --0.02 $\pm$ 0.11 & 299148 &   0.02 $\pm$ 0.16 & 299148 \\
         \hline\hline
    \end{tabular}
\end{table*}

We compared the FGKM sample parameters with the ones of the major spectroscopic surveys, using a cross-match computation specifically performed, using the \gdr{2} cross-match software \citep{marrese17,marrese19}, for the Survey of Surveys project \citep[SoS][]{tsantaki22}. The used surveys are APOGEE \citep[DR16,][]{ahumada20}, GALAH \citep[DR2,][]{buder18}, Gaia-ESO \citep[DR3, hereafter GES,][]{gilmore12}, RAVE \citep[DR6,][]{steinmetz20}, and LAMOST \citep[DR5,][]{deng12}. For each survey, we applied the quality selection criteria suggested in the relevant survey papers and summarized by \citet{tsantaki22}\footnote{The SoS is based on \gdr{2}, thus we used the cross-match between DR2 and EDR3 to find the updated source IDs. We removed all sources with a DR2-DR3 magnitude difference higher than 0.5~mag, angular difference higher than 0.5", and all sources with more than one neighbour or mate.}. We further removed all the confirmed and candidate spectroscopic binaries identified in the surveys \citep{merle17,birko19,qian19,price20,tian20,traven20,kounkel21}. The summary of the number of FGKM stars from the golden sample found in each survey {is given} in Table~\ref{tab:sos}, where the median differences of the main parameters, computed in the sense \gaia\ minus the surveys, are reported together with their MAD (median absolute deviation). A graphical comparison for the main parameters can be found in Figure~\ref{fig:sos}.

The \teff comparison shows agreement with all surveys, both in \gspphot and \gspspec, within uncertainties. The median offsets for \gspspec are generally negative and of the order of --50---100~K, and the same is true for the \gspphot offsets. The spreads range from roughly $\pm$70 to $\pm$120~K, in line with expectations. We note that the surveys agree with each other within a few tens of K, at least in the central portion of the \teff range. Figure~\ref{fig:sos} shows some systematic substructures in the comparisons. For \gspspec, we find good agreement in \teff. 
At the extremes of the \teff range, some discrepancies occur between \gspphot and the comparison with LAMOST, which has the lowest resolution among the surveys.

The \gspspec \logg comparison shows an offset of about --0.3~dex, which is a known feature, as reported in Section 9.1.1 by \citet[][see their equations 1 and 2]{DR3-DPACP-186}, while the \gspphot comparison shows excellent agreement with the surveys. When applying the recommended correction to the \gspspec \logg (dotted lines in Figure~\ref{fig:sos}), the offsets and main trends are highly mitigated. The spreads in the comparisons are roughly around $\pm$0.1~dex in \gspphot and up to $\pm$0.2--0.3~dex in \gspspec (before correction). The \gspphot estimates show good agreement for the subgiants, and most of the dwarfs, and disagreements at the level of up to 0.3 dex is found for the very high ($>4.5$) and low ($<2$) \logg\ stars. Again, we note that the surveys agree with each other to 0.1--0.2~dex, approximately over most of the \logg range.

For metallicity, we use [Fe/H] as an indicator, to be more in line with what is commonly measured by the surveys, which we computed from [M/H] and [$\alpha$/Fe] using the formula by \citet{salaris93}. Again, we note a better agreement of the \gspphot parameters with the surveys than for the \gspspec ones in terms of median offset, which is about zero~dex for \gspphot and 0.1~dex for \gspspec. This was also reported by \citet[][see their equations 3 and 4]{DR3-DPACP-186}. The spreads are in both cases of about 0.10--0.15~dex, which is more than reasonable. We note that the surveys themselves tend to agree with each other to 0.1~dex or better. There is a tendency of both the \gspspec and \gspphot parameters to overestimate the [Fe/H] of metal-poor stars and to underestimate it for metal-rich ones. This effect has been commonly observed in several other projects where the parameters were derived from low- or medium-resolution spectroscopy or photometry.

In conclusion, the overall agreement with the main spectroscopic surveys is good, but there are substructures in the comparisons that need to be kept in mind. Additionally, depending on the type of stars, we note that the \gspspec parameters do not necessarily produce a better agreement with the survey results compared with the \gspphot ones, and the use of the \gspspec \logg and [M/H] corrections \citep{DR3-DPACP-186} is recommended.
This is in part due to the fact that the RVS spectral range extent and resolution is limited, but also the fact that we are  dealing with a high S/N regime, free from major systematic problems, where both the \gspspec\ and \gspphot\ perform close to optimal.

\subsubsection{Validation with the PLATO input catalogue}
\begin{figure}
    \centering
    \includegraphics[width=0.5\textwidth]{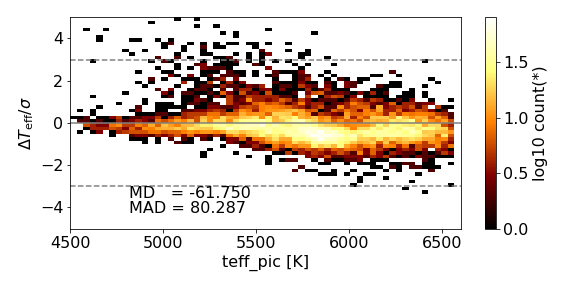}
   \includegraphics[width=0.5\textwidth]{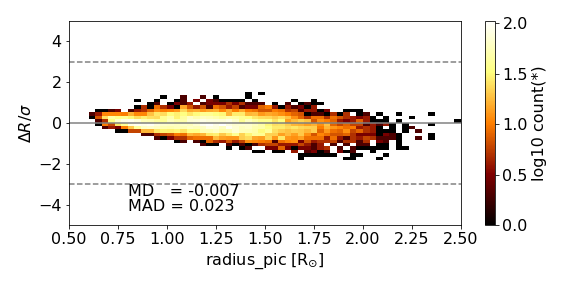}
    \includegraphics[width=0.5\textwidth]{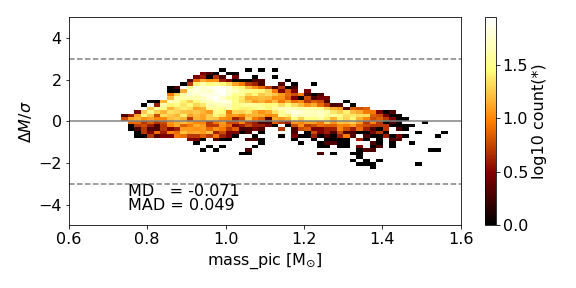}
    \caption{Difference between \teff (top), \radius (middle), and \mass (bottom) from \gaia\ for the FGKM  sample and the PICv1.1 catalogue values normalised by their combined uncertainties for stars in common.  We overlay the $\pm 3\sigma$ lines.
    On each panel we also give the median difference ('MD') and 
    the median absolute deviation ('MAD') in K, $R_\odot$ and $M_\odot$, respectively. }
    \label{fig:fgkm_compare_pic}
\end{figure}

We cross-matched our source list with the PLATO input catalogue (PIC) version 1.1 \citep{2021yCat..36530098M} and obtained 10\,828 common sources.  In Fig.~\ref{fig:fgkm_compare_pic} we compare  \teff, \radius, and \mass (in the sense \gaia\  -- PIC) normalised by the combined uncertainties added in quadrature.  
We also show the $\pm 3 \sigma$ lines, which shows good agreement between the catalogues, but some insignificant artefacts for the comparison of masses.
While we show the differences in terms of $\sigma$, we report the median  and the median absolute deviation of their differences in absolute values on each panel.
The agreement with \teff\ is similar to that reported in the previous sections, where the \gspphot one is on average 50~K smaller.
There are no matches with the \gspspec\ sources.
Radius and mass differences are on the order of 1\% and 6\%, respectively.

In conclusion, we have made a clean sample of 3\,273\,041 FGKM stars, comprising main sequence, subgiant, and giant stars.  This sample was selected using many \gaia\ based indicators along with \gspphot- and \gspspec-based astrophysical parameters.   
The APs of interest are \teff, \logg, \mh, \ag, \ebpminrp, \lum, \radius, \mass, age and spectral type, and we provide a separate table of these parameters in the \gaia\ archive.  We have not applied any calibration or correction to the values in \gdr{3}, but we have filtered some parameters for some sources.    
We validated our selection by comparing with parameters from clusters and other surveys which show typical offsets of $<100$ K in \teff\ with other surveys.  
In Sect.~\ref{sec:exoplanets} we exploit this sample's \teff, radius, and mass to analyse known exoplanet systems, and in
Sect.~\ref{sec:ucd_unseen} we analyse the ages of 11 unseen UCD-companions. 
A user could further filter by selecting in a specific \teff\ range, or by excluding distances further than a certain distance, or by providing an upper limit to the amount to extinction between the observer and the star.

\section{UltraCool dwarfs}\label{sec:ucd}

\subsection{Scientific motivation}
Ultracool dwarfs (UCDs) are objects at the faint end of the main sequence. They were defined in \cite{1997AJ....113.1421K} as sources with spectral types M7 or later. This definition includes the coolest  hydrogen burning stars and brown dwarfs. Even though brown dwarfs can sustain lithium or deuterium fusion at their cores for a short period of time in the early phases of their evolution \citep{2001RvMP...73..719B}, the nuclear reactions stop by the time they reach the main sequence and they keep cooling and fading thereafter. Despite the fundamental differences in the internal structure across the stellar/sub-stellar regime, the atmospheric properties overlap at this boundary  and it becomes very difficult to distinguish between the two regimes based on photometric or spectroscopic properties. In this section we define a high quality sample of UCDs which we propose as excellent candidates to advance our knowledge of these low mass objects.
To complement the 
\teff\ of UCDs in \gdr{3} we provide a catalogue of radii and luminosities by complementing the \gaia\  data with infra-red photometry and we explore
the existence of a minimum in the mass-radius relation slope \citep[e.g.][]{2014AJ....147...94D,10.1093/mnras/sty2520,2020A&A...642A.115C}.

\subsection{Sample selection}

Our initial sample of UCD candidates is from the \aptable\ table 
where a total of 94\,158 sources have been  processed as UCD candidates and estimate \linktoapparam{astrophysical_parameters}{teff_espucd}\footnote{
We remind readers that the main difference with respect to existing compilations of UCD candidates \citep[for example][]{2018A&A...619L...8R} is the use of the \gdr{3} RP spectra to produce \teff\ and to help define the selection criteria  
as described in the \linksec{sec_cu8par_apsis/ssec_cu8par_apsis_espucd.html}{\gdr{3} online documentation.}}.
We also imposed that the first digit of \linktoapparam{astrophysical_parameters}{flags_espucd} = 0 or 1 (the most reliable categories), which gives a total of 67\,428 candidates. 
We then require that the \gaia\ astrometric flags fulfil the following conditions: 
\linktogsparam{gaia_source}{ruwe} $< 1.4$, 
\linktogsparam{gaia_source}{ipd_frac_multi_peak} = 0 and 
\linktogsparam{gaia_source}{ipd_gof_harmonic_amplitude} $< 0.1$ to reduce contamination by unresolved binaries. 
We then select sources with a cross-match (as provided in the \gaia\ archive) in the 2MASS \citep{2006AJ....131.1163S} and AllWISE \citep{2010AJ....140.1868W,2011ApJ...731...53M} catalogues,
with available measurements in the $J$, $H$ and $K_s$ 2MASS bands, and the $W1$ and $W2$ AllWISE bands, all with quality A flags. The $W3$ band was not included as a requirement because the lack of measurement uncertainties reduces drastically the number of sources. Finally, we remove sources above the $G+5\log_{10}(\varpi)+5 = 3+2.5~(G-J)$ line to avoid including suspected low gravity UCDs that have not yet contracted and reached equilibrium.  This gives a total of 31\,822 candidates for this study.

We used the virtual observatory VOSA \citep{2008A&A...492..277B} to calculate the minimum reduced $\chi^2$ fit to the spectral energy distributions constructed using the \gaia\ $G$ and $G_{\rm RP}$ bands and the near- and mid-infrared photometry listed above to CIFIST 2011\_2015 BT Settl models \citep{2012RSPTA.370.2765A}.  
We retain the sources whose reduced $\chi_r^2 < 100$. We allow for rather large values of $\chi_r^2$ in order to account for the known discrepancies between the models and observations, and the discrete nature of the model library. The distribution of $\log_{10}(\chi_r^2)$ is approximately normal 
and 96.5\% of all the values are below the imposed threshold which therefore only removes obvious pathological fits.  
The final sample has a total of 21\,068 sources.

\subsection{Combining \gaia\ with external data to derive \radius and \lum}

\begin{figure}[!htb]
\center{\includegraphics[width=0.49\textwidth]{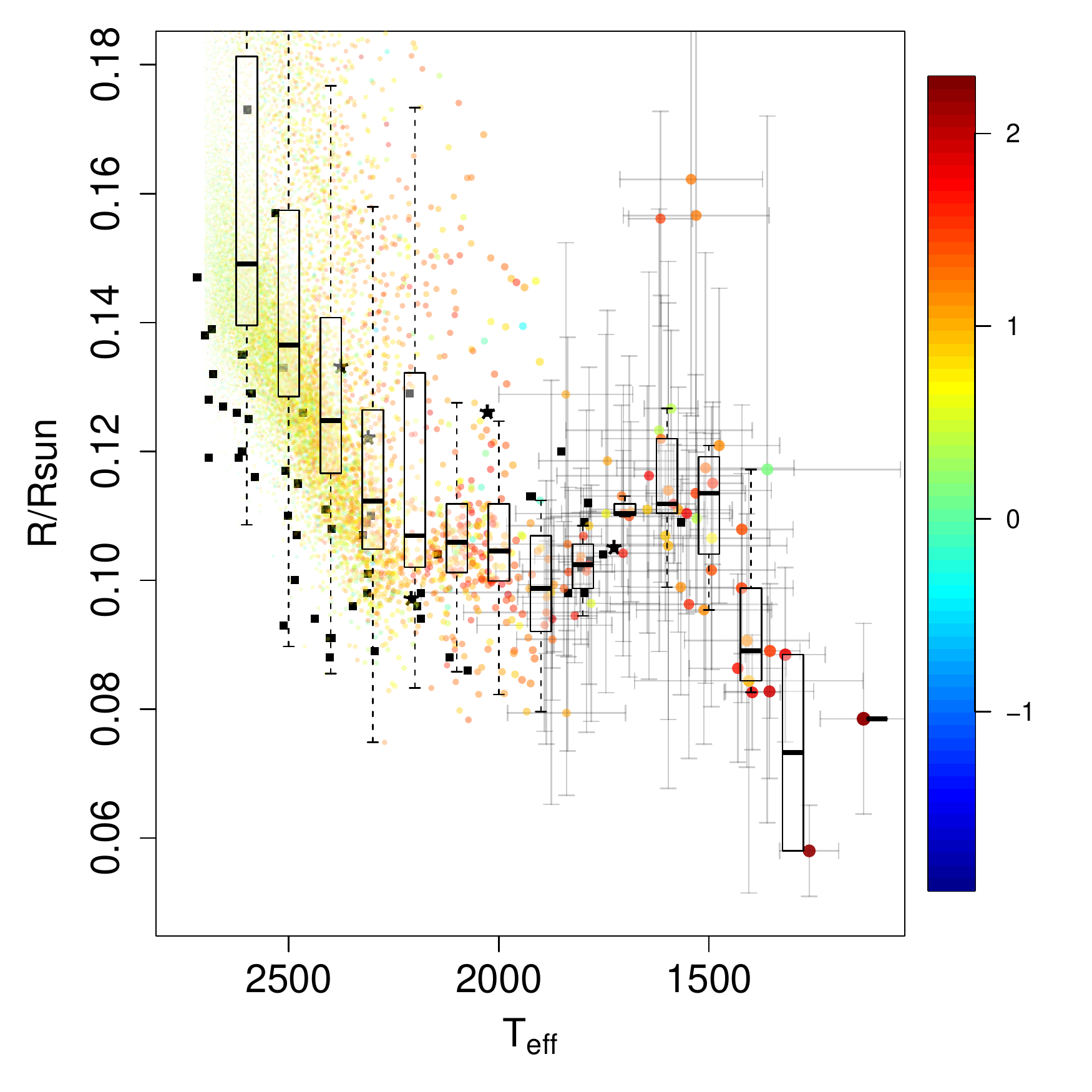}}  
\caption{\label{fig:ucds-radii}
Radii of candidate UCDs in the \gaia\ golden sample. The colour code indicates the logarithm of the VOSA fits $\chi^2$ values, squares represent the data points in Table 1 of \cite{2014AJ....147...94D} and black asterisks denote unresolved binaries therein. The boxplots are calculated within bins of 100 K.}
\end{figure}

\begin{figure}[!htb]
\center{\includegraphics[width=0.49\textwidth]{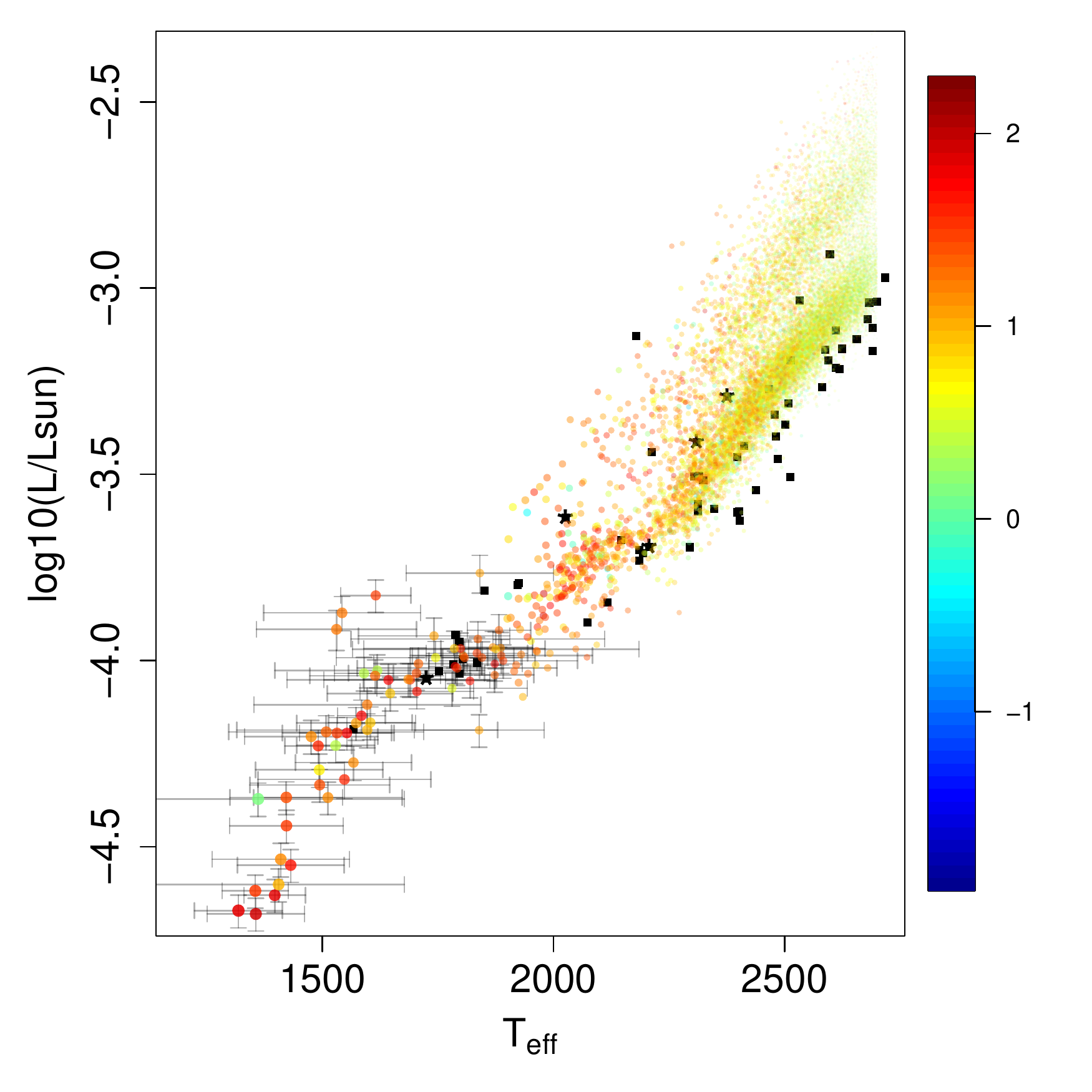}}  
\caption{\label{fig:ucds-lum}
Bolometric luminosities of candidate UCDs in the \gaia\ golden sample. The colour code indicates the logarithm of the VOSA fits $\chi^2$ values, squares represent the data points in Table 1 of \cite{2014AJ....147...94D} and black asterisks denote unresolved binaries therein.}
\end{figure}

\radius\ and \lum\ are parameters that are also calculated by the \flame\ module and available in the \aptable\ table, however these are only available for sources with $\teff > $ 2\,500 K. A comparison of the values for the sources in common is discussed in the next section. 
We computed bolometric fluxes using \gaia\ and IR photometry.  To account for the unobserved flux outside the observed wavelength bands we needed to calculate bolometric corrections.
We used again the CIFIST 2011\_2015 BT Settl models in order to calculate the ratio of observed to total flux for the aforementioned set of photometric bands\footnote{See the help and documentation of VOSA for an updated description of how the band wavelength overlaps are handled in computing the observed flux}. This produces a theoretical flux correction factor $f_{obs}/f_{total}$ for the \teff range between 1200 and 2700 K in steps of 100\,K. 
For each of the UCD candidates with full photometry we obtain the correction factor by interpolating the \teff value derived by the \espucd module in this grid. 
The resulting corrections are in the range between 0.48 and 0.54 mag with a median value of 0.53 mag. We use this ratio to infer the total flux that would be observed at the Earth and derive the bolometric luminosity using the \gaia\ parallax measurement. Finally, using the \espucd \teff estimate and the bolometric luminosity, we inferred radii for the UCD candidates using the Stefan-Boltzmann law.  Figures \ref{fig:ucds-radii}  and \ref{fig:ucds-lum} show the scatter plot of the inferred radii and luminosities as a function of the \espucd \teff. The uncertainties (represented as error bars only for sources cooler than 1900 K to aid readability) were calculated using a simple Taylor expansion and neglecting correlations amongst the intervening variables.

To fully exploit this UCD golden sample, we provide an accompanying table in the \gaia\ archive \pvptable{gold_sample_ucd} which lists {\tt source\_id}, the correction factor to calculate the bolometric flux, radius, luminosity, and uncertainties, along with the $\chi^2_r$ value. This table can be used with the \linktoapparam{astrophysical_parameters}{teff_espucd} 
provided in the \aptable\ table.

\begin{figure}[!htb]
\center{\includegraphics[width=0.49\textwidth]{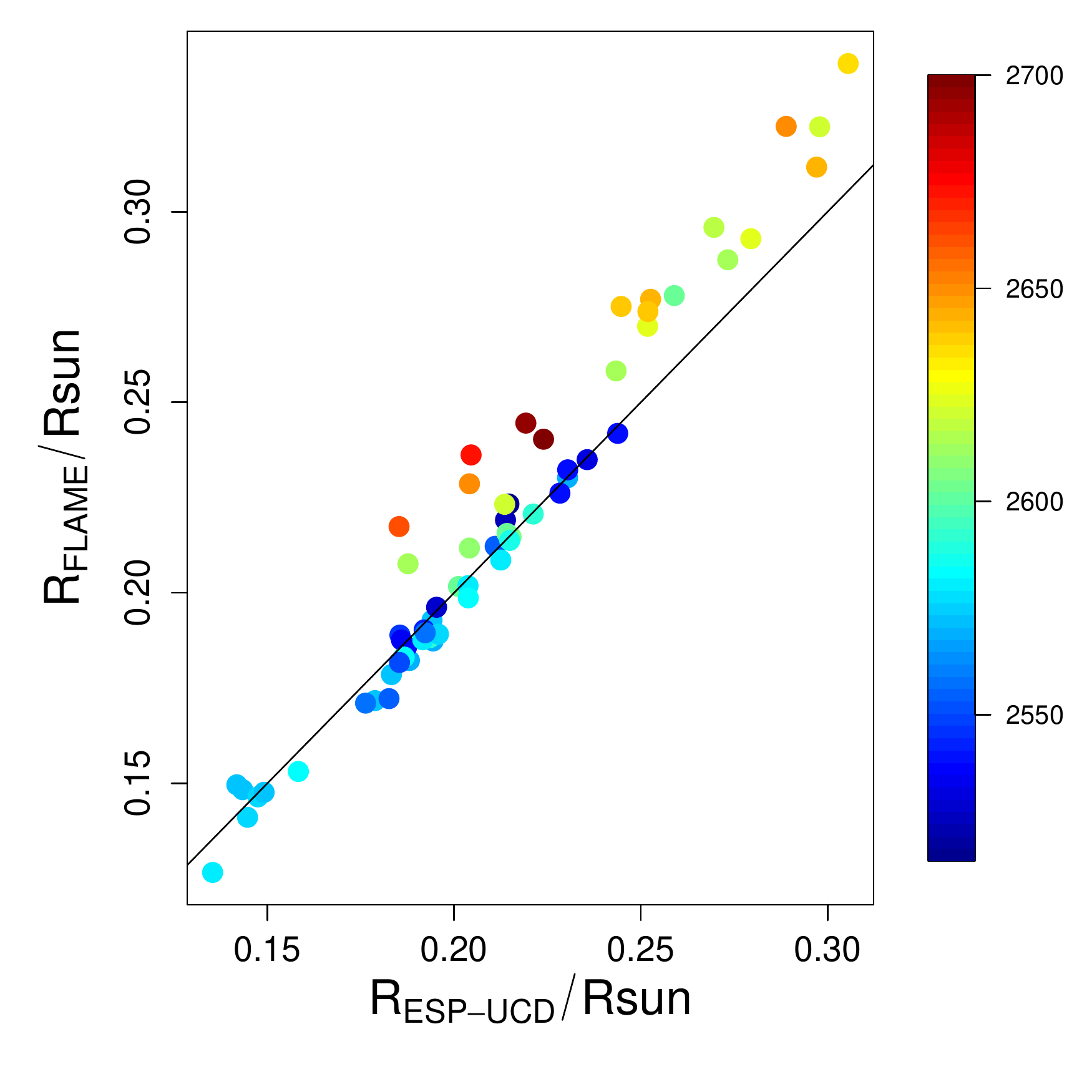}}  
\caption{\label{fig:ucds-flame} Comparison of the radii estimated for the UCD sample by the  \espucd ($x$ axis) and \flame ($y$ axis) modules for the sources in common. The colour code reflects the effective temperature used by the \flame module to estimate the radii.}
\end{figure}

\subsection{Validation}

In Figure \ref{fig:ucds-flame} we compare the radii values of sources with estimates from the \flame and \espucd modules. It shows a remarkable agreement for the lowest temperature regime ($T_{\rm eff} < 2600$ K) but it also shows evidence for a systematic difference in the sense of larger \flame radii above. This is due in part to a difference of approximately 85 K in the temperatures used for the derivation of radii (in the sense of the \teff\ used by \flame -- from \gspphot --
are hotter than the ones produced by the \espucd module).

Figure \ref{fig:ucds-radii} shows the expected decrease in radius as the temperature decreases down to temperatures of the order of $\approx 2200-2000$ K. Then, the radii increases for even cooler temperatures until \teff$\approx 1400$ K where the trend reverses and the slope becomes positive again.

In Fig.~\ref{fig:ucds-lum} we can see a systematic difference between the luminosities estimated by \cite{2014AJ....147...94D} (represented by the black squares) and the ones from this work,  in the range of temperatures $T_{\rm eff} > 2000$ K. 
This difference translates into an offset in radii in Figure \ref{fig:ucds-radii}. The offset in luminosity can be due either to (1) a difference in the $T_{\rm eff}$ estimates if our temperatures were systematically cooler than those of \cite{2014AJ....147...94D} in that regime and/or (2) a difference in the calculation of the bolometric correction (BC) if BCs derived by \cite{2014AJ....147...94D} produce bolometric luminosities systematically fainter than the ones derived here. We examine the two alternatives more closely in the following paragraphs.

\begin{figure}[!htb]
\center{\includegraphics[width=0.49\textwidth]{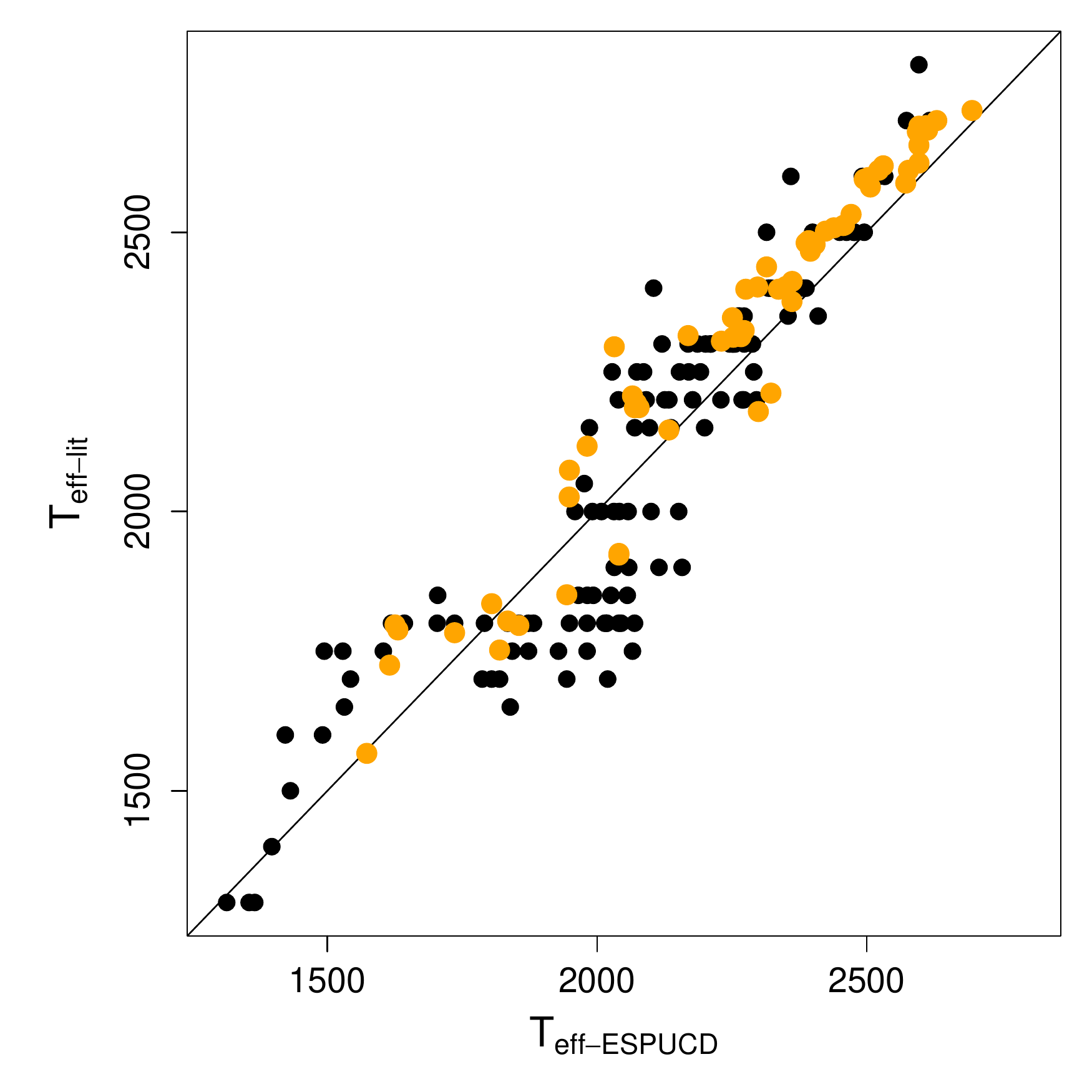}}  
\caption{\label{fig:ucds-cmp-lit}
Comparison of the effective temperatures used to derive radii in this work ($x$ axis) and those used in the literature ($y$ axis) for the UCD sample. Black filled circles denote sources from \cite{2020A&A...642A.115C} and orange filled circles, those from \cite{2014AJ....147...94D}.}
\end{figure}

Figure \ref{fig:ucds-cmp-lit} shows a comparison of the temperatures used by \cite{2014AJ....147...94D} and \cite{2020A&A...642A.115C} to infer radii with those estimated by the \espucd module. It shows hints of a systematic difference of approximately 65 K above $T_{\rm eff} \approx 2200$ K. This is different from, but consistent, with the difference encountered in the comparison with the \flame outputs.

The ESP-UCD \teff\ are based on an empirical training set built from the the \gaia\  UltraCool Dwarf Sample \citep[GUCDS;][]{2017MNRAS.469..401S,2019MNRAS.485.4423S} and the spectral type-$T_{\rm eff}$ relation by \cite{2009ApJ...702..154S}. The values derived by the ESP-UCD regression module were calibrated as described in the \linksec{sec_cu8par_apsis/ssec_cu8par_apsis_espucd.html}{\gdr{3} online documentation} to account for a discrepancy that was found with respect to the regression module trained on BT-Settl models. The RP spectra, simulated from the BT-Settl library of synthetic spectra, were found to reproduce well the observed RP spectra in this $T_{\rm eff}$ regime. Also, the calibrated temperatures were found to produce relatively good agreement with the SIMBAD spectral types where available (again, using the relations by \citealt{2009ApJ...702..154S}) as illustrated in the validation of the ESP-UCD module in \cite{DR3-DPACP-160}. However, in view of the comparisons described above, it is not implausible that the correction applied in the calibration of the results from the empirical training set was overestimated by an amount of the order of 65 K. In any case, the systematic difference in effective temperatures explains part but not all of the discrepancy in the luminosities/radii. Hence, we suspect that this discrepancy may also be caused by differences in the corrections applied to the observed fluxes to derive bolometric luminosities. Our procedure to estimate the bolometric luminosity is different from that used by \cite{2014AJ....147...94D} and this can be the source of the systematic difference in the luminosities above 2000 K apparent in Figure \ref{fig:ucds-lum}. While we interpolate directly the fraction of the total flux emitted in the photometric bands on a grid of BT Settl models, \cite{2014AJ....147...94D} apply a wavelength dependent correction to the BT Settl models such that they agree with the observed photometric magnitudes before that fraction is estimated. Since a direct comparison is not possible due to the unavailability of their correction factors, we cannot discard this different procedure as a potential explanation of the difference. 

The overall trends of decreasing radii down to $\sim$ 2000 K and slowly increasing radii for even cooler temperatures are confirmed with the \gaia\ data although the associated uncertainties are large. The final positive slope in the regime \teff $< 1400$ K is also compatible with that shown in \cite{2020A&A...642A.115C} but not predicted (to the best of our knowledge) by theoretical studies. The sample of UCDs used here can be expected to be a combination of different ages, masses and metallicities (all of them with an impact on the effective temperatures and radii) and hence, no direct conclusion about these fundamental parameters can be easily drawn from Figure \ref{fig:ucds-radii} .   

In summary, we provide a catalogue of 21\,068 UCDs that we consider to be of very high quality from the available sources in the \aptable.   We derive their luminosities and radii by calculating bolometric corrections and make these new parameters available in the accompanying \pvptable{gold_sample_ucd} table.

\section{Carbon stars}\label{sec:carbon}

\subsection{Scientific motivation}

A high number of Asymptotic Giant Branch (AGB) stars have carbon enriched atmospheres and show C$_\mathrm{2}$ and CN molecular bands stronger than usual in  stars cooler than 3\,800~K (i.e.\ $\bpminrp\geq$ 2). The origin of the enrichment can be due to mass transfer in binary systems or due to the pollution by nuclear He fusion products from the inner to the outer layers. Because they belong to a late stage of stellar evolution where mass loss occurs and which precedes the formation of the planetary nebula, carbon stars are important contributors to the interstellar medium and provide good reference cases to study the physical processes affecting the end of the life of low mass stars. During the \gdr{3} development and processing, no synthetic spectra showing such high carbon abundances were included in the simulations that are used in the \apsis\ software to produce APs \citep{DR3-DPACP-157}. Hence, the spectral libraries used as templates to derive the astrophysical parameters from BP and RP, as well as those adopted to measure the radial velocities, are not fully adapted to analyze the data of carbon stars. Therefore, an attempt was made by the ESP-ELS module to flag suspected carbon stars.

\subsection{Sample selection\label{sec:cstars_selection}}

The identification of candidate carbon stars by ESP-ELS is based on a random forest classifier trained on the synthetic BP and RP spectra as well as on the observed \gaia\  data obtained for a sample of galactic carbon stars \citep{2020A&A...633A.135A}. {This identification} is saved in the \linktoapparam{astrophysical_parameters}{spectraltype_esphs} field of the \aptable\ table. In total, 386\,936 targets received the ``CSTAR” tag. While most of these stars are M stars, only a smaller fraction of the sample {exhibit} significant C$_\mathrm{2}$ and CN molecular bands. To identify these cases, we measured the band head strength as follows:

\begin{equation}
R_{\lambda_2} = \frac{f({\lambda_2})}{g_{\lambda_1, \lambda_3}({\lambda_2})}
\label{eq:bandhead_strength_cstars}
\end{equation}

\begin{table}
\caption{Molecular band head strength used to identify the most probable carbon stars}\label{tab:cstars_bandheads}
\begin{tabular}{crrr}
\hline
Strength (molecule) & ${\lambda_1}$ [nm] & ${\lambda_2}$ [nm]& ${\lambda_3}$ [nm]\\
\hline
R$_\mathrm{482.3}$ (C$_\mathrm{2}$) & 462.2345 & 482.3455 & 505.3195 \\
R$_\mathrm{527.1}$ (C$_\mathrm{2}$) & 505.3195 & 527.1080 & 546.5995 \\
R$_\mathrm{773.3}$ (CN) & 716.5865 & 773.2905 & 810.7805 \\
R$_\mathrm{895.0}$ (CN) & 806.8910 & 894.9855 & 936.6820 \\
\hline
\end{tabular}
\end{table}

\noindent where $f({\lambda_2}$) is the flux measured at the top of the band head of the molecular band, and $g_{\lambda_1, \lambda_3}$ the value linearly interpolated between wavelengths $\lambda_1$ and $\lambda_3$. The four band heads we considered are described in Table\,\ref{tab:cstars_bandheads}. These were computed for a random sample of 27\,528 stars having $G_\mathrm{BP}-G_\mathrm{RP}$ (not dereddened) colours uniformly distributed between 1 and 5, in order to locate the range of R$_\mathrm{773.3}$ and R$_\mathrm{895.0}$ values occupied by non-carbon stars. The upper limit of the interquantile dispersion (2.7 \% and 97.3 \%) is the threshold below  which the targets providing the weakest values are excluded (i.e. it provides one lower threshold on R$_\mathrm{773.3}$, and one on R$_\mathrm{895.0}$).

\begin{figure}
    \centering
    \includegraphics[width=\linewidth]{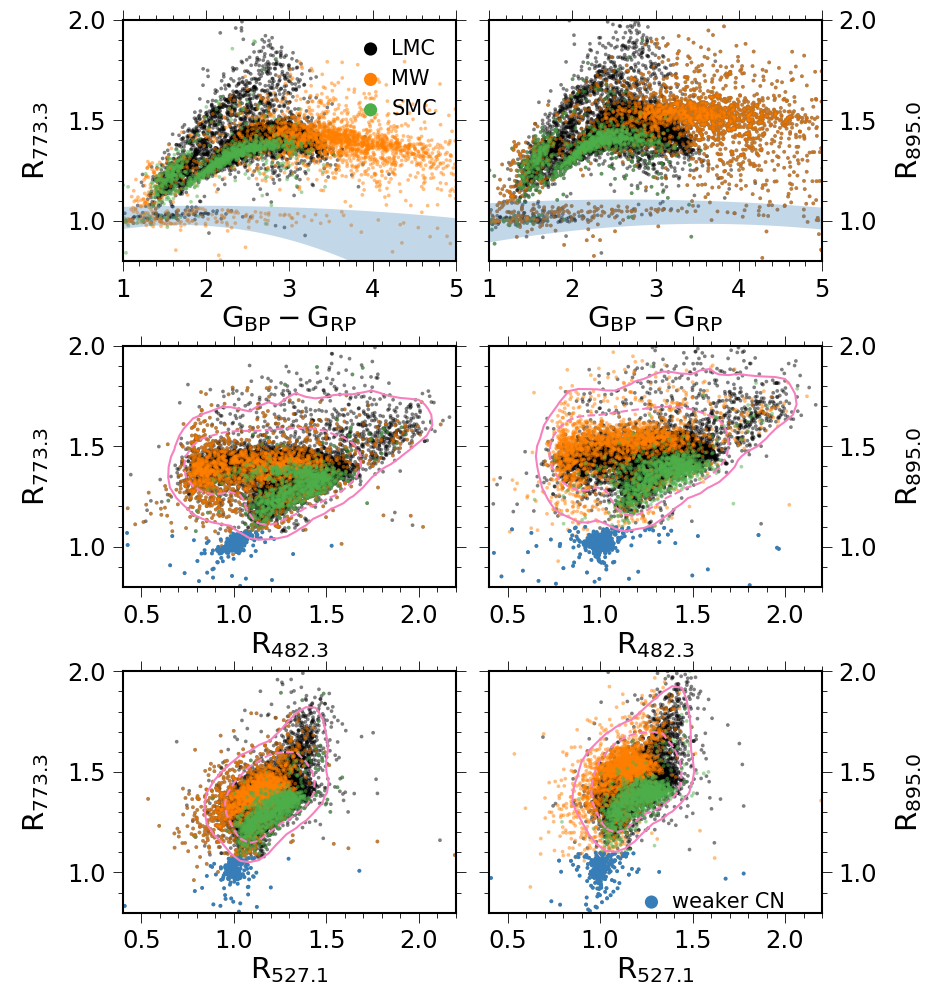}
    \caption{Band head strengths (see Eq.\,\ref{eq:bandhead_strength_cstars} and Table\,\ref{tab:cstars_bandheads}) measured in the BP and RP spectra of known Galactic \citep[MW, orange points,][]{2001BaltA..10....1A}, Large Magellenic Cloud \citep[LMC, black points,][]{2001A&A...369..932K}, and Small Magellenic Cloud  \citep[SMC, green points,][]{1995A&AS..113..539M} carbon stars. Only targets within 1 arcsecond of a \gaia\  DR3 source\_id are taken into account. Upper panels: the locus occupied by non-carbon stars is represented by the blue shaded area. Middle and lower panels: targets with weaker or non-existing CN features are shown with blue points (i.e. they fall in the shaded areas of the upper panels). The pink broken and full lines delimit the domain occupied by 87 \% and 98 \% of the carbon stars with strong CN features.}
    \label{fig:cstars_bands_ref}
\end{figure}

\begin{figure}
    \centering
    \includegraphics[width=\linewidth]{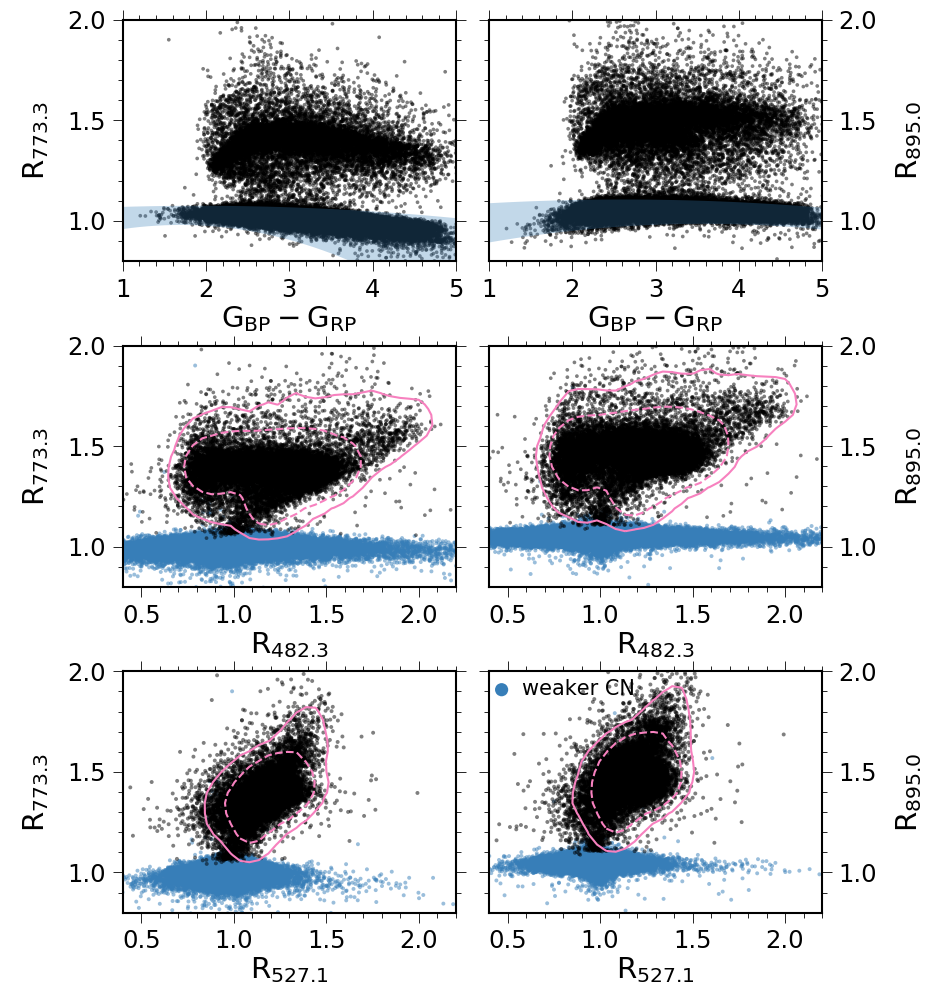}
    \caption{Same as Fig.\,\ref{fig:cstars_bands_ref} but for the 386\,936 candidate carbon stars flagged by ESP-ELS. Pink curves represent the domain occupied by the carbon stars found in the literature (Fig.\,\ref{fig:cstars_bands_ref}, and Sect.\,\ref{sec:cstars_selection}).}
    \label{fig:cstars_bands_espels}
\end{figure}

In Figs.\,\ref{fig:cstars_bands_ref} and \ref{fig:cstars_bands_espels}, the results obtained for known carbon stars, and for the candidate carbon stars flagged by the ESP-ELS module are reported, respectively.
Most of the 386\,936 candidate carbon stars (upper panels of Fig.\,\ref{fig:cstars_bands_espels}) flagged by the algorithm have $G_\mathrm{BP}-G_\mathrm{RP}$ $>$ 2 mag, and have colors consistent with M stars. However, the known carbon stars, especially in the Magellanic clouds have colours down to $\sim$1 mag. A significant fraction of these have therefore not been detected and are not part of the golden sample. Our proposed sample of carbon stars is obtained after applying the lower thresholds on both R$_\mathrm{773.3}$ and R$_\mathrm{895.0}$ ratios.

\subsection{Validation of sample}\label{ssec:hotvalidation}

\begin{figure}
    \centering
    \includegraphics[width=\linewidth]{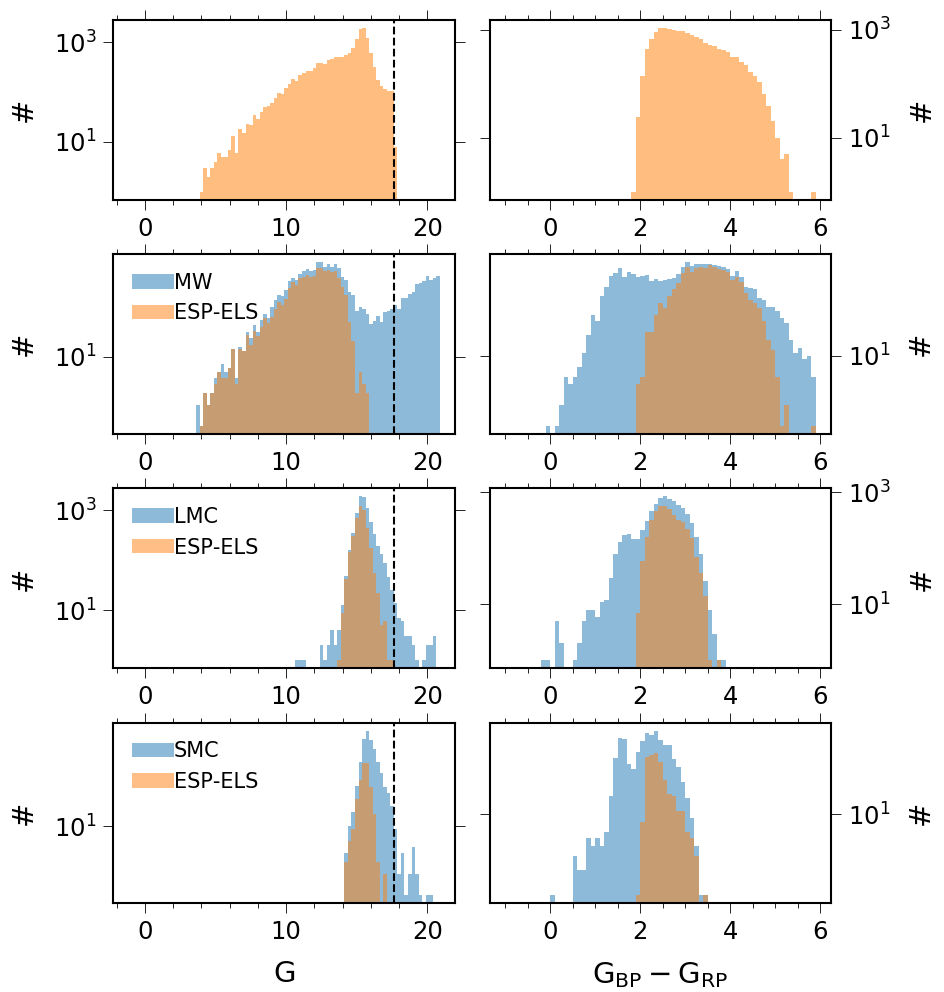}
    \caption{Magnitude and color distribution of carbon stars. Left panels: The vertical black dashed line shows the upper magnitude limit of the data processed by ESP-ELS. Upper panels: All the targets belonging to the golden sample of carbon stars are taken into account. Other panels: Distributions obtained for the known MW  \citep{2001BaltA..10....1A}, LMC \citep{2001A&A...369..932K}, and SMC \citep{1995A&AS..113..539M} carbon stars are shown in blue. In orange, we show the distribution of the targets in common with the sample we propose in this work.}
    \label{fig:cstars_mags}
\end{figure}

\begin{figure}
    \centering
    \includegraphics[width=\linewidth]{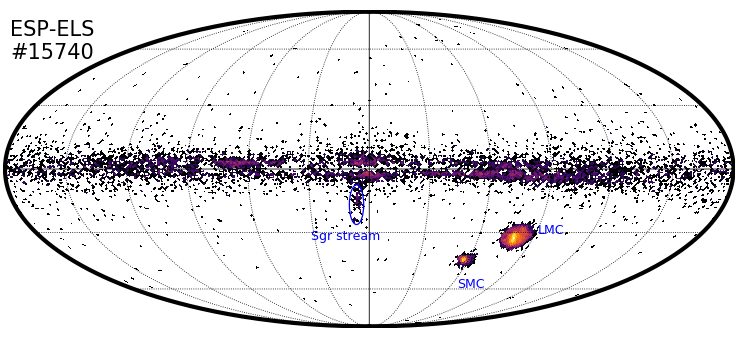}
    \caption{Mollweide view in Galactic coordinates of the carbon stars sample described in this work. The locations of the Magellanic Clouds, and of the Sagittarius stream are shown in blue.}
    \label{fig:cstars_map}
\end{figure}

The sample we propose includes 15\,740 stars exhibiting the strongest CN molecular bands. Their spatial distribution is shown in Fig.\,\ref{fig:cstars_map}. As previously noted, most of the remaining carbon stars have \bpminrp $>$ 2 mag, which is consistent with what is expected from M-type stars (Fig.\,\ref{fig:cstars_mags}). From a cross-match with the 3 main catalogues of carbon stars, about two thirds are known cases. The magnitude and color distributions of the targets found in the literature and in common with the proposed sample are shown in Fig.\,\ref{fig:cstars_mags}. Most of the carbon stars that have not been identified correspond to targets bluer than $G_\mathrm{BP}-G_\mathrm{RP} = 2$ or/and fainter than $G=17.65$ mag. Taking magnitude and color/\teff constraints into account, the fractions of detected known carbon stars are shown in Table\,\ref{tab:cstars_fractions}. 

\begin{figure}
    \centering
    \includegraphics[width=\linewidth]{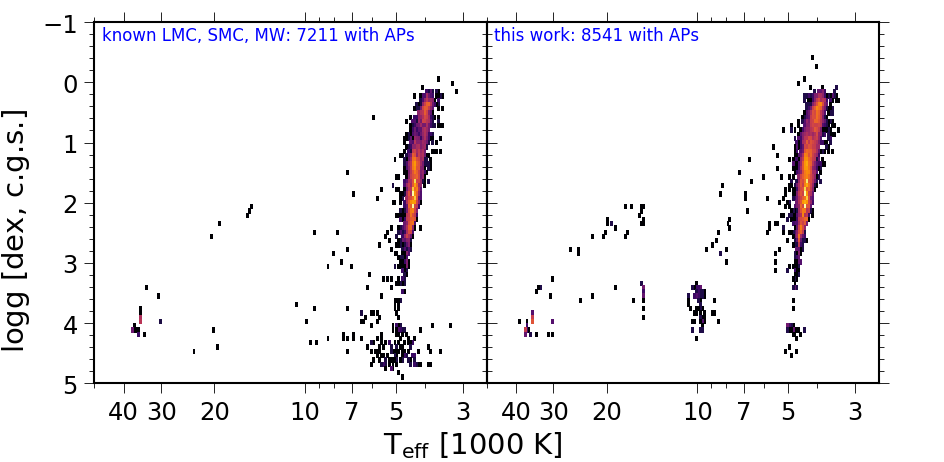}
    \caption{Kiel diagram of carbon stars with published \gspphot\ parameters. Left panel: density plot for known MW, SMC, and LMC C stars. Right panel: density plot for the carbon stars in our sample.}
    \label{fig:cstars_kiel}
\end{figure}

\begin{figure}
    \centering
    \includegraphics[width=\linewidth]{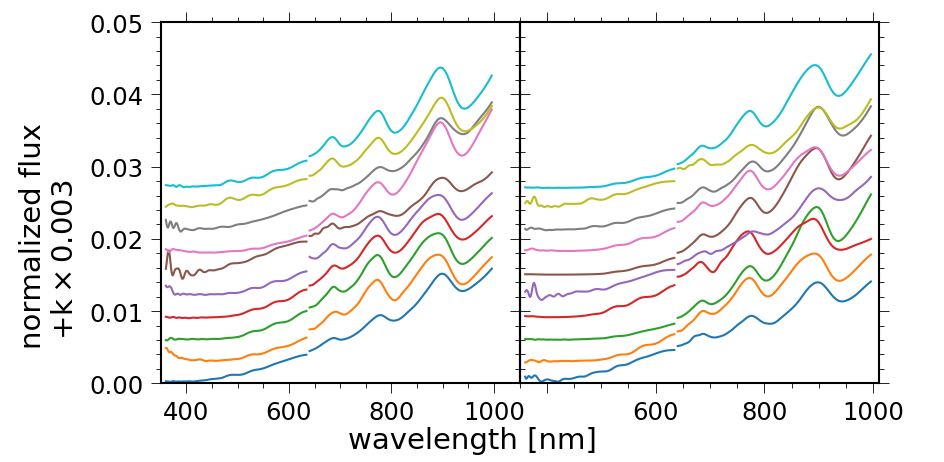}
    \caption{BP and RP spectra of the 20 randomly chosen (amongst 254) carbon stars (this work) with \teff$_\mathrm{GSP-Phot} > $ 6000~K. The ordinate axis provide the flux normalized to the total flux, and shifted by $k \times 0.003$ (where $k$ is an integer that varies from 0 to 9 from the bottom to the top spectrum).}
    \label{fig:cstars_hot_xp}
\end{figure}

Carbon stars are located at the very cool edge of \gspphot's \teff\ domain ($\teff>2500$~K). In addition, no synthetic spectra adapted for the accurate AP determination of carbon stars were available, and only a fraction of the carbon stars have their astrophysical parameters published in GDR3. Hence, it is not surprising that the \teff\ that is obtained tends to be overestimated (by 500 to 1500~K) and should be considered with caution. However, the Kiel diagrams obtained for the known carbon stars (Fig.\,\ref{fig:cstars_kiel}, left panel) and those from our list (same Figure, right panel) are consistent with each other. Notwithstanding the estimated \teff\ and  their location in the diagram is also consistent with AGB stars. Note that a few targets (254) have \teff\ hotter than 6\,000~K, while the corresponding SEDs are typical of AGB carbon stars (showing typical CN bands in the RP) as shown in Fig.\,\ref{fig:cstars_hot_xp}.

To exploit this sample, the list of {\tt source\_id} are made available as a separate table in the archive \pvptable{gold_sample_carbon_stars} for 
the 15\,740 bone fide carbon stars, which were also flagged in the main \aptable\ table (see \linktoapparam{astrophysical_parameters}{flags_esphs} for details).
The initial set of 386\,936 carbon-candidate stars can still be found in the same table, as these remain tagged ``CSTAR” in the \linktoapparam{astrophysical_parameters}{spectraltype_esphs} field.

\begin{table}
\caption{Fractions of detected known carbon stars.}\label{tab:cstars_fractions}
\begin{tabular}{lrr}
\hline
galaxy & G $\leq$ 17.65  & G $\leq$ 17.65 \\
 & \& \bpminrp $\geq$ 2 &  \\
\hline
MW & 0.82 & 0.70 \\
LMC & 0.61 & 0.54 \\
SMC & 0.41 & 0.27 \\
\hline
\end{tabular}
\end{table}

\section{Solar analogues}\label{sec:solaranalogues}

\subsection{Scientific motivation}

The Sun is the reference point in much of stellar astronomy and astrophysics. Solar analogues are stars which in a restricted set of parameters resemble the Sun. In contrast to the Sun, they can be observed in the night sky and with the very same instruments used to study stars in the Milky Way. There is no strict definition of what constitutes a solar analogue, both the set of parameters and allowed parameter ranges vary in the literature. For astrophysical purposes one often aims to constrain the photometrically/spectroscopically accessible parameters \teff, $\log g$ and the overall metallicity \mh\ to within typical measurement uncertainties. Depending on data quality and analysis technique applied, uncertainties as small as 10 K in \teff, 0.03 in $\log g$ and 0.01 in [M/H] are achievable\footnote{Stars with parameters indistinguishable from the Sun are sometimes also referred to as solar twins.} \citep{2021MNRAS.504.1873Y}, but 50 K, 0.15 and 0.05 are more typical values. These small errors are the result of line-by-line differential analyses relative to the Sun, a technique which cancels many of the systematic sources of errors that stellar analyses otherwise often suffer from.

The most accurate analyses have revealed systematic differences of the chemical composition of the Sun relative to solar analogues in the solar neighbourhood: When selected to be good matches in [Fe/H] (iron abundance), the Sun is among the 10-15\% of stars rich in volatile elements  \citep{2009ApJ...704L..66M}. A tight (broken) trend of abundance with condensation temperature of the various elements is found with an amplitude of 0.08 dex (20\% in linear abundance). The reason for this effect is still unknown, but is speculated to be related to selective accretion of gas over dust due to the presence of planets. This finding potentially opens up new avenues for systematic evolutionary studies of solar-type stars and their planets.

Solar analogues have also been used to identify the abundance ratios which depend most sensitively on stellar age and can thus serve as precise spectroscopic clocks. One such study identifies the [Y/Mg] abundance ratio as particularly age-sensitive \citep{2015A&A...579A..52N}. Working with ages rather than metallicity as a proxy for age puts chemical-evolution studies on a much firmer footing. Loosening constraints on the stellar parameters, one can also study ``the Sun as a star'' and its evolution.

Finally, solar analogues also serve a purpose in the study of minor bodies of the solar system. In this context, they are used to subtract the solar spectrum (and earth-atmospheric contributions in the case of ground-based observations) from reflectance spectroscopy of e.g.\ asteroids with the aim of a more uniform classification, see for example the \ssotable{sso_reflectance_spectrum} table in \gdr{3} and \cite{DR3-DPACP-89}. Note that this type of science case asks for stars whose spectral energy distributions resemble that of the Sun as closely as possible. This requirement does not necessarily ask for a perfect match in stellar parameters, especially if one considers fainter G dwarfs which may suffer from extinction and associated reddening.

\begin{figure}[!htb]
\center{\includegraphics[width=0.49\textwidth]{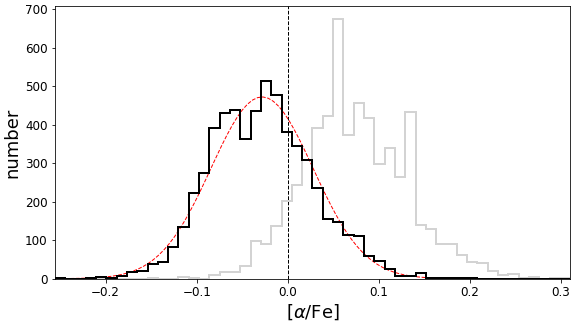}}  
\caption{\label{fig:solar-analogs-alpha-distribution}
Distribution of [$\alpha$/Fe] abundances from \gspspec\ for solar-analogue candidates. Grey shows the raw \texttt{alphafe\_gspspec} values and black shows the calibrated values \citep{DR3-DPACP-186}. The dashed red line shows a Gaussian distribution with mean of --0.028 and standard deviation of 0.056.}
\end{figure}

A scientific application of solar analogues is presented in Section \ref{sec:exploitation}.

\subsection{Candidate selection}
In order to identify candidates for solar analogues from the full \gaia\  sample, we need to define selection criteria. We apply two general criteria:
\begin{enumerate}
\item Apparent magnitude brighter than $G<16$ since fainter sources would be difficult to follow up efficiently with ground-based spectroscopy.
\item Excellent parallax quality, $\varpi/\sigma_\varpi>20$, in order to reliably place sources in the HR diagram.
\end{enumerate}
From these basic criteria, we continue to select candidates from \gspspec\ results. On a known sample of solar analogues and twins  \citep{2014A&A...563A..52P,2014A&A...572A..48R,2015A&A...579A..52N,2016A&A...587A.131M,2016A&A...590A..32T,2018A&A...619A..73L,2019A&A...629A..33G,2020A&A...639A.127C,2021MNRAS.504.1873Y}, \gspspec\ on average deviates from solar values by 14.4K in \teff, by $-0.071$ in $\log g$ and by $-0.05$ in [M/H] \citep[][Sect.~4.1 therein]{DR3-DPACP-160}. Taking those average differences into account, we require that \gspspec\ results agree with 5772~K to within 100~K, to $\log g=4.44$ to within 0.25, and to $\textrm{[M/H]}=0$ to within 0.1. Furthermore, we require good \gspspec\ flags\footnote{\texttt{gspspec\_flags} equal to 0 in characters 1 to 13, except for 8, and equal to 0 or 1 in character 8. These flag characters are related to the fundamental spectral parameters, see Sect.~\ref{ssec:fgkmsampleselection_spec} for details. All other flag characters relate to specific elemental abundances and we ignore them in this context.}.  Finally, we combine \gspspec\ results with \flame\ estimates to further weed out possible contamination: First, we require that \texttt{mass\_flame\_spec} is between 0.95 $M_\odot$ and 1.05 $M_\odot$. Second, we require that \texttt{radius\_flame\_spec} is between 0.8 $R_\odot$ and 1.2 $R_\odot$. This results in a total of 5863 \gspspec\ candidates for solar analogues, of which 916 have RVS spectra published in \gdr{3}\footnote{\gspphot\ results can also be used to select candidates. Here, we only consider PHOENIX and MARCS in the context of solar analogues. We require that \gspphot\ results \textit{for both libraries} agree with 5772~K within 100~K, to $\log g=4.44$ within 0.25, and to $\textrm{[M/H]}=0$ within 0.1, where we correct results from each library for its mean differences to known solar analogues \citep[][Sect.~4.1 therein]{DR3-DPACP-160}. This results in a total of 234\,779 \gspphot\ candidates for solar analogues, 7884 of which have RVS spectra. However, we do not publish this candidate list. Interested readers may contact the authors.}.
The list of \gdr{3} source IDs for the 5863 solar-analogue candidates from \gspspec\ is provided in the \gaia\ archive as a separate table \pvptable{gold_sample_solar_analogues}.

Due to the selection on very high parallax quality ($\varpi/\sigma_\varpi>20$), the candidates tend to be nearby and thus scatter more or less uniformly over the whole sky. Yet, the sky distribution shows the imprint of the \gaia\ scanning law, because high parallax quality is easiest to achieve for sources with many transits.

In Fig.~\ref{fig:solar-analogs-alpha-distribution}, we check and verify that the solar-analogues candidates have [$\alpha$/Fe] abundances that are statistically consistent with the solar value of zero. The standard deviation of [$\alpha$/Fe] for this particular subset of solar-like stars is 0.056, which is lower than the global uncertainty reported for all stellar types in \citet{DR3-DPACP-186}.

\subsection{RVS spectra of candidates}

\begin{figure*}[!htb]
\center{\includegraphics[width=\textwidth]{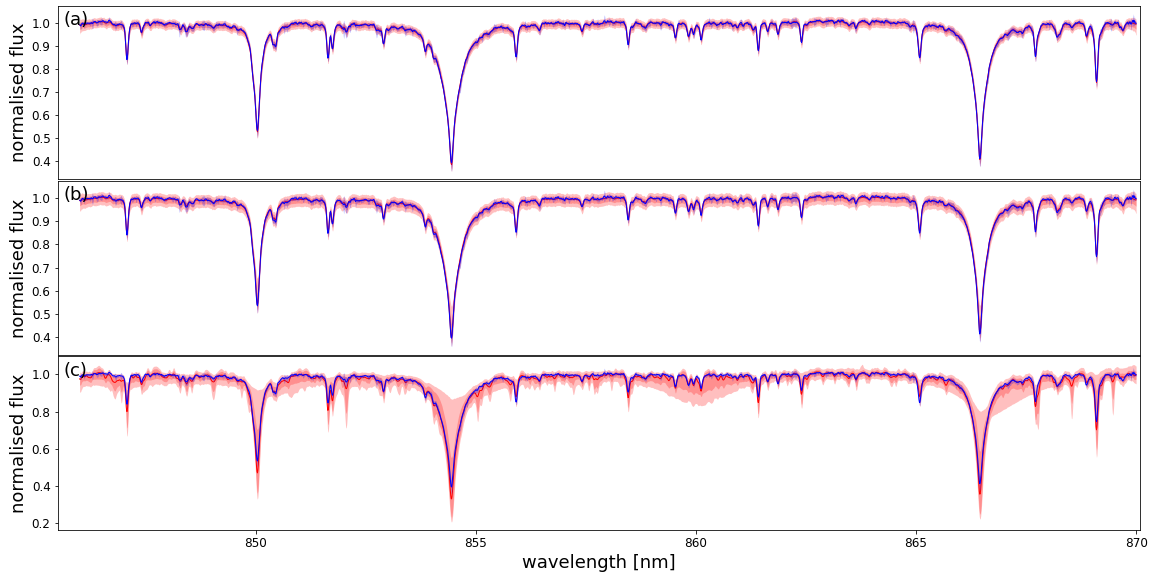}}  
\caption{\label{fig:solar-candidate-RVS-spectra}
RVS spectra of 916 solar-analogue candidates from \gspspec\ (panel a) where 95\% of \gspspec candidates satisfy $G<11.7$. We also show the solar-analogue candidates obtained from a possible selection from \gspphot\ results in panels b, but we only show 1985 \gspphot\ candidates with $G<11.7$. For comparison, panel c shows RVS spectra of 7589 randomly selected stars (i.e.\ no solar-analogue candidates) also with $G<11.7$. In all panels, the red line shows the median in each pixel and the shaded red contours show the pixel-wise central 68\% and 90\% intervals. The solid blue line is identical in all three panels and shows the mean RVS spectrum of 13 solar analogues known from the literature.}
\end{figure*}

As a visual confirmation of the candidate selection, we inspect the published RVS spectra of the candidates. For comparison, we also take the RVS spectra of 13 known solar analogues which have RVS spectra published in \gdr{3}. Figure~\ref{fig:solar-candidate-RVS-spectra}a shows that the 916 \gspspec\ candidates with RVS spectra are in excellent agreement with the mean RVS spectrum of known solar analogues. Most of these 916 \gspspec\ candidates are brighter than $G<11.7$. The 1985 candidates with RVS spectra one would obtain from \gspphot\ and that are similarly bright ($G<11.7$) are shown in Fig.~\ref{fig:solar-candidate-RVS-spectra}b. They show equally good agreement with the known solar analogues as Fig.~\ref{fig:solar-candidate-RVS-spectra}a. This demonstrates that \gspphot\ results are also very reliable under these selection criteria. For orientation, Fig.~\ref{fig:solar-candidate-RVS-spectra}c shows RVS spectra of 7589 random stars with $G<11.7$ and here we see clear differences, e.g.\ the Ca lines vary in depth, where for hot stars in particular the Ca lines are usually weak and instead Paschen lines start to appear. In Fig.~\ref{fig:solar-candidate-RVS-spectra}c, we can also see the DIB around 860nm \citep{DR3-DPACP-144}.

\subsection{Candidates with extinction}
\label{ssec:solaranalogs-with-extinction}

\begin{figure}[!htb]
\center{\includegraphics[width=0.49\textwidth]{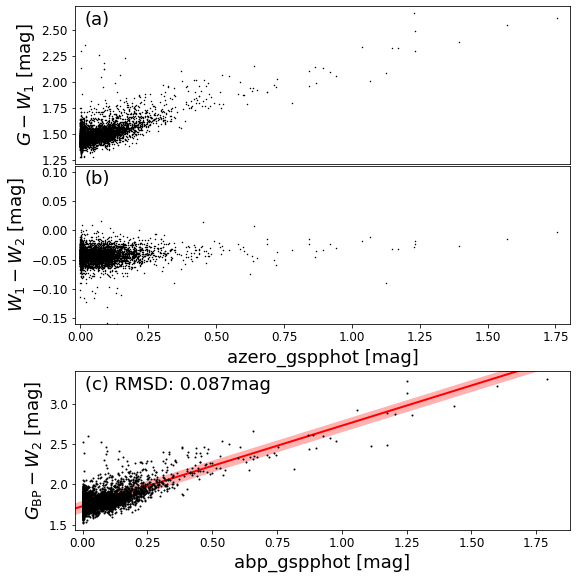}}
\caption{\label{fig:solar-analogs-colours-vs-A0}
Colours of \gspspec\ solar-analogue candidates as a function of \gspphot\ extinction estimates. $W_1$ and $W_2$ denote AllWISE photometry \citep{2014yCat.2328....0C}. We restrict to candidates with $W_1>8$~mag, since AllWISE photometry suffers from saturation for brighter sources. In panel c, the red line is a linear increase with \texttt{abp\_gspphot} offset by the mean $\gbp-W_2$ colour of 589 stars where $\azero<0.001$~mag according to \gspphot. The red interval marks the uncertainty from the standard deviation of the mean. The quoted root-mean-square (RMS) difference is between the $\gbp-W_2$ colour and \texttt{abp\_gspphot} plus the mean.}
\end{figure}

Solar analogues with notable extinction would be of particular scientific interest, e.g.\ for inferring the extinction law. In Fig.~\ref{fig:solar-analogs-colours-vs-A0}, we show colours of \gspspec\ candidates including photometry from \gaia\ and AllWISE \citep{2014yCat.2328....0C} as a function of \gspphot's \azero\ estimate. The $G-W_1$ colour clearly reddens with increasing \azero\ in Fig.~\ref{fig:solar-analogs-colours-vs-A0}a whereas the $W_1-W_2$ colour remains virtually constant in Fig.~\ref{fig:solar-analogs-colours-vs-A0}b.\footnote{We also inspected the variation of these colours with \gspspec's DIB measurements \citep{DR3-DPACP-144} and find qualitatively similar results. Unfortunately, only very few DIB measurements are available for our candidates.}
In Fig.~\ref{fig:solar-analogs-colours-vs-A0}c, we further investigate the reddening of the $\gbp-W_2$ colour, which has the largest wavelength coverage from the near ultra-violet (320-670~nm for \gbp) to the near infrared (4.6 $\mu$m for $W_2$). In particular, \gbp\ will be much more affected by extinction than $W_2$, in fact $A_\textrm{BP}\gg A_{W2}$, such that we can take \gspphot's $A_\textrm{BP}$ estimate as an approximation for the reddening of the $\gbp-W_2$ colour. Indeed, Fig.~\ref{fig:solar-analogs-colours-vs-A0}c shows a linear relation with a low RMS deviation of 0.087 mag across an $A_\textrm{BP}$ range of 1.75 mag. This attests to the quality of the $A_\textrm{BP}$ estimate from \gspphot (at least for bright sources with high-quality parallax measurements).

\begin{figure}[!htb]
\center{\includegraphics[width=0.49\textwidth]{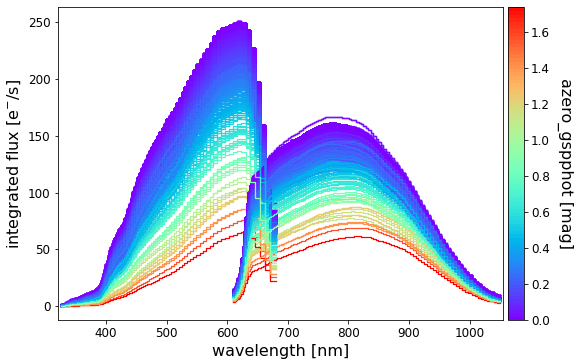}}  
\caption{\label{fig:solar-analogs-XP-spectra-variation-with-A0}
Variation of low-resolution BP and RP spectra of \gspspec\ solar-analogue candidates with \gspphot's \azero\ estimate. In order to make the BP and RP spectra comparable, they have been rescaled to an apparent magnitude of $G=15+A_G$ with $A_G$ taken from \gspphot.}
\end{figure}

Having established in Fig.~\ref{fig:solar-analogs-colours-vs-A0} that the \gspphot\ extinction estimates agree with the reddening of colours of the candidates, we inspect how the low-resolution BP and RP spectra themselves vary with \gspphot\ extinction. For the 5863 \gspspec\ candidates, Fig.~\ref{fig:solar-analogs-XP-spectra-variation-with-A0} shows that the BP and RP spectra clearly redden and dim as the \gspphot\ estimate of \azero\ increases. While BP and RP spectra at low extinction show much more flux in BP than in RP, BP and RP spectra at $\azero\sim 1.5$ mag already show equally high peak fluxes in both BP and RP while their overall flux is reduced by a factor of $\sim$5 in BP and $\sim$3 in RP with respect to a zero-extinction solar-like BP/RP spectrum.

\section{SPSS}\label{sec:spss}

\begin{figure*}[!htb]
\center{
\includegraphics[width=0.305\textwidth]{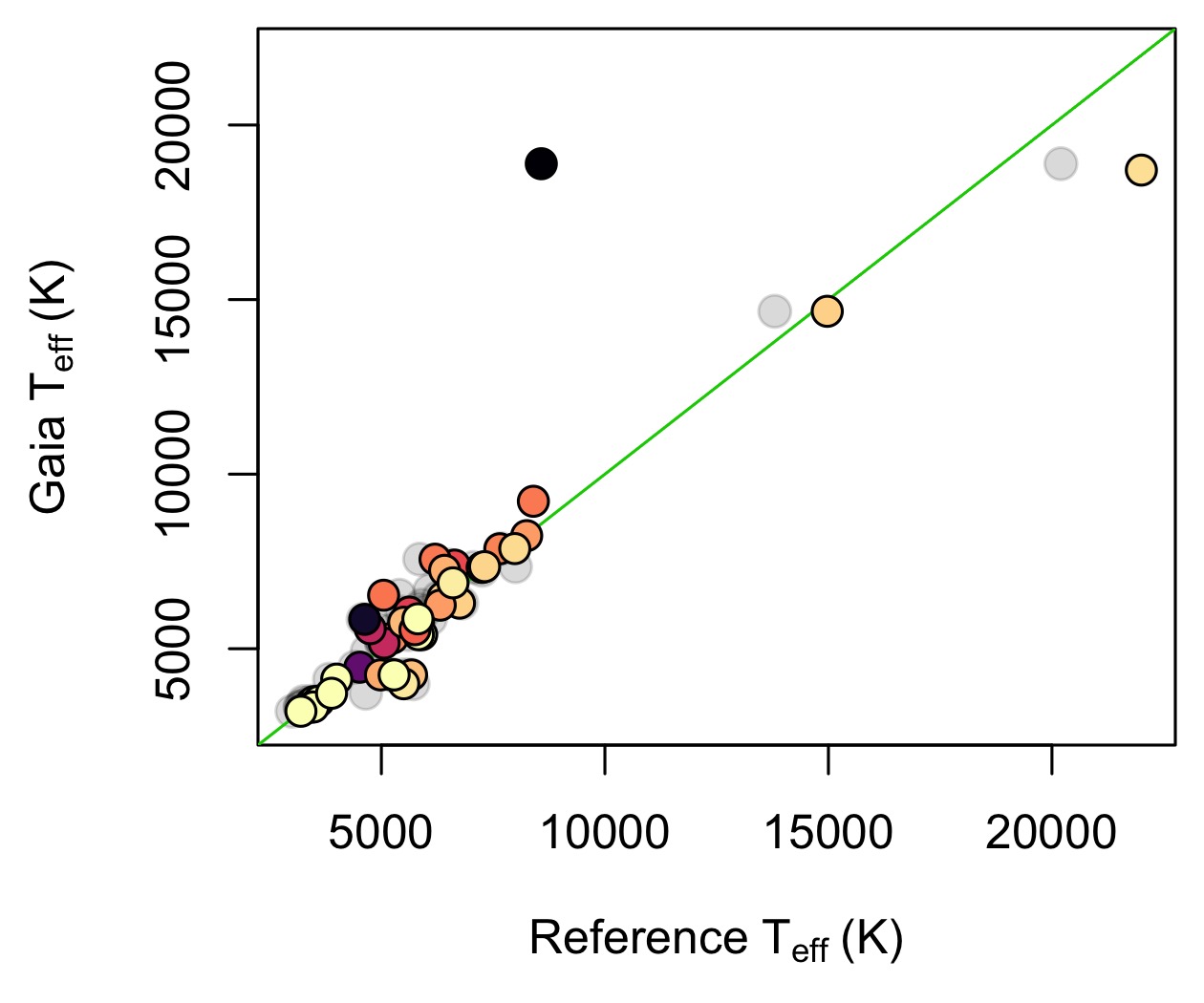}
\includegraphics[width=0.305\textwidth]{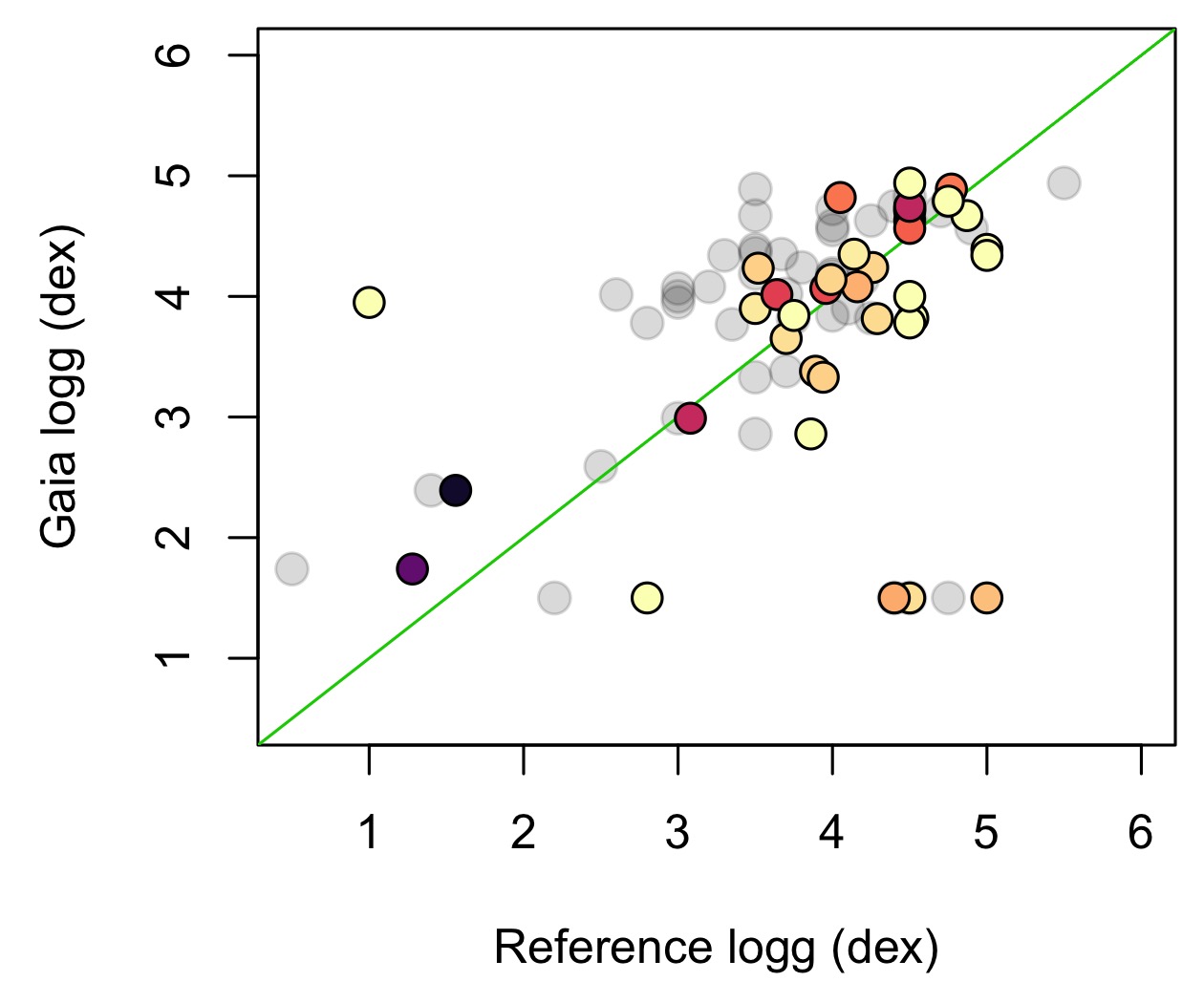}
\includegraphics[width=0.38\textwidth]{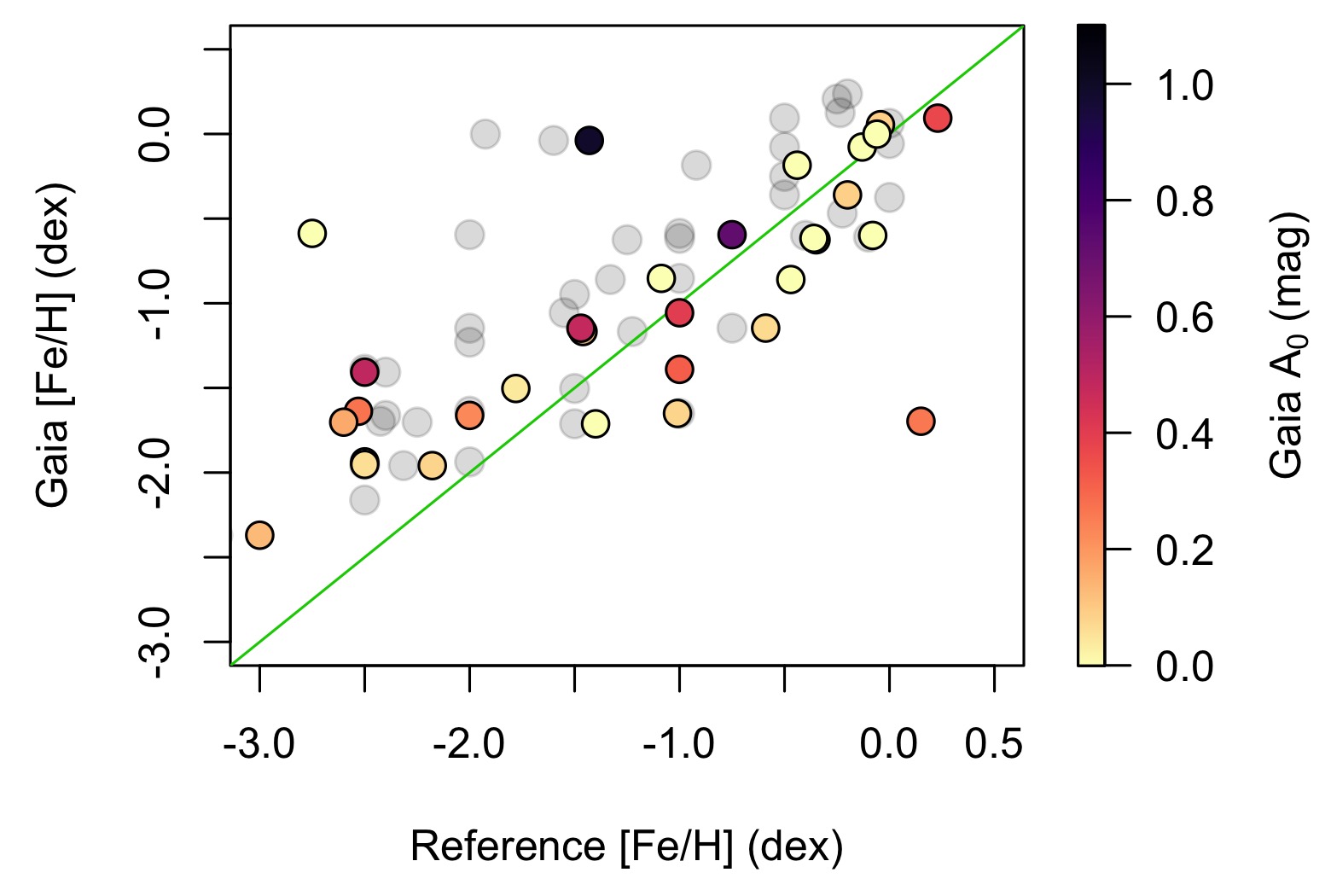}
} 
\caption{\label{fig:spss} Comparison of the SPSS sample main parameters derived here with the two reference sets by \citet{pancino21}: the best-fit parameters to the SPSS flux tables are shown in grey, while a collection of literature spectroscopic estimates is colored according to the interstellar absorption A$_0$ obtained here. The left panel shows the case of \teff, the middle one of log$g$, the right one of [Fe/H]. The 1:1 line is shown in green in all panels.}
\end{figure*}

The \gaia\ Spectro-Photometric Standard Stars\footnote{\url{http://gaiaextra.ssdc.asi.it:8900/}} \citep[SPSS, ][]{pancino12,pancino21} are a grid of flux tables specifically designed to calibrate \gaia\ photometry and BP and RP spectra. They are the result of a dedicated set of ground-based observing campaigns to collect spectrophotometry \citep{altavilla15}, light curves for constancy monitoring \citep{marinoni16}, and absolute photometry for validation \citep{altavilla21}, over more than ten years. The latest version of the grid, SPSS V2, was used to calibrate the \gaia\ photometry in EDR3 and the BP and RP spectra in DR3. It contains 111 stars\footnote{\citet{pancino21} list 112 stars, but one (SPSS\,192, see their Figure~11) was found to have a close companion at about 0.25" with SEApipe \citep{harrison11}.}, based on $\simeq$1\,500 spectra, and it is calibrated on the 2013 version of the CALSPEC\footnote{\url{https://www.stsci.edu/hst/instrumentation/reference-data-for-calibration-and-tools/astronomical-catalogs/calspec}} grid \citep{1995AJ....110.1316B,bohlin14,bohlin19} with a zero-point accuracy of better than 1\%. The SPSS grid is designed to cover those areas of the stellar parameter space that are not well sampled by CALSPEC, in particular the FGKM star types, and to cover the entire \gaia\ wavelength range (330--1050~nm). The final release, SPSS V3, will be used to calibrate \gaia\ DR4, it will contain about 200 stars and will make full use of all the $\simeq$6\,500 spectra collected in the observing campaigns. It will be calibrated on the latest version of the CALSPEC grid \citep{bohlin14,bohlin19}, which differs by about $<$0.5\% from the 2013 one in the grid zero-point. The S/N ratio of the ground-based SPSS spectra is generally well above 100, with the exception of the blue and red extremes of the \gaia\ wavelength range. The SPSS flux tables were thus extended with theoretical spectra, adjusted to match the central, high-S/N ratio region of the observed spectra \citep[see][for details]{pancino21}. It is therefore of the utmost importance, for the next SPSS release, to have a robust estimate of the spectral type, atmospheric parameters (\teff, log$g$, [Fe/H], and [$\alpha$/Fe]), and of the interstellar absorption for as many of the SPSS as possible. 

To this aim, we explored and selected relevant information from the \aptable\ table of \gdr{3} for the SPSS V2 stars.
In particular, whenever available, we selected the \gspspec\ parameters over the \gspphot\ ones for the \teff, log$g$, [Fe/H]\footnote{To obtain [Fe/H] from the \gspphot\ [M/H] estimates, we used the formula by \citet{salaris93} and assumed an $\alpha$-enhancement of +0.35 for metal-poor stars, +0.15 for intermediate metallicities, and zero for solar or higher metallicity. Note that we did not apply any re-calibration to the \logg and [Fe/H] \gspphot\ and \gspspec\ values.}. Similarly, for the choice of the FLAME parameters, i.e., mass, age, luminosity, and radius, we always selected the corresponding FLAME-spec determinations when available (in the \apsupptable\ table). Parameters were available for  all the SPSS in the sample, except for the 56 white dwarfs. For hot stars, a handful of parameters from \esphs\ were available that were not parametrized by \gspphot and \gspspec. The two binarity estimators available ({\tt specmod} and {\tt combmod}) agreed in indicating SPSS\,028 (SA105-663) as a binary, while 15 different SPSS were indicated as photometrically variable ({\tt phot\_variable\_flag}) and will be carefully re-evaluated in the preparation of the SPSS V3 release.

To explore the quality of the results, we compared them with the two sets of parameters presented by \citet{pancino21}: (1) a collection of literature estimates and (2) the best-fit parameters obtained by extending the SPSS V2 flux tables with model libraries. First, we compared the spectral type determinations and found that only 16 SPSS out of 111 had discrepant spectral types, and in all cases the discrepancies never spanned more than one spectral class (e.g., an F star classified as G). For one star, SPSS\,313 (M5--S1490), discordant previous literature spectral type determination (from A to F) was available, and we found it to be a K giant, about 500~K cooler than the coolest literature determination. We then compared the three main atmospheric parameters (Figure~\ref{fig:spss}) with both reference sets. As can be seen, apart from very few outliers, the agreement with the two sets of reference parameters appears good, especially when considering the heterogeneity of the literature estimates. There is an indication that a few stars with A$_0\gtrsim1$~mag have problems in some of the parameters. However, for the majority of stars, the agreement for \teff and \logg is excellent, with median differences of $\Delta$\teff\,=\,--4\,$\pm$\,322~K and $\Delta$log$g$\,=\,--0.04\,$\pm$\,0.59~dex. The comparison of [Fe/H] is still good if one includes metal-poor stars, with $\Delta$[Fe/H]\,=\,0.15\,$\pm$\,0.61~dex. When excluding stars below [Fe/H]\,$\simeq$\,--2~dex, which appear to have an overestimated iron metallicity, the comparison improves, with $\Delta$[Fe/H]\,=\,--0.09\,$\pm$\,0.44~dex. We note that an overestimate for metal-poor stars is a common problem when metallicities or iron abundances are derived from photometric data or low resolution spectra \citep[see also][]{miller15,anders21,xu21}.

In \gdr{3} we present a table \pvptable{gold_sample_spss} which contains the 111 SPSS stars, and for each source, their \gdr{3} source\_id, name, spectral type, binary and variability flags, along with the stellar parameters, extinction, distance, radial velocity, and $v \sin i$ (where available) for the 52 non-subdwarf/white dwarf stars of the sample.


\section{Summary of golden samples}\label{sec:summarytables}

In Sect.~\ref{sec:oba} to Sect.~\ref{sec:spss} we defined several samples of stars, carefully selected to be homogeneous and with the highest quality, that can be used in many different astrophysical contexts.  Complementary data tables  have been made available in the \gdr{3} archive to help exploiting these samples, see \href{https://gea.esac.esa.int/archive/documentation/GDR3/Gaia_archive/chap_cu9pvp/sec_cu9pvp_apsis/}{here in the online documentation}.   In Table~\ref{tab:summarytable} we summarise the names, sizes, and contents of these tables, and here we provide an overview. 

The six tables are all entitled {\tt gold\_sample\_}{\it name} where {\it name} is specific to the sample, i.e. {\tt oba\_stars}, {\tt fgkm\_stars}, {\tt carbon\_stars}, {\tt solar\_analogues}, {\tt spss}, and {\tt ucd}.  These can be called in an ADQL query in \gdr{3} as {\tt gaiadr3.gold\_sample\_}{\it name}.

The tables for the solar analogues and the carbon stars contain the {\tt source\_id} only.  
The OBA table also includes a flag that allows one to apply a kinematic filter.  
The table for the UCDs contains, along with {\tt source\_id}, the newly derived radii and luminosities from the analysis of the \gaia\ and infrared data, and the bolometric flux correction. 
The SPSS sample table contains all 111 SPSS sources along with information such as binary and variability flags, radial velocity, and $v \sin i$.  The stellar parameters and extinction are given for the non-white dwarf stars, some of which are based on \gspspec\ parameters and others on \gspphot\ or on \esphs, {this is indicated by the {\tt notes} in that table}.  
Finally, for the FGKM sample, a table with {\tt source\_id}, the atmospheric parameters, the evolutionary parameters and the spectral type is provided, where specific parameters for some sources have been removed (compared to the \aptable\ table).

\begin{table*}
    \centering
        \caption{Summary of the tables in the \gdr{3} archive to help in the exploitation of the samples presented in this work. }
    \label{tab:summarytable}

    \begin{tabular}{lllllllllll}
    \hline\hline
    star type & table name & N & field contents of tables & section & notes \\
    \hline 
    OBA & \pvptable{gold_sample_oba_stars} & 3\,023\,388 & source\_id, flag & \ref{sec:oba}&\\
    FGKM & \pvptable{gold_sample_fgkm_stars} & 3\,273\,041 & source\_id, all APs & \ref{sec:fgkm} & Table~\ref{tab:fgkm_table}\\
    UCD & \pvptable{gold_sample_ucd} & 21\,068 & source\_id, \radius, \lum & \ref{sec:ucd}\\
    Carbon & \pvptable{gold_sample_carbon_stars} &  386\,936 & source\_id & \ref{sec:carbon}\\
    Solar analogues & \pvptable{gold_sample_solar_analogues} & 5863 & source\_id& \ref{sec:solaranalogues}\\
    SPSS & \pvptable{gold_sample_spss} & 111 & source\_id and all APs & \ref{sec:spss} \\ 
         \hline\hline
         
    \end{tabular}
\end{table*}

\section{Exploitation of the golden samples}\label{sec:exploitation}
In this section we demonstrate four applications of the golden samples presented in this paper.  For the first application we exploit the OBA sample to derive the parameters of the Milky Way rotation curve and the peculiar motion of the Sun. We then use the FGKM sample to characterise known transiting exoplanets. This is followed by an exploitation of the solar analogue sample to derive the colours of the Sun, and finally we use the stellar  companions of unseen UCDs  to explore the ages of these substellar systems.

\subsection{Milky Way rotation curve}\label{sec:rotationcurve}

\begin{figure}
    \centering
    \includegraphics[width=0.48\textwidth]{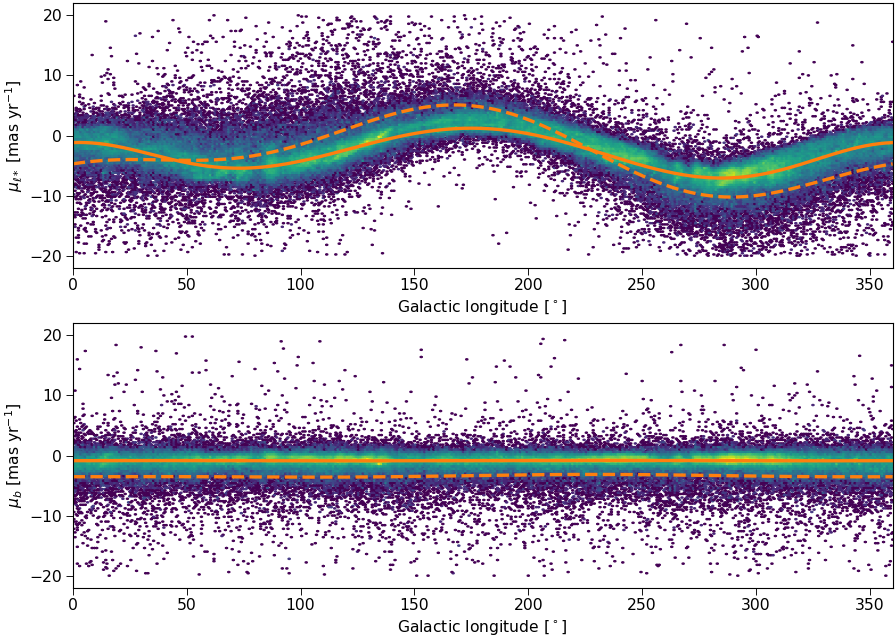}
    \caption{Proper motions in Galactic longitude (top) and latitude (bottom) as a function of Galactic longitude for the sample of $385\,423$ B-stars described in the text. The lines show the proper motions predicted from the rotation curve model parameters, resulting from the fit to the data, for stars at 500~pc from Sun (dashed line) and at 2000~pc (solid line, close to the median distance of stars in the sample).}
    \label{fig:pms_bstars}
\end{figure}

A classical application for the OBA star sample is to infer the parameters of the Milky Way rotation curve near the Sun. Young disk stars have often been used for this purpose because of the low dispersion of their velocities around the overall differential rotation of stars in the thin disk \citep[for a recent example based on \gaia~EDR3 data see][]{bobylev2022}. We illustrate this application with a very simple modelling of the proper motions in terms of the Milky Way disk rotation curve. The rotation curve is described with the circular velocity and the slope of the circular velocity as a function of Galactocentric cylindrical distance $R$, both evaluated at the position of the Sun \citep[or equivalently, the Oort constants $A$ and $B$ for an axisymmetric Milky Way, see e.g.][]{2003ApJ...599..275O}. We use a sub-sample of the OBA stars, namely those with \linktoapparam{astrophysical_parameters}{spectraltype_esphs} equal to `B', with $\plxsnr>10$ and $v_\mathrm{tan}<180$~\kms. The sample is further restricted to $(1000/\varpi)\times\sin b<250$~\pc and $6.5<R<15$~kpc. Over this range in $R$ the above approximation to the rotation curve is reasonable \citep[see e.g.][their Fig. 3]{2019ApJ...871..120E}. \figrefalt{fig:pms_bstars} shows the proper motions in $\ell$ and $b$ as a function of Galactic longitude for the $385\,423$ B-stars in this sample. The figure beautifully reveals the variation of $\pml$ with $\cos(2\ell)$, a consequence of Galactic differential rotation, and shows the slight offset of the proper motions in latitude from zero, reflecting the Sun's motion perpendicular to the Galactic plane. The width of the proper motion distributions mainly reflects the range of distances to the stars in the sample.

To derive the rotation curve parameters we use a Bayesian model for the proper motions in \figref{fig:pms_bstars}. The model has parameters similar to the simple kinematic model described in \secref{sec:oba}: the circular velocity at the position of the Sun \vcircsun, the slope of the circular velocity curve $d\vcirc/dR$, the peculiar motion vector of the Sun $\mathbf{v}_{\odot,\mathrm{pec}}=(\USun, \VSun, \WSun)$, and the velocity dispersions in the plane and perpendicular to the plane, $\sigma_{xy}$ and $\sigma_z$. The position of the Sun is fixed at a Galactocentric distance of $8277$~pc \citep{2022A&A...657L..12G}, while the height above the Galactic plane is taken as the median of $-(1000/\varpi)\times\sin b$ for the B-star sample, which is $16$~pc. The model velocities $\mathbf{v}$ of the stars are calculated from the azimuthal component of the velocity in Galactocentric cylindrical coordinates, $V_\phi = -(\vcircsun+d\vcirc/dR\times(R-R_\odot))$, as $\mathbf{v} = (-V_\phi\sin\phi, V_\phi\cos\phi, 0)$\footnote{Here we use a right handed coordinate system, so $V_\phi$ is negative for the stars in the disk of the Milky Way}. The model proper motions $\boldsymbol{\mu}_\mathrm{pred}$ are then calculated from $\mathbf{v}-\mathbf{v}_\odot$, the parallaxes, and celestial positions of the stars (with $\mathbf{v}_\odot = (\USun, \VSun+\vcircsun, \WSun)$). The parallaxes were used as error-free observables. The velocity of the local standard of rest with respect to the rotational standard of rest is assumed to be $0$~\kms \citep{2016ARA&A..54..529B} and not included in the model.

The 7 model parameters are optimized through a Markov-Chain Monte Carlo sampling of the posterior. The likelihood for the observed proper motions is a normal distribution centered on $\boldsymbol{\mu}_\mathrm{pred}$ with a covariance matrix that accounts for the covariance matrix of the observed proper motions and the velocity dispersions, using the appropriate form of equation (16) in \cite{2000A&A...356.1119L}. The priors on the model parameters are broad normal distributions centred on $220$, $11$, $12$, $7$~\kms for \vcircsun, \USun, \VSun, \WSun, respectively. The prior on $d\vcirc/dR$ is a normal distribution centred on $0$~\kmskpc. The priors on the velocity dispersions are Gamma distributions with parameters $\alpha=2$ and $\beta=0.1$. The model was implemented in Stan \citep{Stanmanual}, using the CmdStanPy interface. The posterior was sampled with 4 Markov Chain Monte Carlo (MCMC) chains for 1500 steps each, and the first 500 steps were discarded as `burn-in'. To keep the required computational resources within bounds, the Stan model was run using a random subset of $20\,000$ stars chosen from the B-star sample above.

The resulting model parameters are: $\vcircsun=234\pm0.5$~\kms, $d\vcirc/dR=-3.6\pm0.1$~\kmskpc, $\USun=8.1\pm0.1$~\kms, $\VSun=11.2\pm0.2$~\kms, $\WSun=8.1\pm0.1$~\kms, $\sigma_{xy}=14.2\pm0.1$~\kms, and $\sigma_z=7.3\pm0.1$~\kms. These numbers are consistent with results from the literature \citep[e.g., as compiled by][]{2016ARA&A..54..529B}. The corresponding Oort parameters are $A=16$~\kmskpc, $B=-12$~\kmskpc, and $A-B=28$~\kmskpc. The total velocity of the Sun translates to an apparent proper motion at the position of Sgr A$^*$ of $-6.25$~\maspyr along the plane and $-0.21$~\maspyr perpendicular to the plane. This is consistent with the most recent evaluation of the proper motion of Sgr A$^*$ by \cite{2020ApJ...892...39R}.

\begin{figure}
    \centering
    \includegraphics[width=0.48\textwidth]{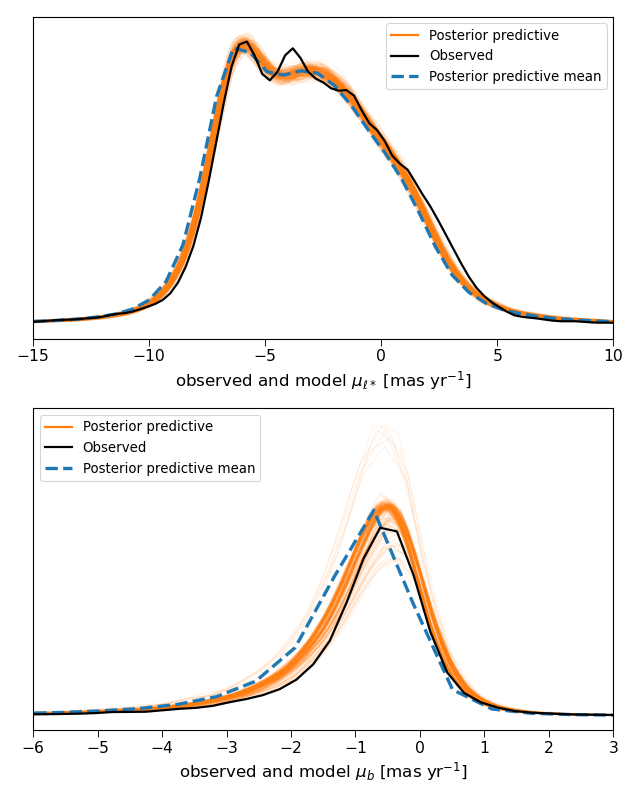}
    \caption{Distributions of the observed and model proper motions for the sample of $385\,423$ B-stars described in the text, with proper motions in Galactic longitude and latitude in the upper and lower panels, respectively. The black lines show the observed proper motion distributions. The thin orange lines are the predicted proper motion distributions for 200 randomly sampled MCMC model parameters. The mean of all such sample distributions is indicated by the thick dashed line.}
    \label{fig:bstars_postpred}
\end{figure}

The uncertainties quoted for the above results should be interpreted as the precision achieved in the context of the model and the sub-sample used. The uncertainties are underestimated. They do not account for the variance due to the choice of the specific random sub-sample of $20\,000$ B stars. More importantly, the obvious model deficiencies are not accounted for, such as ignoring the effects of the Milky Way disk warp and the motions induced by spiral arms \citep{2003ApJ...599..275O}, as well as deviations of the true rotation curve from the simple model. The `mode-mixing' effect discussed in \cite{2003ApJ...599..275O} is not an issue here because of the precise knowledge of the parallaxes of the stars in the sample. The model deficiencies are apparent in \figref{fig:bstars_postpred} which shows a comparison between the observed and model proper motion distributions. As noted above the modelling here is a mere illustration of the possibilities offered in analyzing the proper motions for a sample of young disk stars covering a large range in $R$. For a much more in-depth look at a sample of young disk stars, selected slightly differently from what we presented in \secref{sec:oba}, we refer to the \gdr{3} paper on mapping the asymmetric disk of the Milky Way \citep{DR3-DPACP-75}. The paper presents maps showing rich structure in the velocity field of OB stars which can be traced to the spiral arms, something which the above model obviously does not capture. On the other hand the average $V_\phi$ curve shown in that paper for the OB stars (calculated from the proper motions, parallaxes, and radial velocities) shows that the description of the rotation curve used above is accurate in an average sense.

\subsection{Exoplanet characterisation}\label{sec:exoplanets}
The search for and characterisation of exoplanet systems is at the forefront of scientific research, with many current and future ground- and space-based projects dedicated to this quest, e.g. \cite{jwst2006,kepler2008,plato2014,tess2014,ariel2018}.
The characterisation of the planet itself relies on the knowledge of the planet host.  In particular, the 
planet's radius and mass depends directly on the stars's radius and mass through the following equations
\begin{equation}
M_p \sin (i) =  \frac{M_\star^{2/3} P^{1/3} K (1-e^2)^{0.5}}{(2 \pi G)^{1/3}} 
    \label{eqn:planetmass}
\end{equation}
and 
\begin{equation}
d_{\rm tr} = \left ( \frac{R_p}{R_\star} \right )^2
    \label{eqn:planetradius}
\end{equation}
where $M_p, M_\star$ are the mass of the planet and star, respectively, $i$ and $e$ are 
the inclination and eccentricity of the orbital system,  $P$ the orbital period, $K$ is the semi-amplitude of the radial velocity curve, and $d_{\rm tr}$ is the transit depth due to the planet with radius $R_p$ passing in front of the star with radius $R_\star$ and blocking a part of its light.
 In reality the relationship between the transit depth and relative radii is a little more complicated than Eq.~\ref{eqn:planetradius}, see e.g \cite{heller2019}, but for this illustration purpose we keep things simple.  
 Additionally, we consider only transiting systems, so the inclination of the system is very close to 90$^\circ$ and $\sin (i) \sim 1$.

We obtained a list of the known transiting planets and light curve parameters from \href{http://exoplanets.org/}{\textcolor{blue}{\tt exoplanets.org}}\footnote{For four of the planets we adopted the values from the reference paper due to errors or inconsistencies:  XO-6b from \cite{crouzet17}, KELT-8b from \cite{kelt8}, {\it Kepler}-407b from \cite{kepler407}, and {\it Kepler}-68b \cite{kepler68}.}.  This catalogue contains (as of March 2022) 2651 confirmed transiting exoplanets.
We cross-matched these sources with the FGKM sample and obtained {593} planet matches.  Of these, {354} contain transiting parameters to estimate the planetary radius while  {108} entries contain parameters to estimate both the planet mass and radius, but only 94 have a valid stellar mass in our sample.  

\begin{figure}
    \centering
    \includegraphics[width=0.48\textwidth]{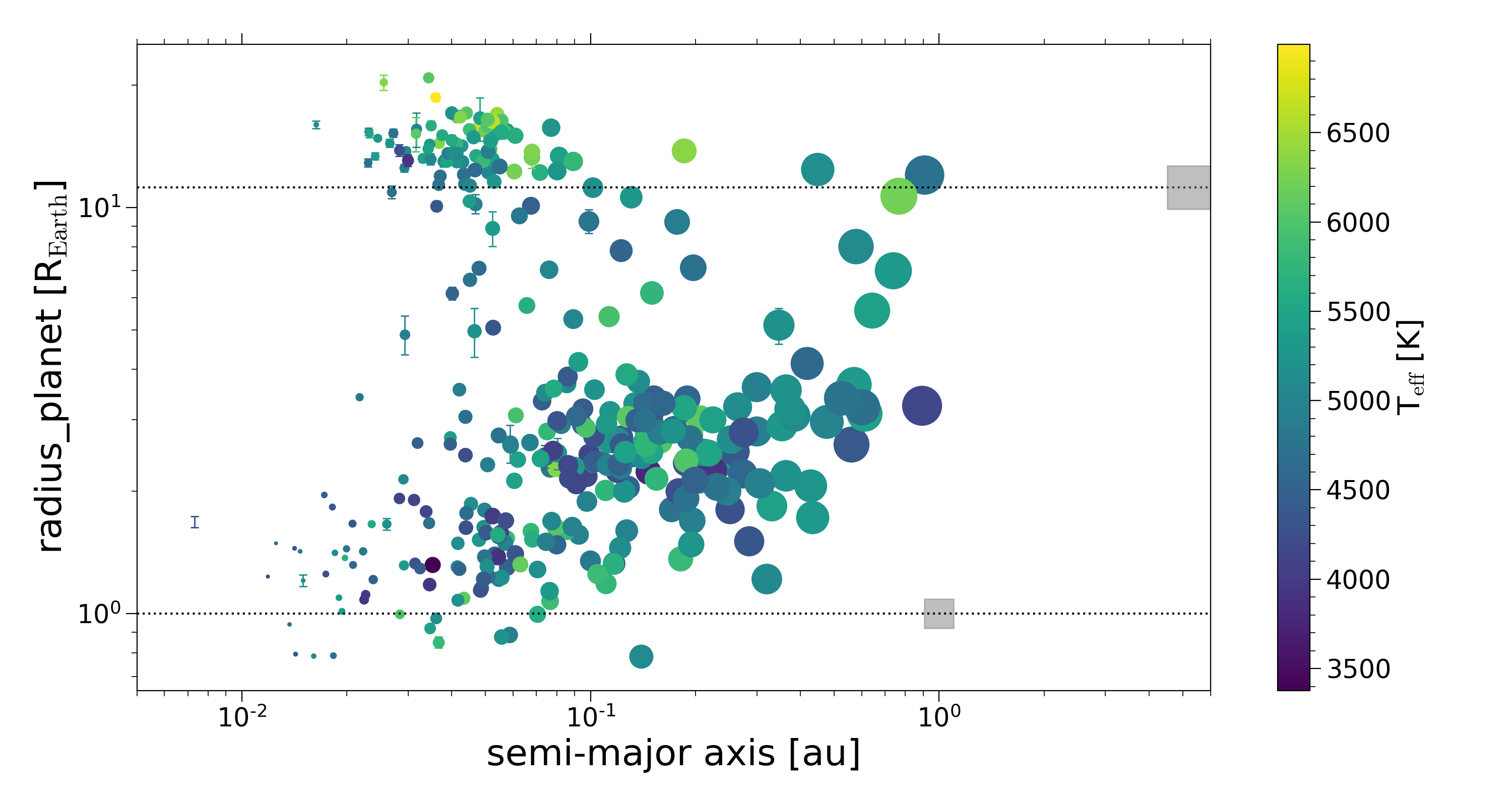}
    \caption{Distribution of planetary radii compared to the separation from their host star (orbital semi-major axis $a$) for planetary systems in the FGKM sample.  Colour-code indicates the \teff\ of the host star, while the symbol size indicates the orbital period in $\log_{10}$ scale (range = 0.57 - 364.8 days). 
    The dotted lines indicate 1 R$_{\rm Earth}$ and 1 R$_{\rm Jup}$ and the Earth and Jupiter are denoted by the square symbols.}
    \label{fig:exoplanet}
\end{figure}

We calculated the radii of the exoplanets using {\tt radius\_flame}, along with the available transit depth parameter.  To evaluate the uncertainties we performed a bootstrap method where we perturbed each of the input observations 1000 times and used the resulting standard distribution of the evaluated parameters to estimate the uncertainties.
We show the distribution of the planetary radii as a function of orbital separation of the planet-star system in Fig.~\ref{fig:exoplanet}.
We colour-coded the planet symbols according to {\tt teff\_gspphot} and the symbol size indicates the orbital period of the system, which ranges from 0.57 days to just under 365 days.   We also show the position of the Earth and Jupiter as grey squares, which highlights how different other planetary systems are to our own.  In
particular, many of these planets are well inside the inner limit 
of the habitable zone and the Jupiter-sized planets are equally close to their host star.

We furthermore calculated the mass of the planets for the 94 sources with radial velocity parameters and stellar masses.  Of these, four did not have a reported eccentricity, and of the other 90, only 24 have non-zero values.  For the planets with no reported eccentricity we assumed circular orbits.
10 of the planets also did not have a reported inclination and so we assumed $i = 90^\circ$ ($\sin i = 1$) which is reasonable for a transiting system.  The median value of the inclinations of the other 84 planetary systems is 87.2$^\circ$ ($\sin i = 0.9988 \sim 1$). 

\begin{figure}
    \centering
    \includegraphics[width=0.48\textwidth]{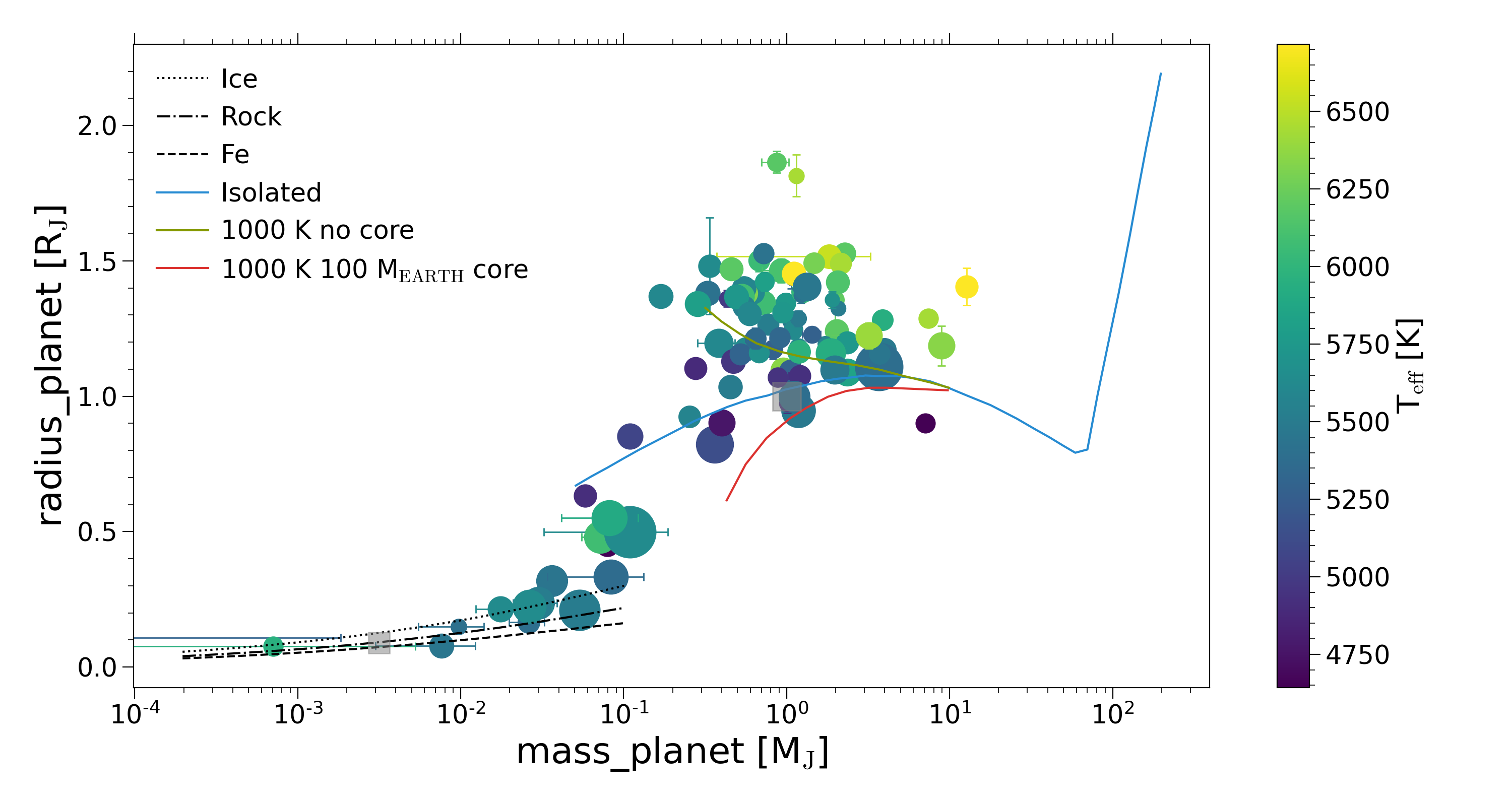}
    \caption{Planet mass and radius (in Jupiter units) of 94 planets with radial velocity and stellar parameters in the FGKM sample. Colour-coding  is as in Fig.~\ref{fig:exoplanet}.
    The symbol size corresponds to the semi-major axis.
    Some radius--mass models of Earth-like \citep{zeng2016} and Jupiter-like \citep{guillot2015} planets are also shown,
    see text for details.
}
    \label{fig:massradius_planet}
\end{figure}

We show our results in the planet radius -- planet mass diagram in Fig.~\ref{fig:massradius_planet}.
We also show some models corresponding to model mass--radius relationships for different Earth-like planet compositions from  \citet{zeng2016} and 
Jupiter-like planet compositions from \cite{guillot2015}.
The black lines represent Earth-like planet mass--radius relations assuming an ice-like (dashed), rocky Earth-like (dashed-dotted) and iron (dashed) composition. 
The coloured lines represent models of an isolated planet of solar composition at 5 Gyr (like for Jupiter, blue),  a heavily irradiated planet with 
an equilibrium temperature of 1000 K with no core (red) and one with a 100 M$_{\rm Earth}$ central core (green). 

The precision on our results (the error bars are shown although they are not always visible) does allow one to distinguish between different bulk compositions of these planets provided we have full control of the potential systematic errors.   
We provide the mass, radius and age properties of the planet and its host in Table~\ref{tab:planetproperties}.

As these figures highlight, there is a dearth of knowledge of Earth-size exoplanets in the habitable zone, along with their accurate characterisation.   
The upcoming ESA PLATO mission promises to populate the habitable zone by observing (at least) one large field over a two-to-three year period, which will allow us to detect and confirm Earth-like orbits around Earth-like planet hosts. 

\begin{table*}
    \centering
    \caption{Mass, radius, and age of known exoplanets and their host stars in the FGKM sample.  The full table is made available online as an electronic table.}
    \begin{tabular}{llllllllllll}
    \hline\hline
       & \multicolumn{3}{l}{planet properties} & 
       \multicolumn{3}{l}{host properties} \\
        \gaia\ DR3 source\_id & planet & radius & mass  & radius & mass & age \\
        & & [R$_{\jupiter}$] &  [M$_{\jupiter}$] & [R$_{\odot}$] & [M$_{\odot}$]& [Ga]\\
         \hline
1424011082893734272 & WASP-92 b &  1.423 $\pm$  0.034 &  0.731 $\pm$  0.069&  1.306 $\pm$  0.027 &  1.042 $\pm$  0.045&  7.74 $\pm$  1.13\\  
2101243789577188736 & Kepler-103 c &  0.498 $\pm$  0.017 &  0.110 $\pm$  0.083&  1.562 $\pm$  0.032 &  1.037 $\pm$  0.040&  9.30 $\pm$  0.90\\ 
4285511294172309504 & CoRoT-11 b &  1.527 $\pm$  0.037 &  2.274 $\pm$  0.336&  1.466 $\pm$  0.033 &  1.225 $\pm$  0.041&  3.49 $\pm$  0.55\\ 
    \hline\hline
    \end{tabular}
    \label{tab:planetproperties}
\end{table*}

\subsection{The colours of the Sun}\label{sec:solarcolours}

The colours of the Sun are not as well known as one would think and desire, neither observationally nor from modelling. Solar analogues offer the possibility to validate, and if necessary calibrate, our understanding of the solar flux as a function of wavelength \citep{2006MNRAS.367..449H,2018MNRAS.479L.102C}. They have also been used to estimate the solar bolometric correction in \gaia's photometric system. Below we make a new attempt to determine precise and accurate solar colours.

We use the sample of solar-analogue candidates selected from \gspspec\ from Sect.~\ref{sec:solaranalogues} in order to estimate the colours of the Sun. As we demonstrated in Sect.~\ref{ssec:solaranalogs-with-extinction}, these stars have reliable extinction estimates from \gspphot. Consequently, the BP/RP spectra with very low extinction (according to \gspphot) can be used to indirectly estimate the intrinsic continuum shape of the Sun. Among \gspspec's solar-analogue candidates, there are 682 with \gspphot's $\azero<0.001$~mag. Given these, we obtain an absolute magnitude of
\begin{equation}
M_{G,\odot} = (4.614  \pm  0.179) \,\textrm{mag}
\end{equation}
For this, we adopt the inverse parallax as a distance estimator because our candidate selection requires very high parallax quality ($\frac{\varpi}{\sigma_\varpi}>20$).  For comparison, a value of $M_{G,\odot}=4.66$ is adopted for \flame \citep[][Sect.~4.3 therein]{DR3-DPACP-157}. Given the 682 candidates with $\azero<0.001$~mag, we also obtain mean colours and standard deviations of
\begin{eqnarray}
(\gbp-\grp)_\odot = (0.818  \pm  0.029)\,\textrm{mag} \\
(\gbp-G)_\odot = (0.324  \pm  0.016)\,\textrm{mag} \\
(G-\grp)_\odot = (0.494  \pm  0.020)\,\textrm{mag} \\
(G-J)_\odot = (0.969  \pm  0.578)\,\textrm{mag} \\
(G-H)_\odot = (1.292  \pm  0.401)\,\textrm{mag} \\
(G-K_s)_\odot = (1.371  \pm  0.351)\,\textrm{mag} \\
(G-W_1)_\odot = (1.449  \pm  0.066)\,\textrm{mag} \\
(G-W_2)_\odot = (1.405  \pm  0.065)\,\textrm{mag}
\end{eqnarray}
where we restrict the AllWISE comparison to cases with $W_1>8$~mag in order to avoid saturation.  These colours are in excellent agreement with the values $(\gbp-\grp)_\odot =0.82$~mag, $(\gbp-G)_\odot =0.33$~mag and $(G-\grp)_\odot =0.49$~mag obtained in \citet{2018MNRAS.479L.102C} from \gaia\ DR2 passbands and synthetic as well as observed spectra for the Sun.  Their absolute magnitude of $M_{G,\odot}=4.67$~mag is also consistent with our estimate.  In order to do this comparison with \gaia\ DR3 passbands and also include near-infrared photometry, we take the Kurucz model \texttt{sun\_mod\_001.fits} from the CALSPEC library\footnote{\url{https://www.stsci.edu/hst/instrumentation/reference-data-for-calibration-and-tools/astronomical-catalogs/calspec}} \citep{1995AJ....110.1316B,2014PASP..126..711B,2020AJ....160...21B} and simulate its photometry using the \texttt{pyphot} package\footnote{\url{http://mfouesneau.github.io/docs/pyphot}}. We obtain synthetic colours of $(\gbp-\grp)_\odot =0.813$~mag, $(\gbp-G)_\odot =0.324$~mag and $(G-\grp)_\odot =0.490$~mag, which are again in excellent agreement with our estimated colours.
For color combinations with 2MASS, we obtain
$(G-J)_\odot =0.992$~mag,
$(G-H)_\odot =1.320$~mag, and
$(G-Ks)_\odot =1.360$~mag, which are again in excellent agreement with our candidates.
Concerning AllWISE \citep{2014yCat.2328....0C}, we obtain $(G-W_1)_\odot =1.380$~mag and $(G-W_2)_\odot =1.301$~mag. These values are slightly bluer than the values we estimate from the \gspspec\ candidates, but are still within $1\sigma$ and $1.6\sigma$, respectively.

\subsection{Ages of UCDs not seen by \gaia}\label{sec:ucd_unseen}

\begin{table*}
    \caption{\label{UCD14_table} Identification and parameters of UCDs without \gaia\ solutions in binary systems with full-solution companions.  The mass and ages indicate the median, lower and upper confidence intervals, while the parallax shows the value and uncertainty.}
    \begin{tabular}{lllllllll}
        \hline
        UCD                           & Parallax                & SpT           & Mass                 & Companion           & SpT    & Age                           & Lit. age             \\
        Code, Name                    & [mas]                   &               & [$M_{\jupiter}$]         & Source\_ID          &        & [Ga]                         & [Ga]                \\
        \hline
        A,   2MASSI J0025036+475919   & 22.8 $\pm$  0.9$^{2}$   & L4:+L4:$^{2}$ & $84.4_{83.7}^{87.4}$ & 392562179817077120  & F8     & $ 5.1_{ 4.2}^{ 5.9}$          & $ 4.4_{ 1.6}^{ 5.3}$ \\ 
        B,   2MASS J02233667+5240066  & 12.2 $\pm$  1.1$^{1}$   & L1.5$^{3}$    & $80.6_{80.3}^{80.7}$ & 452046549154458880  & F5     & $ 2.9_{ 2.5}^{ 3.4}$          & $ 2.0_{ 1.7}^{ 2.8}$ \\ 
        C,   2MASS J06462756+7935045  & 53.6 $\pm$  2.2$^{1}$   & L9$^{4}$      & $59.4_{51.9}^{64.4}$ & 1141280704422528128 & F7V    & $ 2.2_{ 1.6}^{ 2.9}$          & $ 4.3_{ 2.5}^{ 6.0}$ \\ 
        D,   HD 49197B                & 29.9 $\pm$  2.1$^{1}$   & L4$^{5}$      & $79.3_{78.9}^{79.6}$ & 952346742338146176  & F5     & $ 4.0_{ 3.1}^{ 4.9}$          & $ 2.8_{ 0.3}^{ 4.7}$ \\ 
        E,   2MASS J12173646+1427119  & 16.1 $\pm$  0.7$^{1}$   & L1$^{3}$      & $82.3_{82.3}^{84.5}$ & 3921176983720146560 & F8     & $ 4.9_{ 4.2}^{ 5.7}{\dagger}$ & $ 1.8_{ 0.5}^{ 3.8}$ \\ 
        F,   HD 118865B               & 21.7 $\pm$  0.9$^{1}$   & T5$^{6}$      & $67.5_{65.2}^{69.1}$ & 3663438298389132416 & F7V    & $ 4.6_{ 3.9}^{ 5.4}{\dagger}$ & $ 3.2_{ 2.5}^{ 3.5}$ \\ 
        G,   2MASS J14165987+5006258  & 22.1 $\pm$  0.8$^{7}$   & L5.5$^{8}$    & $75.6_{75.3}^{75.8}$ & 1508557582834745088 & G5     & $ 6.3_{ 5.3}^{ 7.3}$          & $ 5.8_{ 4.2}^{ 9.0}$ \\ 
        H,   ULAS J142320.79+011638.2 & 29.4 $\pm$  1.0$^{1}$   & T8p$^{9}$     & $34.9_{30.3}^{38.4}$ & 3654496279558010624 & G1.5V  & $ 5.6_{ 4.0}^{ 7.2}$          & $ 5.9_{ 4.6}^{11.2}$ \\ 
        I,   Gl 564 C                 & 74.7 $\pm$ 21.3$^{1}$   & L4:$^{10}$    & $77.4_{77.1}^{77.6}$ & 1265976524286377856 & F9IV-V & $ 5.3_{ 4.4}^{ 6.2}{\dagger}$ & $ 0.9_{ 0.6}^{ 4.7}$ \\ 
        J,   Gl779B 244691            & 47.1 $\pm$  6.1$^{1}$   & L4.5$^{11}$   & $71.0_{39.9}^{73.3}$ & 1821708351374312064 & G0V    & $ 2.8_{ 0.5}^{ 3.8}$          & $ 3.2_{ 1.7}^{ 5.2}$ \\ 
        K,   eps Indi C               & 275.3 $\pm$  3.0$^{12}$ & T6$^{13}$     &  $36.5_{11.0}^{50.2}$ & 6412595290592307840 & K5V    & $ 2.0_{ 0.2}^{ 4.3}$          & $ 4.6_{ 0.9}^{ 5.7}$ \\ 
        \hline                                                                                                                                                                               \\
    \end{tabular}\\
    Notes: 1: Spectrophotometric distance, 2: \cite{2007AJ....133..439C}, 3: \cite{2014APJ...792..119D}, 4: \cite{2011APJ...739...81L}, 5: \cite{2004APJ...617.1330M}, 6: \cite{2013MNRAS.433..457B}, 7: \cite{2012APJ...752...56F}, 8: \cite{2006AJ....131.2722C}, 9: \cite{2012MNRAS.422.1922P}, 10: \cite{2002APJ...567L..59G}, 
    11: \cite{2002APJ...568L.107L}, 
    12: \cite{2016AJ....152...24W}, 
    13: \cite{2006APJ...639.1095B}. The $\dagger$ indicates \flame\ ages derived using the \gspphot-\teff, while other ages are derived using the \gspspec-\teff.
\end{table*}

Another application of the \gaia\ astrophysical parameters is to
constrain the characteristics of faint UCDs that are beyond the
mission magnitude limit but in binary systems with brighter objects
that are observed by \gaia. Once we have identified a multiple system
we assume that the UCD has the same chemical composition, age,
distance and, after allowing for orbital motion, proper motions. In
addition, if the movement due to the orbital motion is detected by \gaia\
this will provide a constraint on the mass of the various
components. Brown dwarfs evolve and cool over time and
their observational properties are degenerate with age, mass and
metallicity; binary systems are therefore benchmarks for understanding these
processes. 
\gaia\ will provide a large homogeneous multi-parametric
sample with intersecting constraints that will tie down the UCD
regime.  
For this illustrative discussion we concentrate on the age parameter\footnote{The  \href{http://basti-iac.oa-abruzzo.inaf.it}{\basti} models \citep{2018ApJ...856..125H} were used to derive the age and they span from the ZAMS to the tip of the red giant branch for stellar masses between 0.5 and 10\Msun.}.  We note however, that more precise ages can be obtained by combining \gaia\ with other observational data such as asteroseismology.

To identify a potential list of objects with a high probability to be in a binary system we used the
positional and kinematical criteria given by Eq. 2 in
\cite{2019MNRAS.485.4423S} and the list of known UCDs from that
study. When the faint UCD did not have a measured parallax we used its
spectro-photometric distance.
 We found {8} UCDs without \gaia\ DR3 5 parameter solutions that are in binary systems in the  FGKM  sample, while also 
 in the regime of reliable ages (see \citealt{DR3-DPACP-160}).
 We added a further three interesting targets with reliable ages here because they were rejected from the FGKM sample for failing on only one of the criteria: A \linktogsparam{gaia_source}{ipd_frac_multi_peak} = 22; and B and F \linktoapparam{astrophysical_parameters}{classprob_dsc_combmod_binary} $> 0.99$.

 These 11 UCDs are listed
in Table\,\ref{UCD14_table} with name, adopted parallax, spectral type and mass along with the companion \gaia\ 
\texttt{source\_id},  age and the median published ages
with 16\% and 84\% percentiles. 

\begin{figure}[!htb]
\center{\includegraphics[width=0.45\textwidth]{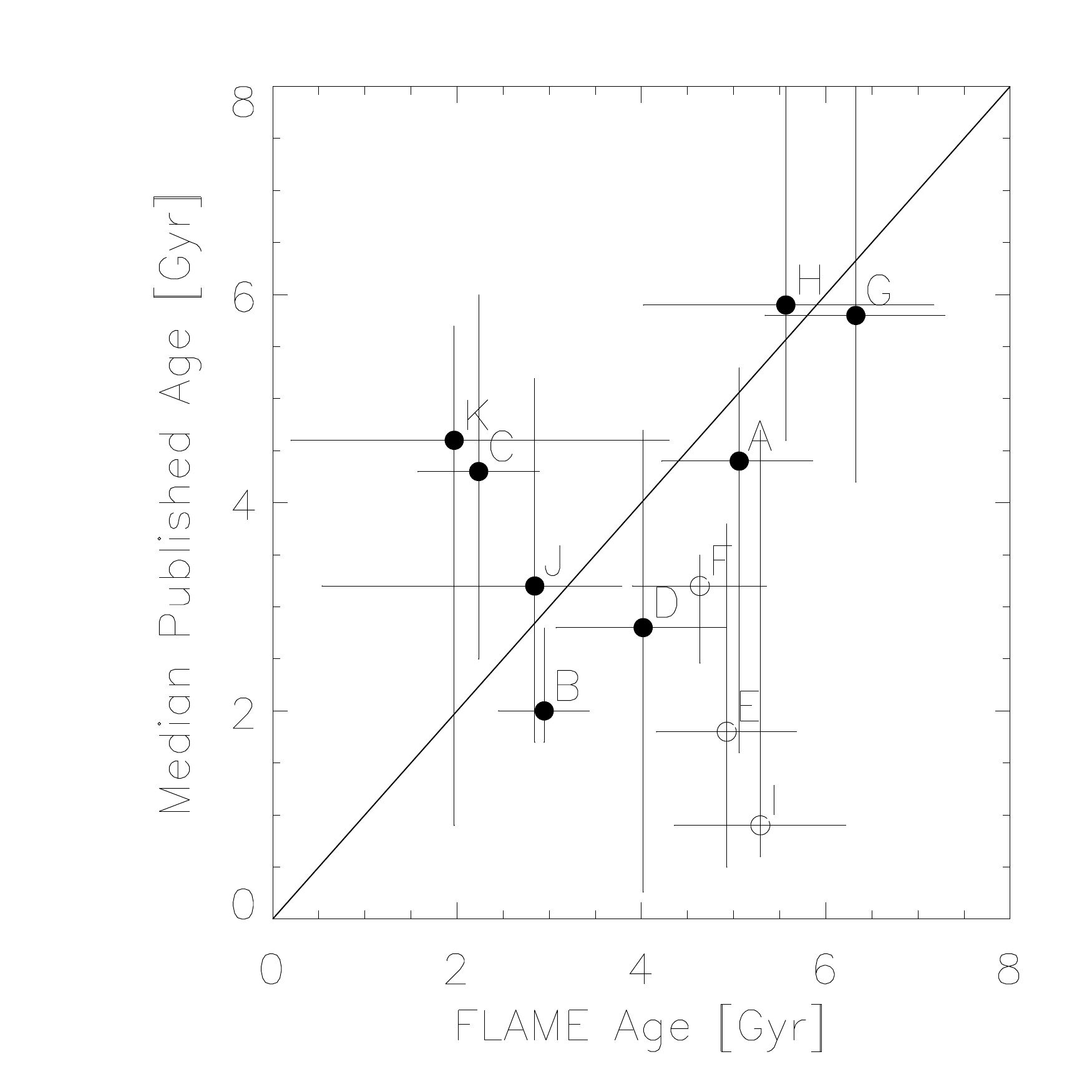}}  
\caption{\label{fig:UCD14_ages}
Literature ages versus \flame\ ages for companions of UCDs not seen by \gaia. Filled circles correspond to \linktoapparam{astrophysical_parameters_spec}{age_flame_spec} (i.e. using the \teff\ from \gspspec) and open circles are \linktoapparam{astrophysical_parameters}{age_flame}. The error bars
represent the 16\% and 84\% percentiles.}
\end{figure}

The number of literature age
estimates vary from 6 to 46 for each target and are from varied
sources: model comparisons
\citep{2009A&A...501..941H,2011A&A...530A.138C}, chromospheric
activity \citep{2013A&A...551L...8P,2004APJ...617.1330M}, or Galactic
kinematics \citep{2012AstL...38..771G}. The published age percentiles
often indicate uncertainties of a factor of 2 or a large portion of
the age of the Galaxy indicating the current difficulty in determining ages for stars. In Figure\,\ref{fig:UCD14_ages} we show the \gaia\ vs
the median published values from Table\,\ref{UCD14_table}.  
When available, we used the values based on the \gspspec\ \teff:
\linktoapparam{astrophysical_parameters_supp}{age_flame_spec} these are denoted by the filled circles.  
The open circles are  \linktoapparam{astrophysical_parameters}{age_flame} which are based on \gspphot\ \teff.  
For most of the stars we find general agreement with the literature, with the worst agreements for systems E, I and K.
For E, the \teff\ from both \gspphot\ and \gspspec\ agree to within 25 K and we would therefore trust its age if the star is within the regime of models that were used.
For I and K we find significant disagreements between the \gspphot\ and \gspspec\ \teff, and this could indicate a possible issue with {\tt age\_flame}.
{We discuss each of the systems individually in the next section. }

\begin{figure}[!htb]
\center{\includegraphics[width=0.5\textwidth]{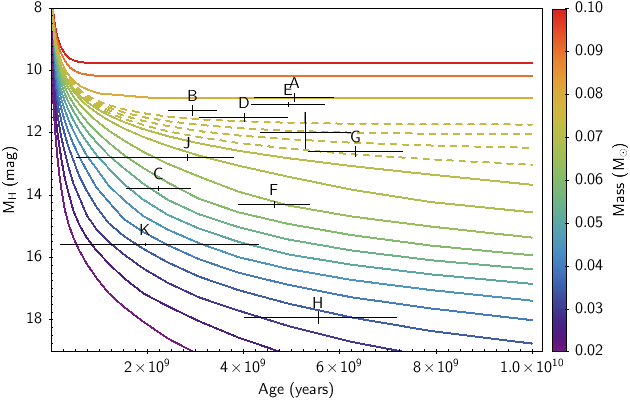}}  
\caption{\label{fig:UCD13_masses}
  Evolutionary tracks and UCD locations in the H-band absolute magnitude versus age diagram, adopting the companion age. The tracks {are colour-coded by mass}.
  The dashed lines indicate the stellar to substellar transition zone (from 0.072 to 0.075 M$_\odot$).}
\end{figure}

The interpolated masses of the UCDs are estimated  from a comparison to the illustrated tracks in Figure\,\ref{fig:UCD13_masses}  taken from \cite{2015A&A...577A..42B} for stars and \cite{2020A&A...637A..38P}  for brown dwarfs assuming the age of the companion star from this work, and these are reported in Table~\ref{UCD14_table}.

\subsubsection{Notes on individual systems}
\object{2MASSI J0025036+475919} (A) is an L4+L4 binary in a multiple system
with the spectroscopic binary HD 2057 \citep[][ and references therein]{2006AJ....132..891R}  and  another component 11\arcsec\ from the primary (\gaia\ EDR3 392562179817297536).  
  Lithium absorption has been detected in the combined spectra of the secondary indicating it has an age less than 1.0\,Ga \citep{2007AJ....133..439C, 2010AJ....139..176F, 2015APJ...810..158F} which is much lower than the primary age indicated here.
This is one of the widest binary systems
($\sim$10\,000\,AU) with an ultracool component but in the range of other
systems of similar total mass. The difference in age estimates of the
primary and secondary is not easily reconcilable. 
One possible solution is that it is not a binary system; the agreement of high proper motions is a strong constraint, however, the spectroscopic distance is
very uncertain as the binary nature of
the secondary requires an assumption of the component flux contributions.
Another possibility is that the primary age estimate is high due to its binary nature.

\object{2MASS J02233667+5240066} (B) was first noted to be in a common
proper motion system with HIP 11161 in \cite{2014APJ...792..119D}.
The primary has been shown to have acceleration terms
\citep{2022A&A...657A...7K,2021APJS..254...42B} but the separation with
the UCD is large (41\arcsec) and the primary has now been resolved by \gaia\ into two components and it is listed as a
spectroscopic binary in the non-single stars orbital solution results.
It also has a very high  \linktoapparam{astrophysical_parameters}{classprob_dsc_combmod_binary} (>0.99).
The observed acceleration is therefore due to the primary
binarity and not the UCD.  Using \linktoapparam{astrophysical_parameters}{age_flame_spec} we find a mass of  $80\,M_{\jupiter}$ which defines the end of the stellar main sequence. 

\object{2MASS J06462756+7935045} (C) was indicated as being in a binary system with HD\,46588 based on a high common proper
motion \citep{2011APJ...739...81L}. 
It is an L9 brown dwarf, one of the few known at the L/T transition in wide binary systems.  These allow constraints on their
astrophysical properties. 
The \texttt{age\_flame\_spec} is lower by 1$\sigma$ 
than the primary literature age. Since this is one of the few L9s where
an independent age is known it is important to clarify this
discrepancy.  Assuming the literature age and distance from the primary, \citeauthor{2011APJ...739...81L} find a $T_{\rm eff}=1360_{1280}^{1410}\,K$, which is an important constraint for the temperature at the the L/T boundary. If we assume  the  lower \texttt{age\_flame\_spec} this will increase the temperature estimate at this boundary.

\object{HD 49197 B} (D) has been studied extensively since its first
discovery by
\cite{2004APJ...617.1330M} using high resolution observing techniques.  It is at a separation of 0.95\arcsec\ {from the primary}. There are ongoing adaptive optics projects
to try to determine a binary solution
\citep{2020AJ....159...63B,2014AJ....147...86T}.  With a 
magnitude difference of greater than 10  \gaia\ will not be able to resolve the system.
If we adopt the low end of the literature age range, HD 49197 B is a brown dwarf,
if we adopt the high end - for example that indicated by \texttt{age\_flame\_spec} - the object becomes a star.
 Since there is also a possibility of finding the mass of this
companion either through high resolution imaging or the detection of
acceleration terms in the \gaia\ primary solution \citep[proper motion
  anomalies between the Hipparcos and \gaia\ results have already been
  detected in][]{2022A&A...657A...7K}, knowing its age will be crucial for
constraining the stellar-substellar boundary.

\object{2MASS J12173646+1427119} (E) was first discovered in the
Pan-STARRS survey as a companion to HIP 59933 at 40\arcsec. The secondary is
detected by \gaia\ (EDR3 3921177219942653696) but with only a
two-parameter solution. The primary, EDR3 3921176983720146560, has a non-single
star solution which indicates a companion of $0.09\msun$ with a period of
1\,yr and a corresponding separation of 1\,AU; given the small separation there must be a third component in the system. Any age above 0.5\,Ga would indicate that this is a stellar object 
but very close to the stellar-substellar boundary as indicated by our 82\,$M_{\jupiter}$. 

\object{HD 118865B} (F) is a T5 in a system with an F4 spectral type first noted in
\cite{2013MNRAS.433..457B} where they find an age range of  1.5--4.9\,Ga and mass of 45--60 $M_{\jupiter}$.  We find a  primary age that is at the top end of their range and hence a slightly larger mass. If confirmed it will provide a
high mass for this T5 compared to other
similar  type brown dwarfs. 

\object{2MASS J14165987+5006258} (G) is noted as a binary system in
\cite{2010AJ....139..176F} with a very large separation of
$\sim$26\,000\,AU. The primary \texttt{age\_flame\_spec} estimate is consistent with published
values and also with blue near-IR colors observed in
\cite{2010AJ....139..176F} where they also re-evaluate its spectral
type from L5.5 to L4. The estimated mass indicates that this object is of stellar and not
sub-stellar type and further characterisation will contribute to our understanding of very old borderline stellar objects. 

\object{ULAS J142320.79+011638.2} (H) is the coolest object in this
sample in a system with an early-G dwarf, HIP 70319. There are a
significant number of age estimates from very young to very old and
the \texttt{age\_flame\_spec} is in agreement with the median. This age is consistent
with a low metallicity primary and also with a broader Y-band peak and
more depressed K-band peak than other T8s
\citep{2021APJS..253....7K}. This is an important benchmark for metal
poor T dwarfs.

\object{Gl 564 B/C} (I) is an L4+L4 binary in a triple system with Gl 564, a G2 V star. 
The
majority of the published age ranges are very young because Gl 564 is
chromospherically active with a high lithium abundance and fast
rotation \citep{2002APJ...567L.133P}. The space motion also puts the
object in the Ursa Major moving group from the Banyan $\Sigma$ tool   \citep{2018APJ...856...23G}, which has an age of around
500\,Myr \citep{2003AJ....125.1980K}.  This is in contrast to the high \texttt{age\_flame} which is difficult to reconcile given the high
lithium abundance and space motion.  A possible explanation for this discrepancy is in the limitations of the models that were used, for example, they do not include rotation.  If the
system is in the first 0.5\,Ga they will be contracting brown
dwarfs. The  orbital period of the UCD binary system is
around 10\,yr \citep{2002APJ...567L.133P} and we will therefore soon  have dynamical masses with \gaia. 
These objects will
provide a well-constrained calibration point for the theoretical
models describing low-mass, ultracool objects.

\object{Gl779B} (J) is an L4.5 UCD at 0.7\arcsec\ from GJ 779, a G0 star. High
levels of chromospheric activity suggest a young age, lithium
abundance indicates a slightly older than the Hyades age but kinematics
indicate an old disk star. The \texttt{age\_flame\_spec} is consistent with the
published estimates. The
orbit is such that it should be visible in the future \gaia\ observations which
will lead to a dynamical mass estimate  \citep{2014APJ...781...29C}. A comparison of the
accelerations found from comparisons of Hipparcos and \gaia\ DR2
results indicate a mass of around $0.07\msun$
\citep{2019AJ....158..140B} and this is therefore on the
stellar-substellar boundary currently defining the end of the main
sequence, and in agreement with our estimated mass.

\object{Eps Ind C} (K) is the second closest brown dwarf binary T1+T6 system in a triple system with the K5V star eps~Ind.  One of the brown dwarfs has a \gaia\ solution (\gaia\ EDR3 6412596012146801152)  which we assume is the T1. Later releases should provide a dynamical solution for the component masses. There is a significant history of publications for both the primary and the secondary system.
With a period of around 11\,yr and an observed separation that
varies from 0.6 to 2\arcsec, it is a defining system for parameters of
early T dwarfs. The \linktoapparam{astrophysical_parameters_supp}{age_flame_spec} is at the low end of the published
age range for the primary, {and the masses of the secondaries from \cite{2018APJ...865...28D} also imply an inconsistency with such a young age}.  {A dynamical} mass determination from the \gaia\ observations should resolve this inconsistency.

We have seen that the results of \gaia\ can be brought to bear on our understanding of objects fainter than its magnitude limit. Indeed there will probably be less than 1000 brown dwarfs brighter than the \gaia\ magnitude limit  \citep{2019MNRAS.485.4423S} while we expect there to be tens of thousands in binary systems or detected from astrometric and radial velocity perturbations. Therefore, the contribution of \gaia\ to brown dwarf studies will be predominantly due to indirectly detected objects rather than direct detections.

\section{Conclusion}\label{sec:conclusions}

In this work we defined homogeneous samples of high quality astrophysical parameters by exploiting many \gaia data products that appear in \gdr{3}, while focusing on the sources and data products in the \aptable\ and the \apsupptable\ tables which were  produced by the \apsis\ software \citep{DR3-DPACP-157, DR3-DPACP-158, DR3-DPACP-160}.  
We considered different regimes of stars all across the HR diagram.  In the first part of this work we considered large samples of young massive disk OBA stars (Sect.~\ref{sec:oba}), FGKM spectral type stars (Sect.~\ref{sec:fgkm}), and faint ultra-cool dwarfs (UCDs, Sect.~\ref{sec:ucd}).
Then we focussed on smaller samples of specific object types;  carbon stars (Sect.~\ref{sec:carbon}), solar analogues (Sect.~\ref{sec:solaranalogues}), and the Spectro Photometric Standard stars (SPSS, \citealt{2021MNRAS.503.3660P}, Sect.~\ref{sec:spss}).  Concerning the latter, this paper provides the first homogeneous determination of the SPSS dataset to date.  
We validated each of the samples using the \gaia\ data itself and external catalogues, and our results are published in six tables that will appear alongside \gdr{3}, see Sect.~\ref{sec:summarytables} and Table~\ref{tab:summarytable}.  

In Sect.~\ref{sec:exploitation}, we demonstrated some use cases of these samples of stars.  We used a subset of the OBA sample to illustrate its usefulness to analyse the Milky Way rotation curve (Sect.~\ref{sec:rotationcurve}).
We then used the properties of the FGKM stars to analyse known exoplanet systems including the determination of planet radii and masses (Sect.~\ref{sec:exoplanets}).   We then predicted the colours of the Sun in various passbands using the solar analogue sample (Sect.~\ref{sec:solarcolours}).
Finally, we analysed the ages of some unseen UCD-companions to the FGKM stars (Sect.~\ref{sec:ucd_unseen}).

The aim of this work was to highlight the science that can be done with \gdr{3}.  We focused on specific types of stars using strict quality criteria on many of the data products, which sometimes included some ad hoc filtering criteria tuned with a particular science case in mind.  We emphasize that our strict personal selections may not be applicable to a user's specific science case, and the user should acknowledge this before exploiting these samples.   We fully encourage all users to exploit all of the astrophysical parameters in \gdr{3} independent of our specific selection criteria highlighted in this work.  Indeed there are up to 470 million stars with stellar parameters derived using the mean BP and RP spectra, up to 6 million stellar parameters and abundances derived from the mean RVS spectra, along with up to 130 million masses and ages, and many other new stellar products that were not the focus of this work, such as DIB estimates, activity index of active stars, and H$\alpha$ emission.  As illustrated in this work, many science cases can be explored with these data.

\begin{acknowledgements}
    We thank the referee for their constructive comments on the manuscript.
    This work has made use of data from the European Space Agency (ESA)
    mission \gaia\ (\url{https://www.cosmos.esa.int/gaia}), processed by the Gaia Data Processing and Analysis Consortium (DPAC, \url{https://www.cosmos.esa.int/web/gaia/dpac/consortium}). Funding for the DPAC has been provided by national institutions, in particular, the institutions participating in the Gaia Multilateral Agreement.  The full list of funding agencies and grants is listed after the references.\\
    This research has used NASA's Astrophysics Data System, and the VizieR catalogue access tool (CDS, Strasbourg, France).\\
    %
    The data processing and analysis made use of
        matplotlib \citep{Hunter:2007},
        NumPy \citep{harris2020array},
        the IPython package \citep{PER-GRA:2007},
        Vaex \citep{Breddels2018},
        TOPCAT \citep{Taylor2005},
        pyphot (\url{http://github.com/mfouesneau/pyphot}),
        R \citep{RManual},
        Astropy \citep{astropy:2013, astropy:2018},
        CmdStanPy (\url{https://github.com/stan-dev/cmdstanpy}),
        and ArviZ \citep{arviz_2019}.\\
    In case of errors or omissions, please contact the \href{https://www.cosmos.esa.int/web/gaia/gaia-helpdesk}{\gaia\ Helpdesk}.
The full list of acknowledgements can also be found \href{https://gea.esac.esa.int/archive/documentation/GDR3/Miscellaneous/sec_acknowl/}{in the official online documentation for \gdr{3}.}
\end{acknowledgements}
\bibliographystyle{aa}
\bibliography{refs.bib}
\appendix
\section*{\gaia funding institutions}
This work presents results from the European Space Agency (ESA) space mission \gaia. \gaia\ data are processed by the \gaia\ Data Processing and Analysis Consortium (DPAC). Funding for the DPAC is provided by national institutions, in particular the institutions participating in the \gaia\ MultiLateral Agreement (MLA). The \gaia\ mission website is \url{https://www.cosmos.esa.int/gaia}. The \gaia\ archive website is \url{https://archives.esac.esa.int/gaia}.

The \gaia\ mission and data processing have financially been supported by, in alphabetical order by country:
\\ -- the Algerian Centre de Recherche en Astronomie, Astrophysique et G\'{e}ophysique of Bouzareah Observatory;
\\ -- the Austrian Fonds zur F\"{o}rderung der wissenschaftlichen Forschung (FWF) Hertha Firnberg Programme through grants T359, P20046, and P23737;
\\ -- the BELgian federal Science Policy Office (BELSPO) through various PROgramme de D\'{e}veloppement d'Exp\'{e}riences scientifiques (PRODEX)
      grants, the Research Foundation Flanders (Fonds Wetenschappelijk Onderzoek) through grant VS.091.16N,
      the Fonds de la Recherche Scientifique (FNRS), and the Research Council of Katholieke Universiteit (KU) Leuven through
      grant C16/18/005 (Pushing AsteRoseismology to the next level with TESS, GaiA, and the Sloan DIgital Sky SurvEy -- PARADISE);  
\\ -- the Brazil-France exchange programmes Funda\c{c}\~{a}o de Amparo \`{a} Pesquisa do Estado de S\~{a}o Paulo (FAPESP) and Coordena\c{c}\~{a}o de Aperfeicoamento de Pessoal de N\'{\i}vel Superior (CAPES) - Comit\'{e} Fran\c{c}ais d'Evaluation de la Coop\'{e}ration Universitaire et Scientifique avec le Br\'{e}sil (COFECUB);
\\ -- the Chilean Agencia Nacional de Investigaci\'{o}n y Desarrollo (ANID) through Fondo Nacional de Desarrollo Cient\'{\i}fico y Tecnol\'{o}gico (FONDECYT) Regular Project 1210992 (L.~Chemin);
\\ -- the National Natural Science Foundation of China (NSFC) through grants 11573054, 11703065, and 12173069, the China Scholarship Council through grant 201806040200, and the Natural Science Foundation of Shanghai through grant 21ZR1474100;  
\\ -- the Tenure Track Pilot Programme of the Croatian Science Foundation and the \'{E}cole Polytechnique F\'{e}d\'{e}rale de Lausanne and the project TTP-2018-07-1171 `Mining the Variable Sky', with the funds of the Croatian-Swiss Research Programme;
\\ -- the Czech-Republic Ministry of Education, Youth, and Sports through grant LG 15010 and INTER-EXCELLENCE grant LTAUSA18093, and the Czech Space Office through ESA PECS contract 98058;
\\ -- the Danish Ministry of Science;
\\ -- the Estonian Ministry of Education and Research through grant IUT40-1;
\\ -- the European Commission’s Sixth Framework Programme through the European Leadership in Space Astrometry (\href{https://www.cosmos.esa.int/web/gaia/elsa-rtn-programme}{ELSA}) Marie Curie Research Training Network (MRTN-CT-2006-033481), through Marie Curie project PIOF-GA-2009-255267 (Space AsteroSeismology \& RR Lyrae stars, SAS-RRL), and through a Marie Curie Transfer-of-Knowledge (ToK) fellowship (MTKD-CT-2004-014188); the European Commission's Seventh Framework Programme through grant FP7-606740 (FP7-SPACE-2013-1) for the \gaia\ European Network for Improved data User Services (\href{https://gaia.ub.edu/twiki/do/view/GENIUS/}{GENIUS}) and through grant 264895 for the \gaia\ Research for European Astronomy Training (\href{https://www.cosmos.esa.int/web/gaia/great-programme}{GREAT-ITN}) network;
\\ -- the European Cooperation in Science and Technology (COST) through COST Action CA18104 `Revealing the Milky Way with \gaia (MW-Gaia)';
\\ -- the European Research Council (ERC) through grants 320360, 647208, and 834148 and through the European Union’s Horizon 2020 research and innovation and excellent science programmes through Marie Sk{\l}odowska-Curie grant 745617 (Our Galaxy at full HD -- Gal-HD) and 895174 (The build-up and fate of self-gravitating systems in the Universe) as well as grants 687378 (Small Bodies: Near and Far), 682115 (Using the Magellanic Clouds to Understand the Interaction of Galaxies), 695099 (A sub-percent distance scale from binaries and Cepheids -- CepBin), 716155 (Structured ACCREtion Disks -- SACCRED), 951549 (Sub-percent calibration of the extragalactic distance scale in the era of big surveys -- UniverScale), and 101004214 (Innovative Scientific Data Exploration and Exploitation Applications for Space Sciences -- EXPLORE);
\\ -- the European Science Foundation (ESF), in the framework of the \gaia\ Research for European Astronomy Training Research Network Programme (\href{https://www.cosmos.esa.int/web/gaia/great-programme}{GREAT-ESF});
\\ -- the European Space Agency (ESA) in the framework of the \gaia\ project, through the Plan for European Cooperating States (PECS) programme through contracts C98090 and 4000106398/12/NL/KML for Hungary, through contract 4000115263/15/NL/IB for Germany, and through PROgramme de D\'{e}veloppement d'Exp\'{e}riences scientifiques (PRODEX) grant 4000127986 for Slovenia;  
\\ -- the Academy of Finland through grants 299543, 307157, 325805, 328654, 336546, and 345115 and the Magnus Ehrnrooth Foundation;
\\ -- the French Centre National d’\'{E}tudes Spatiales (CNES), the Agence Nationale de la Recherche (ANR) through grant ANR-10-IDEX-0001-02 for the `Investissements d'avenir' programme, through grant ANR-15-CE31-0007 for project `Modelling the Milky Way in the \gaia era’ (MOD4Gaia), through grant ANR-14-CE33-0014-01 for project `The Milky Way disc formation in the \gaia era’ (ARCHEOGAL), through grant ANR-15-CE31-0012-01 for project `Unlocking the potential of Cepheids as primary distance calibrators’ (UnlockCepheids), through grant ANR-19-CE31-0017 for project `Secular evolution of galxies' (SEGAL), and through grant ANR-18-CE31-0006 for project `Galactic Dark Matter' (GaDaMa), the Centre National de la Recherche Scientifique (CNRS) and its SNO \gaia of the Institut des Sciences de l’Univers (INSU), its Programmes Nationaux: Cosmologie et Galaxies (PNCG), Gravitation R\'{e}f\'{e}rences Astronomie M\'{e}trologie (PNGRAM), Plan\'{e}tologie (PNP), Physique et Chimie du Milieu Interstellaire (PCMI), and Physique Stellaire (PNPS), the `Action F\'{e}d\'{e}ratrice \gaia' of the Observatoire de Paris, the R\'{e}gion de Franche-Comt\'{e}, the Institut National Polytechnique (INP) and the Institut National de Physique nucl\'{e}aire et de Physique des Particules (IN2P3) co-funded by CNES;
\\ -- the German Aerospace Agency (Deutsches Zentrum f\"{u}r Luft- und Raumfahrt e.V., DLR) through grants 50QG0501, 50QG0601, 50QG0602, 50QG0701, 50QG0901, 50QG1001, 50QG1101, 50\-QG1401, 50QG1402, 50QG1403, 50QG1404, 50QG1904, 50QG2101, 50QG2102, and 50QG2202, and the Centre for Information Services and High Performance Computing (ZIH) at the Technische Universit\"{a}t Dresden for generous allocations of computer time;
\\ -- the Hungarian Academy of Sciences through the Lend\"{u}let Programme grants LP2014-17 and LP2018-7 and the Hungarian National Research, Development, and Innovation Office (NKFIH) through grant KKP-137523 (`SeismoLab');
\\ -- the Science Foundation Ireland (SFI) through a Royal Society - SFI University Research Fellowship (M.~Fraser);
\\ -- the Israel Ministry of Science and Technology through grant 3-18143 and the Tel Aviv University Center for Artificial Intelligence and Data Science (TAD) through a grant;
\\ -- the Agenzia Spaziale Italiana (ASI) through contracts I/037/08/0, I/058/10/0, 2014-025-R.0, 2014-025-R.1.2015, and 2018-24-HH.0 to the Italian Istituto Nazionale di Astrofisica (INAF), contract 2014-049-R.0/1/2 to INAF for the Space Science Data Centre (SSDC, formerly known as the ASI Science Data Center, ASDC), contracts I/008/10/0, 2013/030/I.0, 2013-030-I.0.1-2015, and 2016-17-I.0 to the Aerospace Logistics Technology Engineering Company (ALTEC S.p.A.), INAF, and the Italian Ministry of Education, University, and Research (Ministero dell'Istruzione, dell'Universit\`{a} e della Ricerca) through the Premiale project `MIning The Cosmos Big Data and Innovative Italian Technology for Frontier Astrophysics and Cosmology' (MITiC);
\\ -- the Netherlands Organisation for Scientific Research (NWO) through grant NWO-M-614.061.414, through a VICI grant (A.~Helmi), and through a Spinoza prize (A.~Helmi), and the Netherlands Research School for Astronomy (NOVA);
\\ -- the Polish National Science Centre through HARMONIA grant 2018/30/M/ST9/00311 and DAINA grant 2017/27/L/ST9/03221 and the Ministry of Science and Higher Education (MNiSW) through grant DIR/WK/2018/12;
\\ -- the Portuguese Funda\c{c}\~{a}o para a Ci\^{e}ncia e a Tecnologia (FCT) through national funds, grants SFRH/\-BD/128840/2017 and PTDC/FIS-AST/30389/2017, and work contract DL 57/2016/CP1364/CT0006, the Fundo Europeu de Desenvolvimento Regional (FEDER) through grant POCI-01-0145-FEDER-030389 and its Programa Operacional Competitividade e Internacionaliza\c{c}\~{a}o (COMPETE2020) through grants UIDB/04434/2020 and UIDP/04434/2020, and the Strategic Programme UIDB/\-00099/2020 for the Centro de Astrof\'{\i}sica e Gravita\c{c}\~{a}o (CENTRA);  
\\ -- the Slovenian Research Agency through grant P1-0188;
\\ -- the Spanish Ministry of Economy (MINECO/FEDER, UE), the Spanish Ministry of Science and Innovation (MICIN), the Spanish Ministry of Education, Culture, and Sports, and the Spanish Government through grants BES-2016-078499, BES-2017-083126, BES-C-2017-0085, ESP2016-80079-C2-1-R, ESP2016-80079-C2-2-R, FPU16/03827, PDC2021-121059-C22, RTI2018-095076-B-C22, and TIN2015-65316-P (`Computaci\'{o}n de Altas Prestaciones VII'), the Juan de la Cierva Incorporaci\'{o}n Programme (FJCI-2015-2671 and IJC2019-04862-I for F.~Anders), the Severo Ochoa Centre of Excellence Programme (SEV2015-0493), and MICIN/AEI/10.13039/501100011033 (and the European Union through European Regional Development Fund `A way of making Europe') through grant RTI2018-095076-B-C21, the Institute of Cosmos Sciences University of Barcelona (ICCUB, Unidad de Excelencia `Mar\'{\i}a de Maeztu’) through grant CEX2019-000918-M, the University of Barcelona's official doctoral programme for the development of an R+D+i project through an Ajuts de Personal Investigador en Formaci\'{o} (APIF) grant, the Spanish Virtual Observatory through project AyA2017-84089, the Galician Regional Government, Xunta de Galicia, through grants ED431B-2021/36, ED481A-2019/155, and ED481A-2021/296, the Centro de Investigaci\'{o}n en Tecnolog\'{\i}as de la Informaci\'{o}n y las Comunicaciones (CITIC), funded by the Xunta de Galicia and the European Union (European Regional Development Fund -- Galicia 2014-2020 Programme), through grant ED431G-2019/01, the Red Espa\~{n}ola de Supercomputaci\'{o}n (RES) computer resources at MareNostrum, the Barcelona Supercomputing Centre - Centro Nacional de Supercomputaci\'{o}n (BSC-CNS) through activities AECT-2017-2-0002, AECT-2017-3-0006, AECT-2018-1-0017, AECT-2018-2-0013, AECT-2018-3-0011, AECT-2019-1-0010, AECT-2019-2-0014, AECT-2019-3-0003, AECT-2020-1-0004, and DATA-2020-1-0010, the Departament d'Innovaci\'{o}, Universitats i Empresa de la Generalitat de Catalunya through grant 2014-SGR-1051 for project `Models de Programaci\'{o} i Entorns d'Execuci\'{o} Parallels' (MPEXPAR), and Ramon y Cajal Fellowship RYC2018-025968-I funded by MICIN/AEI/10.13039/501100011033 and the European Science Foundation (`Investing in your future');
\\ -- the Swedish National Space Agency (SNSA/Rymdstyrelsen);
\\ -- the Swiss State Secretariat for Education, Research, and Innovation through the Swiss Activit\'{e}s Nationales Compl\'{e}mentaires and the Swiss National Science Foundation through an Eccellenza Professorial Fellowship (award PCEFP2\_194638 for R.~Anderson);
\\ -- the United Kingdom Particle Physics and Astronomy Research Council (PPARC), the United Kingdom Science and Technology Facilities Council (STFC), and the United Kingdom Space Agency (UKSA) through the following grants to the University of Bristol, the University of Cambridge, the University of Edinburgh, the University of Leicester, the Mullard Space Sciences Laboratory of University College London, and the United Kingdom Rutherford Appleton Laboratory (RAL): PP/D006511/1, PP/D006546/1, PP/D006570/1, ST/I000852/1, ST/J005045/1, ST/K00056X/1, ST/\-K000209/1, ST/K000756/1, ST/L006561/1, ST/N000595/1, ST/N000641/1, ST/N000978/1, ST/\-N001117/1, ST/S000089/1, ST/S000976/1, ST/S000984/1, ST/S001123/1, ST/S001948/1, ST/\-S001980/1, ST/S002103/1, ST/V000969/1, ST/W002469/1, ST/W002493/1, ST/W002671/1, ST/W002809/1, and EP/V520342/1.

The GBOT programme  uses observations collected at (i) the European Organisation for Astronomical Research in the Southern Hemisphere (ESO) with the VLT Survey Telescope (VST), under ESO programmes
092.B-0165,
093.B-0236,
094.B-0181,
095.B-0046,
096.B-0162,
097.B-0304,
098.B-0030,
099.B-0034,
0100.B-0131,
0101.B-0156,
0102.B-0174, and
0103.B-0165;
%
%
and (ii) the Liverpool Telescope, which is operated on the island of La Palma by Liverpool John Moores University in the Spanish Observatorio del Roque de los Muchachos of the Instituto de Astrof\'{\i}sica de Canarias with financial support from the United Kingdom Science and Technology Facilities Council, and (iii) telescopes of the Las Cumbres Observatory Global Telescope Network.
\end{document}